\documentclass{aa} 

\usepackage[varg]{txfonts}

\usepackage{graphicx,url}
\usepackage{numprint,color}
\usepackage{amsmath,amssymb,verbatim}
\usepackage{textcomp}
\usepackage{xspace}
\usepackage[switch]{lineno} 

\usepackage{booktabs}
\usepackage[normalem]{ulem} 

\usepackage{stfloats}

\newcommand{\avg}[1]{\left< #1 \right>} 
\newcommand{\ud}{\mathrm{d}}

\let\oldAA\AA
\renewcommand{\AA}{\text{\normalfont\oldAA}}

\definecolor{amethyst}{rgb}{0.6, 0.4, 0.8}
\definecolor{green}{rgb}{0.55, 0.71, 0.0}
\definecolor{apricot}{rgb}{0.98, 0.81, 0.69}
\definecolor{auburn}{rgb}{0.43, 0.21,0.1}
\definecolor{babyblueeyes}{rgb}{0.63, 0.79, 0.95}
\definecolor{bittersweet}{rgb}{1.0, 0.44, 0.37}
\definecolor{blue(munsell)}{rgb}{0.0, 0.5, 0.69}
\definecolor{oceanboatblue}{rgb}{0.0, 0.47, 0.75}
\definecolor{brightmaroon}{rgb}{0.76, 0.13, 0.28}

\newcommand{\revone}[1]{{\color{black}{#1}}}
\newcommand{\revtwo}[1]{{\color{black}{#1}}}

\newcommand{\mga}[1]{\textcolor{green}{Markus Gaug: #1}} 
\newcommand{\amo}[1]{\textcolor{amethyst}{Abelardo Moralejo: #1}} 
\newcommand{\jsi}[1]{\textcolor{bittersweet}{Julian Sitarek: #1}} 

\usepackage{epstopdf}
\renewcommand{\t}[1]{\mathrm{#1}} 
\npdecimalsign{.}

\newcommand\ddfrac[2]{\ensuremath{\frac{\displaystyle #1}{\displaystyle #2}}}  

\begin{document}

    \title{Correcting Imaging Atmospheric Cherenkov Telescope data with atmospheric profiles obtained with an elastic light detecting and ranging system}

    \author{F. Schmuckermaier \inst{1}\thanks{email: fschmuck@mpp.mpg.de, markus.gaug@uab.cat}
              \and
              M. Gaug \inst{2}\footnotemark[1] 
              \and 
              C. Fruck \inst{1}
              \and
              A. Moralejo \inst{3}
              \and
              A. Hahn \inst{1}
              \and 
              D. Dominis Prester \inst{4}
              \and 
              D. Dorner \inst{5}
              \and
              L. Font  \inst{2}
              \and
              S. Mi\'canovi\'c \inst{4}
              \and
              R. Mirzoyan \inst{1}
              \and
              D. Paneque \inst{1}
              \and
              L. Pavleti\'c \inst{4}
              \and
              J. Sitarek \inst{6}
              \and 
              M. Will \inst{1}
              }

    \institute{
        Max-Planck-Institut f\"ur Physik, F\"ohringer Ring 6, 80805 M\"unchen, Germany \and
        Departament de F\'sica, Universitat Aut\'onoma de Barcelona and CERES-IEEC, Unitat de F\'sica de les Radiacions, Edifici C3, Campus UAB, 08193 Bellaterra, Spain \and
        Institut de F\'isica d'Altes Energies (IFAE), Edifici Cn, Campus UAB, 08193 Bellaterra, Spain \and
        University of Rijeka, Department of Physics, R. Matej\v{c}i\'c 2, 51000 Rijeka, Croatia \and
        Universit\"at W\"urzburg, Lehrstuhl für Astronomie, Campus Hubland Nord, Emil-Fischer-Straße 31, D-97074 W\"urzburg, Germany\and
        University of Lodz, Faculty of Physics and Applied Informatics, Department of Astrophysics, ul. Pomorska 149/153, 90-236 \L{}\'od\'z, Poland
        }

    \date{Received YYY / Accepted XXX}

    \abstract
    {We are operating an elastic light detecting and ranging system  (LIDAR)  for the monitoring of atmospheric conditions during regular observations of the MAGIC telescopes.}
    {We present and evaluate methods for converting aerosol extinction profiles, obtained with the LIDAR, into corrections of the reconstructed gamma-ray event energy and instrument response functions of Imaging Atmospheric Cherenkov Telescopes.} 
    {We assess the performance of these correction schemes with almost seven years of Crab Nebula data obtained with the MAGIC telescopes under various zenith angles and different aerosol extinction scenarios of Cherenkov light.} 
    {
   The methods enable the reconstruction of data taken under nonoptimal atmospheric conditions with aerosol transmissions down to $\sim$0.65 with systematic uncertainties comparable to those for data taken under optimal conditions. For the first time, the correction of data affected by clouds has been included in the assessment. The data can also be corrected when the transmission is lower than 0.65, but the results are less accurate and suffer from larger systematics.} 
    {} 
     

\titlerunning{Correcting IACT data with an elastic LIDAR}
\authorrunning{F. Schmuckermaier \& M. Gaug}
\maketitle

\section{Introduction}

Imaging Atmospheric Cherenkov Telescopes (IACTs) detect gamma rays from the Universe indirectly, through their interaction with the atmosphere~\citep{hillas,hinton2009,hillas2013}. First, the gamma ray gets converted into an electron-positron pair after interacting with an atomic nucleus of the atmosphere. For a typical atmosphere and vertical incidence, one average gamma-ray interaction length is reached at $\sim$21\,km~a.s.l. (above sea level); about half of the impacting gamma rays are converted to an electron-positron pair below $\sim$24\,km~a.s.l. for a vertical incidence. The so created electrons and positrons radiate themselves through bremsstrahlung, a discrete process in which the electron or positron can lose a large fraction of its energy in single interactions, which produces secondary gamma rays. 
The number of particles in the air shower increases exponentially until the shower maximum is reached. The shower maximum is reached, on average, at atmospheric depths of 250\,g/cm$^2$ to 550\,g/cm$^2$, corresponding to $\sim$12\,km and $\sim$5\,km a.s.l. at a vertical incidence, for gamma-ray energies of 20\,GeV and 100\,TeV, respectively. Electrons and positrons in the shower radiate Cherenkov light, the spectrum of which is flat in terms of photon energy (falling with $\ud N / \ud \lambda \propto 1/\lambda^2$ for photon wavelengths $\lambda$). 
Modern photo multiplier tube (PMT)-based IACTs have a sensitive range from $\sim$2\,eV photon energy ($\sim$600\,nm wavelength) to $\sim$4\,eV ($\sim$300\,nm), within which Cherenkov light is detected. The air-shower development itself depends mainly on the molecular density profile along its trajectory whereas the distribution of Cherenkov light on the ground depends on the related variation in the Cherenkov angle~\citep{bernloehr2000,Munar:2019}. The loss of Cherenkov photons on its way to the IACT is affected by absorbing molecules and, more importantly, by the scattering of the photons out of the camera's field of view (FoV), both by molecular and aerosol scattering\citep{bernloehr2000,garrido2013}. Clouds can be located below the air shower or be in the process of crossing it, but they are usually thinner~\citep{LIDAR1} than the several-kilometer-long particle showers and normally affect only a fraction of the emitted Cherenkov light, only a part of which gets scattered out of the camera FoV of an IACT. 

Imaging Atmospheric Cherenkov Telescopes largely employ Monte Carlo (MC) simulations of electromagnetic showers and the related simulation of the Cherenkov light emitted by the shower particles~\citep{Carmona:2008,Bernloehr:2008,Hassan:2017}. The accuracy of such MC simulations is thus limited by the accuracy with which their input parameters have been previously determined, particularly those related to the details of the individual telescope's instrument response and the atmosphere.

Several current IACTs have adopted the strategy of including slow changes in the atmospheric properties, such as seasonal variations in the molecular profile, into the (CPU-intense) air shower simulations. Correction schemes for fast changes, such as those produced by clouds entering and leaving the FoV of an IACT or ground-layer aerosols, are applied  either to the simulation of Cherenkov light propagation toward the telescope~\citep{nolan2010,connolly:phd,Devin:2019} or directly to data~\citep{dorner2009,Sobczynska:2020}. 

The advantage of modifying the simulation of Cherenkov light propagation resides in an accurate representation of the physical processes involved, whereas a data correction scheme permits more flexibility to account for a wide range of possible aerosol profiles without the need of intense simulation efforts. As we show later, an event-wise correction scheme is possible for the particle shower energies, whereas corrections of effective exposure and energy redistribution functions can only be applied on a statistical basis, within an under-determined set of equations, which require additional assumptions to be consistently solved.

In this paper we present two such data correction schemes based on two different plausible assumptions: (i) a relation between event rates and atmospheric aerosol transmission and (ii) the statistical nature of aerosol extinction profiles and their evolution with time. Both schemes are then compared using a large set of Crab Nebula data taken by the Major Atmospheric Gamma-Ray Imaging Cherenkov (MAGIC) telescopes under nonoptimal atmospheric conditions. 
The MAGIC telescopes are located on the island of La~Palma, a site that exhibits a large variety in aerosol extinction profiles continuously characterized by an elastic light detecting and ranging system~\citep[LIDAR;][]{LIDAR1}. We make intensive use of these LIDAR data to investigate the two data correction schemes and develop upon previous analyses~\citep{Fruck:2015,Schmuckermaier:Atmohead2022}.

\section{Data corrections using the atmospheric transmission profile}
\label{sec:data_correction}

\begin{figure}
\centering
\resizebox{\hsize}{!}{\includegraphics{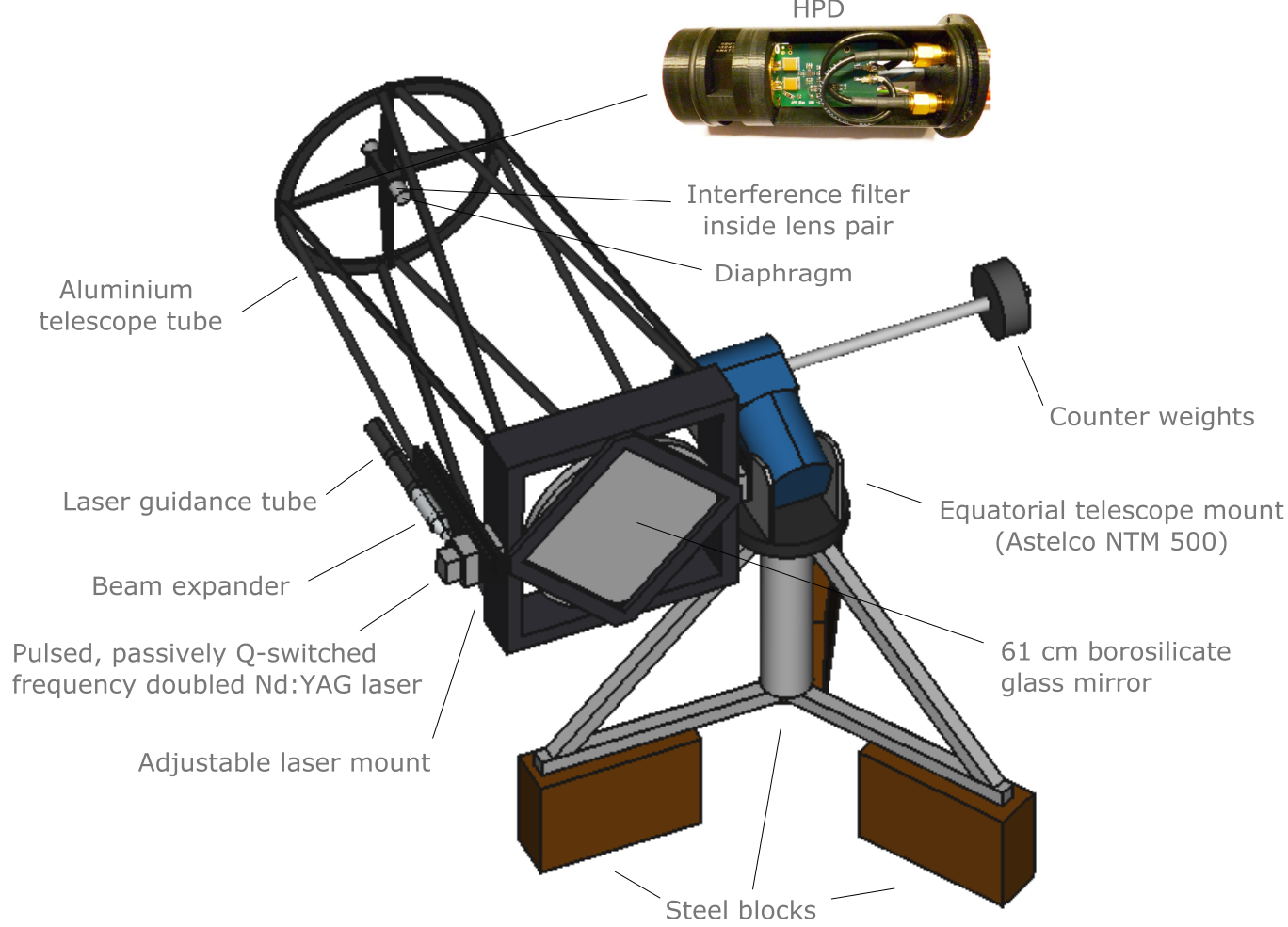}}
\caption{CAD image of the MAGIC LIDAR, showing its hardware components.}
\label{fig:lidar}
\end{figure}

The MAGIC telescopes consist of two IACTs (MAGIC~I and MAGIC~II) located at the Observatorio del Roque de los Muchachos (ORM; 28.762$^\circ$N 17.890$^\circ$W, 2200\,m~a.s.l., corresponding to 800~g/cm$^2$ for vertically incident air showers) on the Canary Island of La Palma. Stereoscopic observations started in 2009 and enabled the detection of gamma rays with energies from about 30$\,$GeV up to $\gtrsim$100$\,$TeV~\citep{aleksic:2016,MAGICCrab100TeV}. 

In order to reduce systematic uncertainties originating from the atmospheric conditions, the MAGIC collaboration operates an elastic LIDAR system on site. A CAD rendering of the LIDAR with all its components is shown in Fig.~\ref{fig:lidar}. The structural base of the LIDAR consists of a commercially available equatorial telescope mount, the Astelco NTM 500. The telescope mount carries a welded aluminum frame functioning as the telescope base, on which all other components are attached. The LIDAR uses a passively Q-switched Nd:YAG laser firing pulses of 25$\,\mu$J energy, operated at a repetition rate of 250$\,$Hz and at a wavelength of 532$\,$nm. The laser beam directly enters a beam expander, which widens the beam width by a factor of 20 to reduce the beam divergence to around 0.6~mrad (full angle). From the beam expander, the laser beam enters a guidance tube containing baffle rings to remove any remaining stray light. The whole laser setup is mounted on a plate that can be adjusted along two axes. The backscattered light gets then  focused onto a diaphragm with an aperture of around 6$\,$mm on the detector module by a 61$\,$cm mirror made from borosilicate glass with a focal length of 150$\,$cm. A lens parallelizes the backscattered light before it passes an interference filter to reduce the light of the night sky background by over a factor of 100. After that, a second lens focuses the light onto a hybrid photo detector (HPD). A Hamamatsu R9792U-40 HPD is used with a quantum efficiency of around 50\% at 532$\,$nm~\citep{HPD_paper}.

A more detailed description of the hardware and data acquisition can be found in~\citet{LIDAR1}. The LIDAR is operated in slave-mode and follows the MAGIC telescope tracking direction within $\sim\!\!5^\circ$ distance in order  not to cross the FoV of MAGIC. Additionally, a veto system is implemented  rejecting  any MAGIC science data event  triggered by the LIDAR during offline analysis. The LIDAR takes quasi-continuous data in intervals of 4~minutes, with an uptime better than 90\% since 2015 until a major mirror upgrade in 2022. The LIDAR is able to characterize tropospheric extinction of green light from ground up to altitudes of more than 20\,km~a.s.l. (for low zenith observations) with a range resolution of $\sim$50\,m (in the lower part of the troposphere) and $\sim$100\,m (in its upper part). The LIDAR has been absolutely calibrated with a combined (statistical and systematic) uncertainty of $<$4\% for the system constant and hence an uncertainty of  only $\pm$0.02 for the aerosol optical depth of complete layers to ground. Only for light emission points {within} an extended aerosol or cloud layer can the uncertainty on the aerosol transmission from that point to the base of the very same layer be larger, due to the often missing knowledge of the ratio of extinction to backscatter coefficient, the so-called LIDAR ratio (LR), within the layer (see~\citet{LIDAR1}). Any additional uncertainty in such cases scales with the uncertainty of the employed LR and can be estimated to be at most 10\% of the total transmission of the layer. Further details on the performance of the LIDAR analysis are available in~\citet{LIDAR1}.

\begin{figure}
\centering
\resizebox{\hsize}{!}{\includegraphics{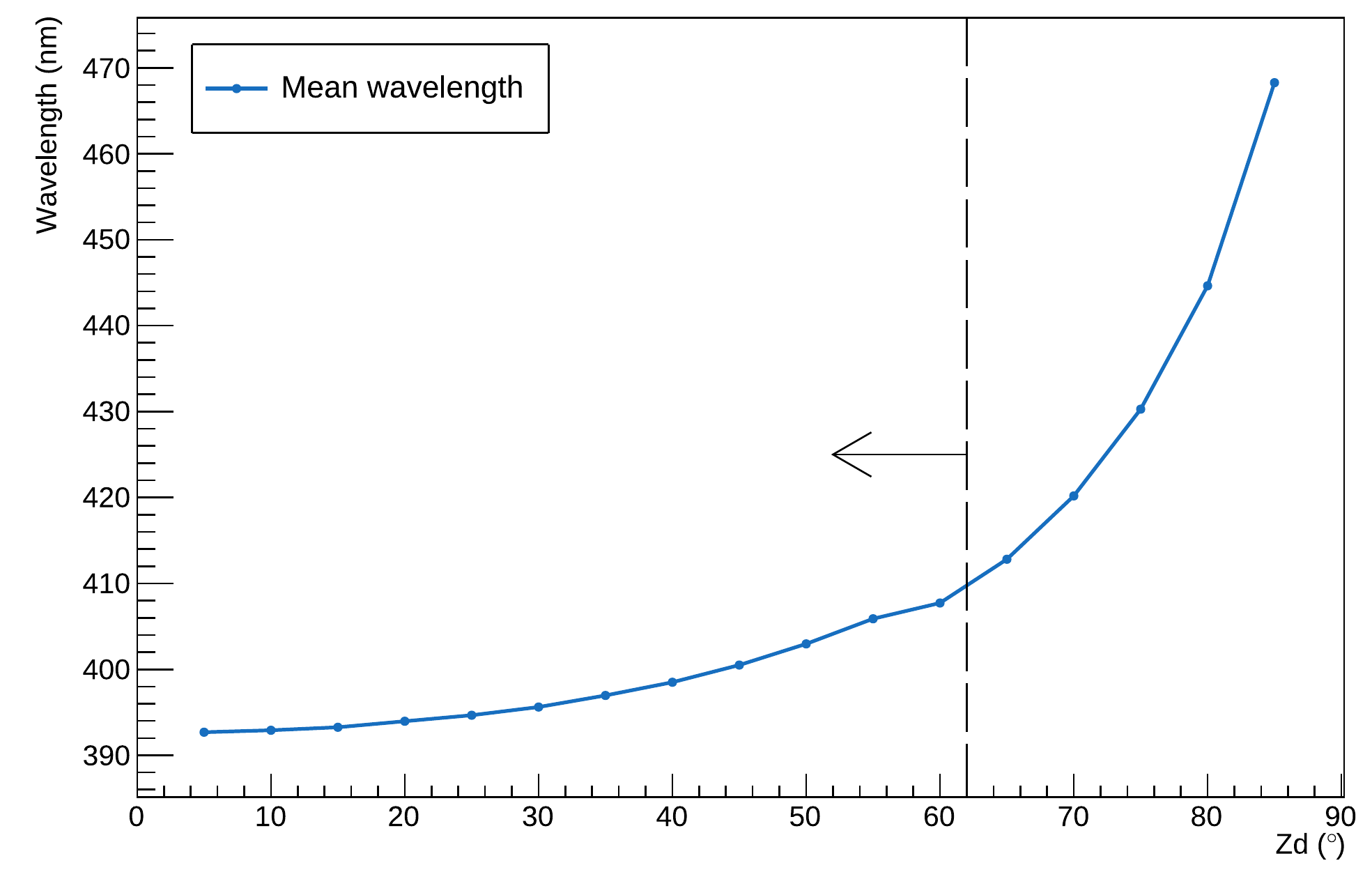}}
\caption{Average Cherenkov wavelength, $\overline{\lambda}_\mathrm{Cher}$, of air showers at given zenith angles, \textit{Zd}, detected by the MAGIC~I camera, estimated from MC simulations. The simulations include clear-night atmospheric extinction, wavelength-dependent mirror reflectivity, and PMT quantum efficiency. The dashed line and arrow mark the zenith range for which LIDAR corrections are applied in the MAGIC standard analysis chain.}
\label{fig:mean_cherenkov_vs_zd}
\end{figure}

The LIDAR analysis provides aerosol extinction profiles $\alpha_\mathrm{LIDAR}(r)$ as a function of distance $r$, for the 532\,nm laser wavelength. These profiles require a correction for the fact that the wavelength spectrum of detected Cherenkov light is not centered at the LIDAR wavelength, but instead shows a strong contribution of UV light. For the extinction $\alpha_\mathrm{\overline{\lambda}_\mathrm{Cher}}$, expected for an average Cherenkov light wavelength, $\overline{\lambda}_\mathrm{Cher}$, we can use the following scaling relation~\citep{angstrom:1929}:
\begin{align}
    \alpha_\mathrm{\overline{\lambda}_\mathrm{Cher}} = \alpha_\mathrm{LIDAR} \cdot \left( \ddfrac{\overline{\lambda}_\mathrm{Cher}}{532~\mathrm{nm}}  \right)^{-\AA} \qquad,
\end{align}
\noindent
where \textit{\AA} denotes the \AA ngstr\"om exponent, which depends on the given aerosol conditions. 

\begin{figure}
\centering
\resizebox{\hsize}{!}{\includegraphics{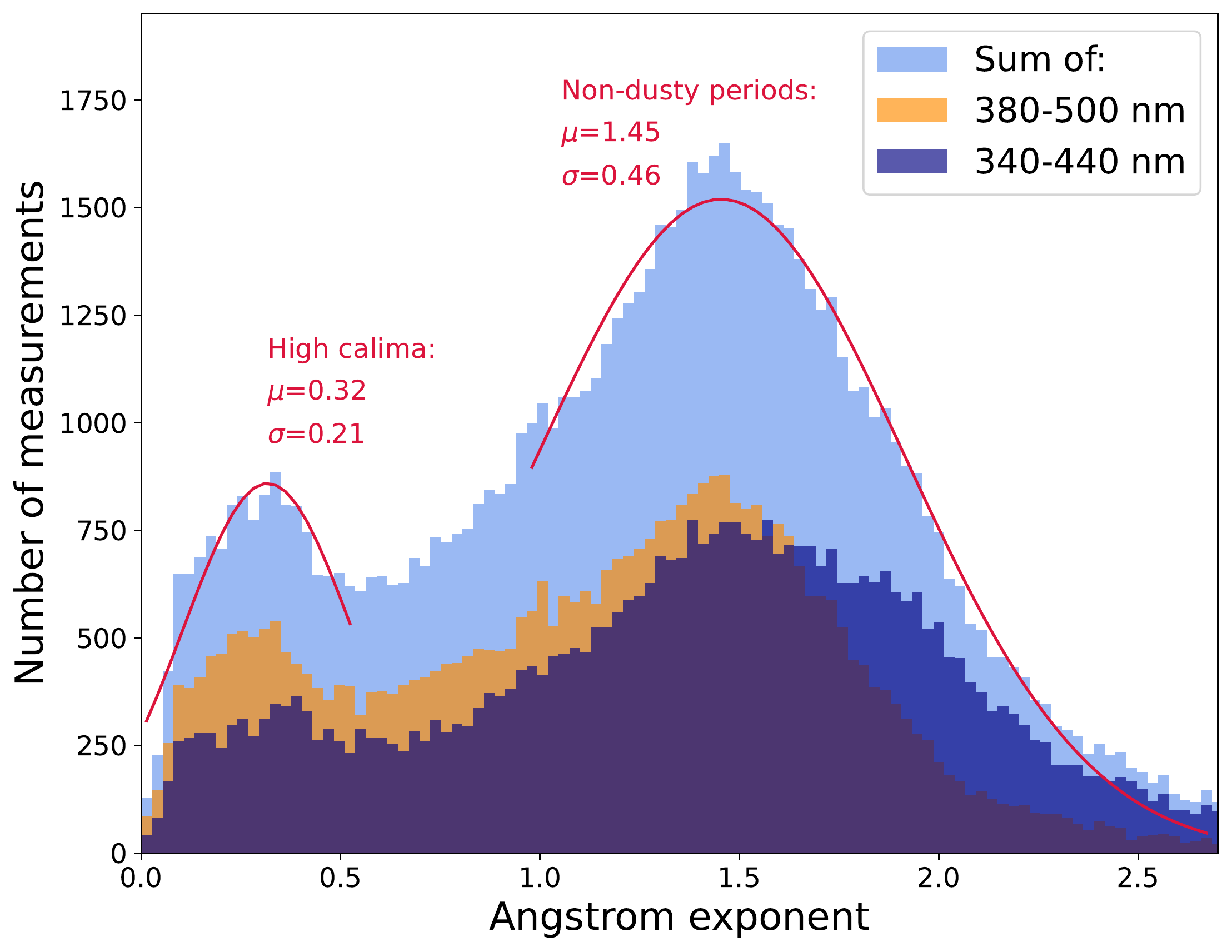}}
\caption{\AA ngstr\"om exponents obtained from the AERONET station at the ORM for two different wavelength regions, as well as the combined histogram fitted with Gaussian distributions. The data cover the period between November 2019 and July 2022.}
\label{fig:angstrom}
\end{figure}

The spectrum of Cherenkov light is flat in photon energy. After convolution with the effective optical bandwidth of the telescope~\citep{GaugMuons:2019}, containing contributions from the mirror reflectance,
the efficiency of light concentrators\footnote{The transparency of the protective camera window has a negligible effect on the distribution of Cherenkov photon wavelengths \citep[see][]{sultanova_dispersion_2009} and has therefore not been included in this study.} and the photon detection efficiency of the used photomultiplier tubes~\citep{Borla:2009}, an average wavelength for the {detected} Cherenkov light of $\overline{\lambda} \approx 400$\,nm is obtained with a slight dependence on the observation zenith angle (see Fig.~\ref{fig:mean_cherenkov_vs_zd}).  
The Aerosol Robotic Network project (AERONET)\footnote{see \url{http://aeronet.gsfc.nasa.gov/}} Sun photometer data inversion products from the location {Roque\_Muchachos} (28.764$^\circ$N, 17.894$^\circ$W, located at a distance of $\sim$400~m from the MAGIC LIDAR and at comparable altitude) provide diurnal \AA ngstr\"om exponents for the wavelength ranges 380\,nm to 500\,nm and 340\,nm to 440\,nm. Data from November 2019 to today are available. Figure~\ref{fig:angstrom} shows the histograms of the obtained \AA ngstr\"om exponents for both ranges. Both histograms  show a bimodal distribution. The first peak corresponds to conditions of higher dust content, especially Saharan dust intrusions locally known as {calima}~\citep{rodrigues-desertdust}. These periods show almost gray (corresponding to $\AA = 0$) extinction~\citep{whittet1987,Sicard:2010} and with that a rather low value of the \AA ngstr\"om exponent, namely $\AA \approx 0.32\pm0.21$. For periods of higher dust content we therefore adopt the following correction scheme:
\begin{align}
\begin{split}
    \alpha_\mathrm{\overline{\lambda}_\mathrm{Cher},\mathrm{calima}} &= \alpha_\mathrm{LIDAR} \cdot \left( \ddfrac{400~\mathrm{nm}}{532~\mathrm{nm}}  \right)^{-(0.32\pm 0.21)} \\
    &= \alpha_\mathrm{LIDAR} \cdot (1.10 \pm 0.07)\qquad.
\end{split}
\end{align}
For non-dusty periods, represented by the second peak in Fig.~\ref{fig:angstrom}, the obtained \AA ngstr\"om exponent values are higher, centered at $\AA \approx 1.45\pm0.46$. The resulting correction is given by\begin{align}
\begin{split}
    \alpha_\mathrm{\overline{\lambda}_\mathrm{Cher},\mathrm{non-dusty}} &= \alpha_\mathrm{LIDAR} \cdot \left( \ddfrac{400~\mathrm{nm}}{532~\mathrm{nm}}  \right)^{-(1.45\pm 0.46)}\\
    &= \alpha_\mathrm{LIDAR} \cdot (1.51 \pm 0.20)\qquad.
\end{split}
\end{align}

 The shown \AA ngstr\"om exponents have been obtained, however, during daytime, whereas it is known~\citep{rodriguez2009} that non-dusty aerosol sizes {decrease} at night. The used values should be considered, therefore, a lower limit to their nighttime characteristics.
 
In order to distinguish between both cases, calima and non-dusty periods, we adopt the criterion used in~\citet{LIDAR1}: a total ground layer aerosol transmission at 532~nm of less than 0.93 gets categorized as calima, whereas the rest is considered a clear night. 

It should be noted that the wavelength corrections introduce an additional systematic uncertainty on the order of 10\% on the aerosol optical depth. This translates to an uncertainty of $\sim$0.5\% for the ground layer transmission on clear nights and $\sim$2\% for the case of nights affected by calima. 

Optically thin cirrus clouds, such as those frequently observed with the MAGIC LIDAR, have particle sizes approaching the geometric optics regime, and hence the spectral dependence of the extinction coefficients will vanish~\citep{Vaughan:2010}. In the following, we assumed that no wavelength correction was necessary for the sample of cirrus found in our LIDAR data sample.

\subsection{Interpolation of LIDAR aerosol transmission profiles}

The MAGIC LIDAR is pointing ahead of the observed FoV of the MAGIC telescopes by an angular offset of $\sim$5$^\circ$. The aerosol extinction profiles, $\alpha_\mathrm{\overline{\lambda}_\mathrm{Cher}}(r)$, are evaluated as a function of the range, $r$, and then converted to integrated aerosol transmissions: 
\begin{align}
\begin{split}
  T_\mathrm{aer}(r) &= \exp\left(-\int_0^{r}
  \alpha_\mathrm{\overline{\lambda}_\mathrm{Cher}}
  (r')\, \ud r' \right) \equiv T_\mathrm{aer}(h,\theta) \\
  &= \exp\left(-\int_0^{h/\cos\theta}
  \alpha_\mathrm{\overline{\lambda}_\mathrm{Cher}}
  (h') /\cos\theta\,\ud h' \right)\quad,
\end{split}
\end{align}
\noindent
where $\theta$ is the zenith angle of the LIDAR pointing. 

\begin{figure}
    \centering
    \resizebox{\hsize}{!}{\includegraphics{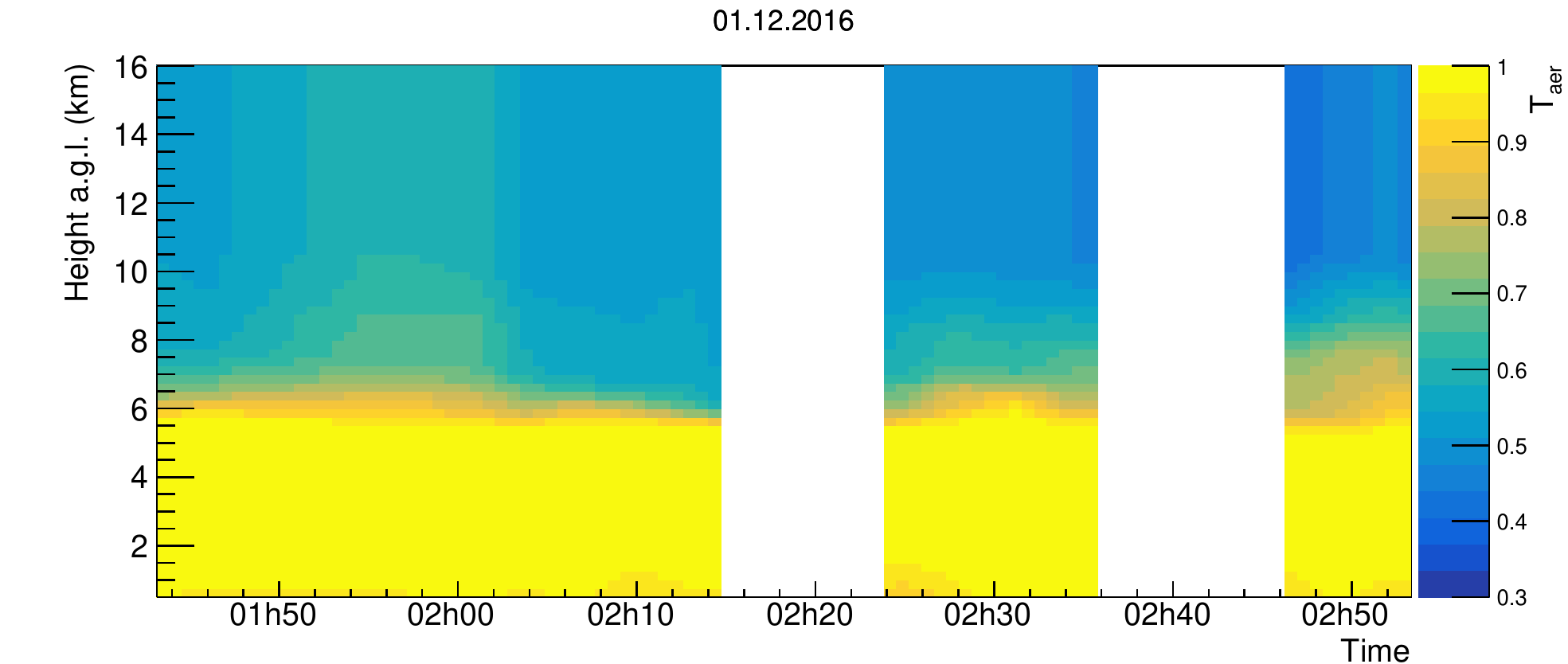}}  \resizebox{\hsize}{!}{\includegraphics{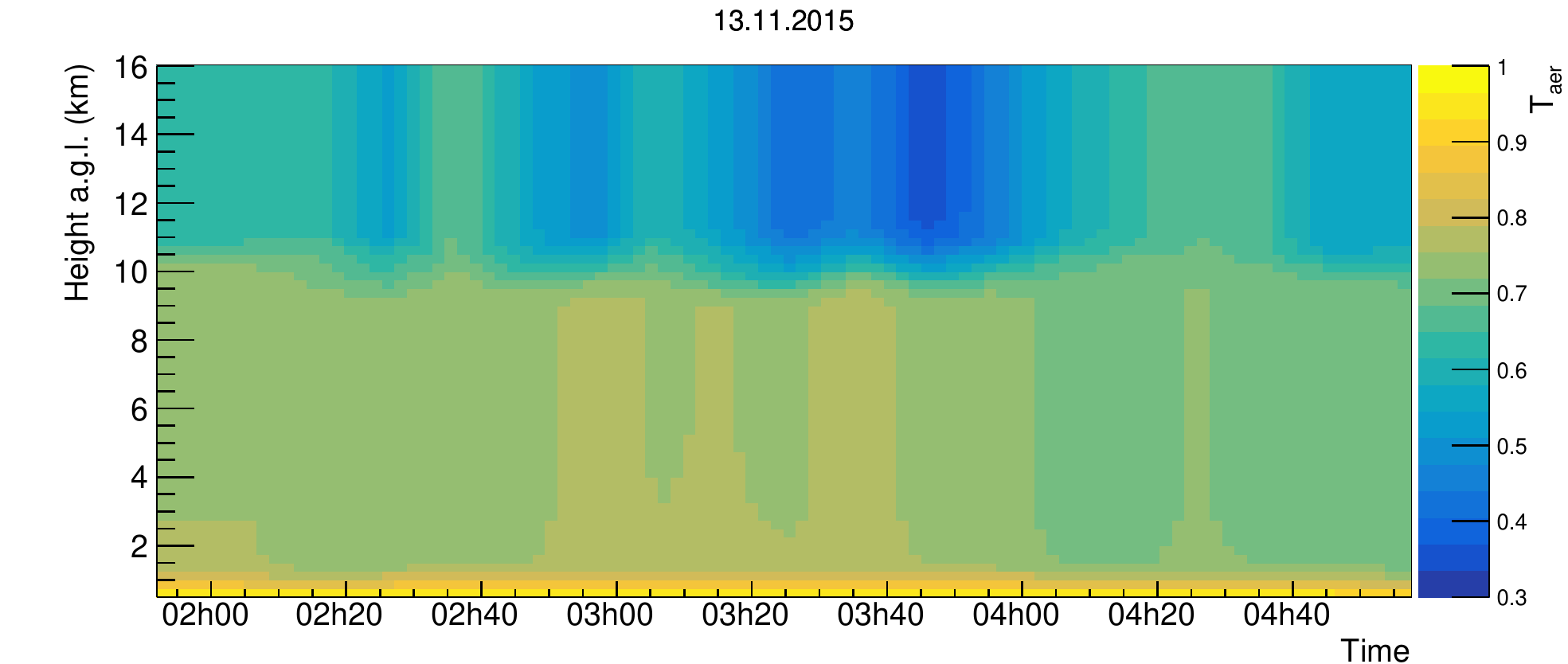}}
    \caption{Interpolated aerosol transmission profile, $T_\mathrm{aer}$, as a function of height and time. Top: Case of intermittent clouds above 5~km above ground. Bottom:  Strong calima in the first kilometer above ground accompanied by a few cirrus clouds above 10~km. The white spaces represent times when the MAGIC telescopes were not observing.}
    \label{fig:TransVsTime}
\end{figure}

Subsequently, the obtained aerosol transmissions, $T_\mathrm{aer}(h)$, are divided into bins of equidistant height, $h$, with a width of 250~m. Every altitude bin gets eventually interpolated in time. A LIDAR profile is available every 4~minutes under normal operation. The transmission profiles are extrapolated in time for MAGIC science observation periods up to 15~minutes outside the time range covered by LIDAR. 

In this manner, an aerosol transmission profile can be assigned to each individual gamma-ray event detected by MAGIC (see Fig.~\ref{fig:TransVsTime} for an example).



\subsection{Energy correction}
\label{data_energy}     

In a first-order approximation, the amount of Cherenkov light reaching the IACT scales linearly with the gamma-ray energy for a given impact parameter~\citep{hillas}. The relative energy bias, caused by aerosol extinction, can therefore be assumed to be proportional to $T_\mathrm{aer}(h)$, where $h$ can be taken approximately as the altitude of the air shower. 

However, it is not sufficient to only scale up the energy of individual air shower events to obtain the correct energy spectrum, since the instrument response functions (IRFs) also also have a strong dependence on energy~\citep{aleksic:2016}. Therefore, the correction of the energy spectrum is performed in two steps: (i) through an event-wise correction of the reconstructed shower energies and (ii) through an \revone{energy-}bin-wise correction of the effective collection areas and energy migration matrices of the system. 

\begin{figure}
  \centering
  \resizebox{\hsize}{!}{\includegraphics{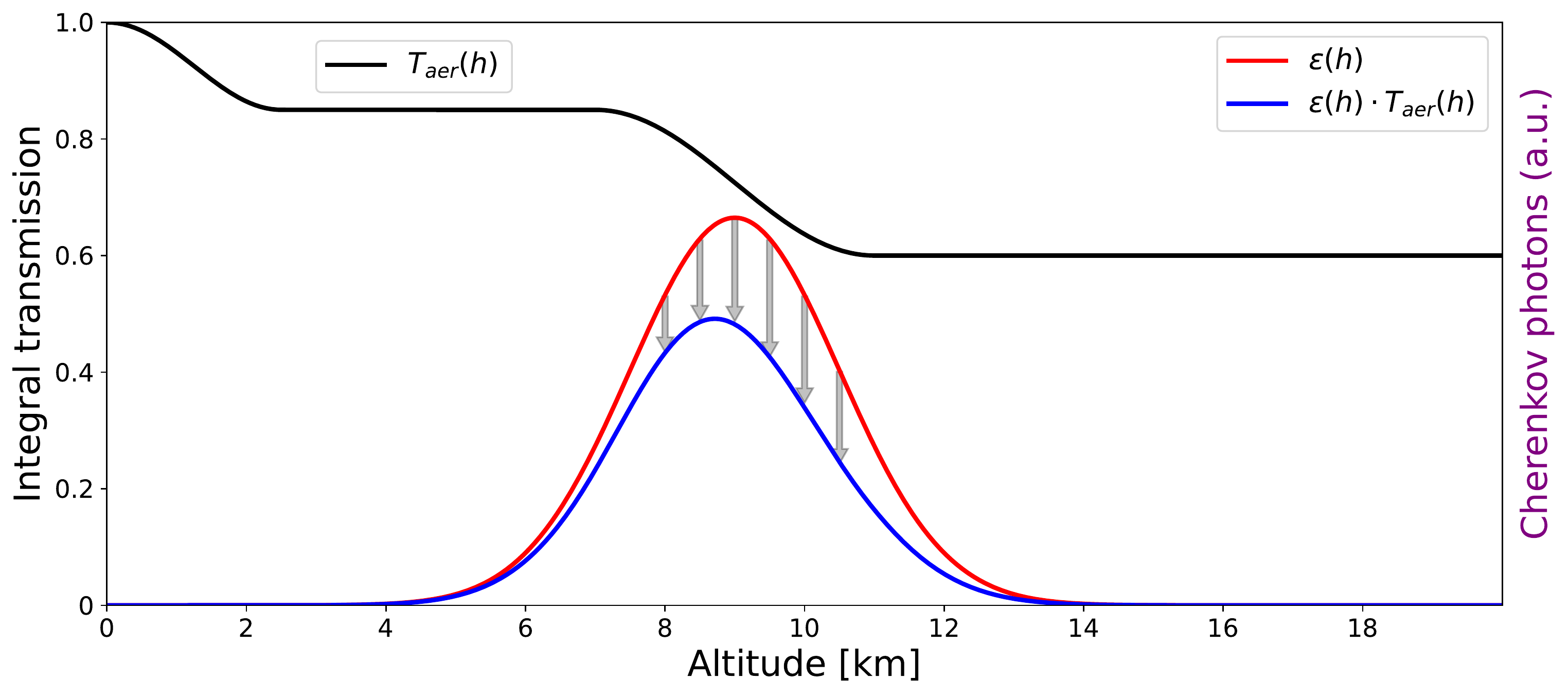}}
  \caption{Sketch of the averaging procedure of the aerosol optical depth, $T_\mathrm{aer}$, over an assumed air-shower light emission profile. In this case, some aerosol extinction due to the ground layer is shown, and a cloud is located at a typical height of 9~km above ground. The left vertical axis shows $T_\mathrm{aer}$, and the right axis denotes the normalized distribution, $\epsilon(h)$, of the number of Cherenkov photons that can reach the IACT telescope and camera. In red, $\epsilon(h)$ is shown and in blue, the product of $\epsilon(h)\cdot T_\mathrm{aer}(h)$. }
  \label{fig-6}       
\end{figure}

Under normal circumstances, and especially in the presence of cloud layers, distinct parts of the Cherenkov light emitting particle shower lie within, above, or below the attenuating layer of particulates. In the case of layers of low and moderate optical depth, it is adequate~\citep{garrido2013,doro2014} to simply scale the reconstructed shower energy with the inverse average transmission seen by the Cherenkov light emitted along the shower: 
\begin{equation}
  \overline{T}_\mathrm{aer} = \int_{0}^{\infty}{\epsilon(h) \cdot T_\mathrm{aer}(h) \ \mathrm{d} h}\quad .
  \label{tau_aer}
\end{equation}
The normalized emission profile of the observable Cherenkov light from each shower, $\epsilon(r)$, is estimated using an energy-dependent Gaussian longitudinal extension \revone{of the emission profile of the subset of those Cherenkov photons, which are detected by the telescopes and end up becoming part of the analyzed shower image~\citep[see, e.g., Fig.~F1 of][]{fruck:phd}}. \revone{The mean emission heights (Gaussian parameter $\mu$)} correspond to the respective reconstructed heights of each shower maximum~\citep{Fegan:1997,aleksic:2016}, obtained from stereo reconstruction, and the width, $\sigma$, has been obtained from toy MC air-shower simulations: 
\begin{align}
    \sigma &= 3.6 - 0.4\log_{10}\left(E_\mathrm{est} [\mathrm{GeV}] \right) \cdot \left(\cos\theta\right)^{0.15}~\mathrm{km} \quad, \label{eq:sigma} 
\end{align}
\noindent
where 
$\theta$ is the observation zenith angle of the telescopes~\citep[see Fig.~F3 of][]{fruck:phd}. \revone{The effect of an aerosol transmission profile folded with  the emission profile of Cherenkov light is sketched in Fig.~\ref{fig-6}.}


By scaling the old estimate, $E_\t{est}$, by the inverse of the average optical depth, the new energy estimate, $E_\mathrm{corr}$, is obtained:
\begin{equation}
  E_\t{corr} = \frac{E_\t{est}}{\overline{T}_\mathrm{aer}}\, . \label{eq:Ebiascorr}
\end{equation}
Since the width of the longitudinal profile depends on the reconstructed energy, this process is iterated until convergence is achieved. 
\revone{It has been shown \citep{Sobczynska:2020} that Eq.~\ref{eq:Ebiascorr} can produce unbiased energy estimates for $\overline{T}_\mathrm{aer} \gtrsim 0.5$.}


\subsection{Instrument response function corrections - Method I}
\label{sec:method1}
\label{data_area}   

 By modifying the energy estimation of an air shower event, the (energy-dependent) IRFs -- effective collection area and energy redistribution probability density functions -- need to be evaluated at a respectively corrected energy. In the case of IACTs, the effective collection area in particular shows a very strong energy dependence below about 100\,GeV  and a moderate one above for low zenith angles~\citep{aleksic:2016}. 
 
\begin{figure}
  \centering
  \resizebox{\hsize}{!}{\includegraphics{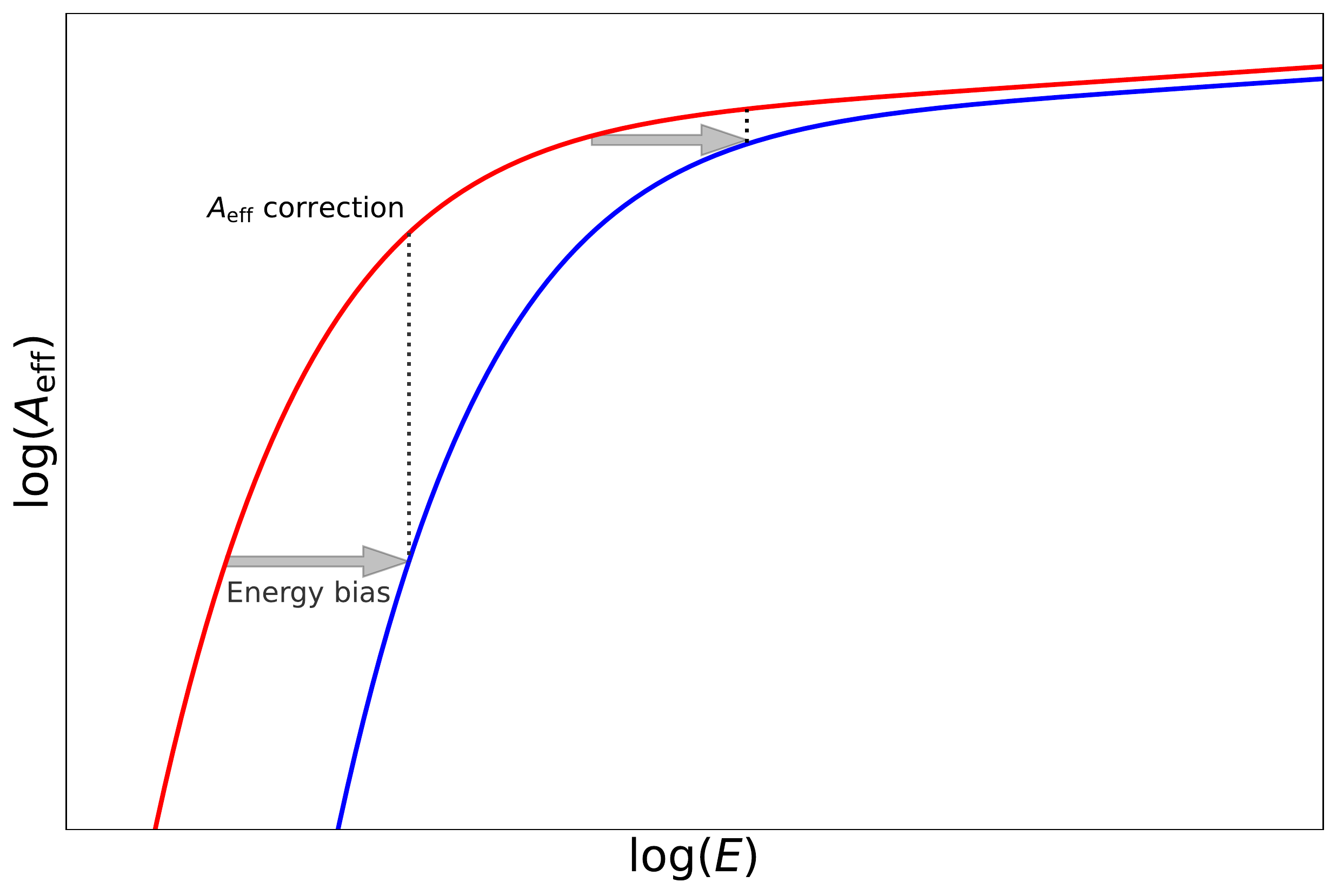}}
  \caption{Illustration of the modification of the effective collection area \revone{due to strong aerosol extinction} as a function of energy. The red line corresponds to the collection area expected for clear nights, whereas the blue line corresponds to the model for unfavorable aerosol conditions. The gray arrows labeled "energy bias" indicate the shift in energy for two example energies. }
  \label{fig-7}       
\end{figure}

 During spectrum reconstruction, IRFs are evaluated in energy bins. Being used to reconstruct gamma-ray fluxes, an {event-wise IRF} is a priori undefined. Nevertheless, atmospheric conditions, and particularly clouds, may change on timescales much faster than the typical duration of an analysis period, which makes it desirable to attribute (different) IRFs to each event, or at least, reshuffle each event into an energy bin that provides the best approximation to its concurrent IRF. That energy bin does not need to be necessarily the one of the corrected energy. Quite on the contrary, we adopted a model according to which 
air showers of energy $E$ during nonoptimal atmospheric conditions 
produce images in our telescopes that resemble those produced by 
showers of energy \revone{$E\cdot\overline{T}_\mathrm{aer}$} on a clear, aerosol-less night.
 That assumption is illustrated for the effective collection area in Fig.~\ref{fig-7}. 
 \revtwo{The gray arrows represent the introduced shift in energy.} \revone{We note that the length of the two arrows is different for Method I}. $\overline{T}_\mathrm{aer}$ is itself also energy-dependent (through the energy-dependent height of shower development relative to the absorbing layer), \revone{which makes the corrected effective collection area (blue curve) in Fig.~\ref{fig-7} not simply a constant shift (for all showers) of the uncorrected effective area (red curve) along the energy axis}. \revone{In Method II, a constant average shift is used. The shift results in a change of effective area shown by the dotted line.} We recall that only the clear-night IRFs are produced by MC simulations here and $\overline{T}_\mathrm{aer}$, and hence also the corrected IRFs, obtained from real data on an event-wise basis. 
 

Furthermore, we do not track the effective on-time that applies to each of the continuously changing aerosol atmospheres. One could make such an effort and split the data set into time bins of equal or approximately equal aerosol conditions and trace the effective on-time that the telescopes have spent in each of the respective "atmospheres." Such a strategy will be pursued for the future Cherenkov Telescope Array~\citep[CTA;][]{Gaug:2017Atmo}.

Instead, we exploit here the assumption that the effective area \revtwo{for gamma-ray events} in each energy bin scales roughly with the event rate observed in that energy bin. This assumption, however, is questionable due to the fact that event rates in IACTs are normally background-dominated, even after cuts. However, there are strong indications that cosmic-ray-dominated trigger rates show a similar behavior~\citep{Hahn2014,CTC:2019}. 
Accordingly, we can assume that the {event rate after cuts} of an IACT, which now contains signal and background, 
scales with the (gamma-ray) effective area, at least if the spectral index of the observed gamma-ray source is not too far from the background (which is the case for the Crab Nebula), and if the behavior of gamma-like background events is sufficiently similar to the signal. An {a posteriori} validation of these assumptions will be made later through the comparison of spectral reconstruction of Crab Nebula data taken during nonoptimal atmospheric conditions with that from clear nights. 



We wanted to calculate an average exposure, $\avg{A(E,t)\cdot t_\mathrm{eff}(t)}$, in bins of energy, where the effective area changes over time and where its dwell time, $t_\mathrm{eff}(t),$ in each state depends on how fast the atmospheric conditions change. 
Hence, 

\begin{align}
    \avg{A(E_i)} &= \ddfrac{\int_0^T A(E_i,t) \ud t}{\int_0^T \ud t} 
    = \ddfrac{\int_0^T A(E_i,t) \ud t}{\int_0^T A(E_i,t) / A(E_i,t) \ud t} \nonumber\\  & \xrightarrow{A \propto \ud N_i/\ud t} = \ddfrac{\int_0^T \ddfrac{\ud N_i}{\ud t} \ud t}{\int_0^T \ddfrac{\ud N_i}{\ud t} / A(E_i,t) \ud t} = \ddfrac{N_i}{\sum_{j=0}^{N_i} \ddfrac{1}{A_j(E_i)}} \quad, \label{eq:aeffaverage} 
\end{align}
\noindent
where $T$ is the overall effective on-time of the entire sample. The index $i$ denotes the energy bin investigated. \revone{The energy within bin $i$,  $E_i$, follows the Lafferty and Wyatt prescription~\citep{1995NIMPA.355..541L}, that is to say, it corresponds to the energy at which an assumed spectral shape has the same value as its average over the bin. The width of the energy bins should not be too small to avoid insignificant excess events in a bin, but also not too large to unnecessarily affect accuracy. Some discussion on the bin width in the following unfolding process is given in~\citet{unfolding}. Conclusions found there are also applicable to this correction method.} $N_i$ are the number of events accumulated in the given energy bin, while $j$ runs over the individual events. The effective area, $A_j$, is now evaluated at the time (and corresponding aerosol conditions) of a given event.

Again, we note that the events $j$ are composed of signal and background after cuts. Constant signal-to-background rates throughout $T$ are assumed, that is, signal and background rates are assumed to be equally affected by changes in pointing and by  atmospheric variations. Different behavior  introduces a systematic error, which will be quantified at the end of this paper.

Averaging of the inverse of $A$, instead of the $A$ may sound odd at first glance, and stems from the fact that the \revone{time between subsequent events} is large when the event rate is {small}. \revone{Reduced aerosol transmission removes events with a small image size: those found at or below the energy threshold, but also events at higher energy with large impact distances~\citep{garrido2013,MAGICmoon}}. 
 Alternative derivations of Eq.~\ref{eq:aeffaverage} can be found in~\citet{fruck2013,Fruck:2015}. 

In the following, we evaluate the 
time-dependent collection area, $A$, at energy $E_\mathrm{corr}\cdot\overline{T}_\mathrm{aer}$ \revone{for an event, $j$, falling into the energy bin $i$:}
\begin{align}
A_j(E_i) &= A\left(E_\mathrm{corr,i}\cdot\overline{T}_{\mathrm{aer},j} \right)\, ,
\label{eq:aeffcorr}
\end{align}
\noindent
where the respective actual aerosol atmosphere found during event $j$ is encapsulated
in the term $\overline{T}_{\mathrm{aer},j}$
(supposing that the energy correction $E_\mathrm{corr,i}$ is unbiased on average).
In case of binned energies and effective areas, the
event-averaged collection area is then
\begin{align} 
  \avg{A_{i}} = \frac{ N_i }{ \sum_{j=0}^{N_i}{\frac{1}{A_{i-\delta_j}}} }\, .
  \label{eq:aeff}
\end{align}
Here, $\delta_j$ denotes the energy bin-shift produced by
the factor $(1-\overline{T}_{\mathrm{aer},j})$,
which itself  gets rounded to an integer number of energy bins. Averaging of the effective area over zenith angles is made implicitly in this approach.
A detailed description of the subsequent unfolding process performed is given in  Appendix~\ref{sec:unfolding}.
 Method~I was the standard LIDAR correction method of the MAGIC Collaboration until 2018, when support for this method was discontinued in favor of Method~II. 

\subsection{Instrument response function corrections - Method II}
\label{sec:method2}

The method was introduced in 2018 and uses the same event-wise energy correction shown in \revone{Eq.~\ref{eq:Ebiascorr}}. For the sample to be analyzed, we create a histogram of the event-wise correction factors weighted by the elapsed time. The histogram indicates how long the atmosphere was under conditions where the energy correction factor (averaged for gamma-like events of all energies) had a certain value, $\overline{T}_\mathrm{aer}$. The histogram is also done in bins of the zenith angle $\theta$. From $t_\mathrm{elapsed}$ in bins of $\overline{T}_\mathrm{aer}$ and $\theta$ we can obtain the average corrected effective area for a given zenith angle as
\begin{equation}
\langle A_\mathrm{eff}(E_\mathrm{true}, \theta) \rangle  = \frac{\sum_{i} A_\mathrm{eff} (E_\mathrm{true} \cdot \overline{T}_{\mathrm{aer}, i}, \theta) \cdot t_{\mathrm{elapsed}, i}(\theta)}{\sum_{i} t_{\mathrm{elapsed}, i}(\theta)}
\label{eq:method2Aeff}
,\end{equation}
where the sum is over the bins  $\overline{T}_{\mathrm{aer},i}$. Thus, it is the average of the energy-shifted effective areas weighted by the time spent in each observation condition. Using elapsed time (instead of effective time) simplifies the calculation, and makes no real difference, since the dead time fraction is  only $\sim$1\% and similar for all data~\citep{magicperformance1}. Finally, averaging in zenith bins gives the average effective area for the entire sample\footnote{While in principle it is better to analyze the data in small time chunks in which observation conditions (e.g., zenith angle or atmospheric transparency) have only small variations, integration over longer times is often convenient to collect enough statistics on the real data.}. 
We note, however, that different combinations of shower energy and cloud layer may end up in the same $\overline{T}_{\mathrm{aer},i}$ bin. For instance, an opaque cloud at high altitudes will absorb only a small fraction of the Cherenkov light of a shower of high energy developing below the cloud~\citep{sobczynska_analysis_2020}, whereas a low, but almost transparent cloud may affect all the Cherenkov light emitted by a shower of lower energy. Nevertheless, both cases are used to scale the energy, at which the effective is evaluated, by a same factor $\overline{T}_{\mathrm{aer},i}$, for all energy bins. The discrepancy can introduce a systematic error, particularly in the case of rapidly changing cloud conditions.

An analogous computation was performed to obtain the average-corrected energy migration matrix. Since the reconstructed energies of real events are already corrected for the aerosol-induced bias, in this averaging the original migration matrix ($E_\mathrm{est}$ vs. $E_\mathrm{true}$) is shifted along the diagonal (toward higher energies) by the factors $1/\overline{T}_\mathrm{aer, i}$, obtaining a new migration matrix ($E_\mathrm{corr}$ vs. $E_\mathrm{true}$). The underlying assumption is that the reconstruction quality of events of energy $E_\mathrm{true}$ is analogous to that of events of energy $\overline{T}_\mathrm{aer, i} \cdot E_\mathrm{true}$ observed under the clear atmospheric conditions used for the MC simulation.
The IRFs averaged in this way can be used for the higher-level analysis of the LIDAR-corrected data in the same way as normal IRFs are used in the analysis of data taken under optimal atmospheric conditions. 

The main simplification of this 
method is the assumption that 
the effective area and migration matrix can be shifted by a global correction factor, namely the average energy correction applied for a given atmospheric condition. It does not take the proper energy dependence of such shifts into account. Therefore, the method should work better the closer the aerosol layers come to the telescopes, because in that case Cherenkov light from showers of all energies suffers the same extinction. 

\section{Evaluation of the performance of the LIDAR-based corrections}

\subsection{The data set used to scrutinize the correction methods}

In order to evaluate the performance of the previously outlined correction algorithms, data from the Crab Nebula taken under optimal as well as suboptimal atmospheric conditions were analyzed. The Crab Nebula normally shows a very bright and stable emission across several decades of high photon energy~\citep{Abdo:2010,aleksic:2015,JourdainCrabNebula:2020} and is considered a standard candle for many energy regimes including gamma rays. Very recently, however, larger and smaller flares have been detected above 100~MeV with high significance~\citep{Tavani:2011,Rudy:2015,Arakawa:2020}, lasting several days up to one month and are probably related with activity of the inner knot~\citep{Tziamtzis:2009,Moran:2013_2,Rudy:2015} and/or the peel off of wisps from the central pulsar region~\citep{Schweizer:2013}.

Apart from the short-lasting possible flares \citep[hitherto undetectable in the very-high energy regime; see][]{Abramowski:2014,Scherpenberg:2019}, the observed spectra by MAGIC do not vary significantly more than what can be attributed to the instrument's systematic uncertainties~\citep{aleksic:2016} \revone{and are therefore assumed to be stable in the context of this study.} Hence,  the reconstructed spectrum should not depend on the aerosol conditions or the observational zenith angle, if correctly accounted for. 

In this work, Crab data from mid 2013 until early 2020 were analyzed. The Crab Nebula can only be observed with MAGIC from September to April. The majority of the data were hence taken during the colder winter months, when the influence of clouds above the MAGIC site is stronger. The phenomenon of Saharan dust intrusions ("calima") usually, but no exclusively, occurs from July until September and is therefore under-represented in our data set.

The data set includes observations made at zenith angles between 5$^\circ$ and 62$^\circ$. At higher zeniths, it becomes increasingly difficult for the LIDAR to reach the distance ranges required, necessitating an alternative measurement of atmospheric extinction~\citep{VLZA:2020}. Because of the very different systematics, we decided not to consider these "very large zenith angle" data sets. Also, only data taken under dark conditions have been used, corresponding to the night sky background levels 1-2 as defined in~\cite{2017APh....94...29A}. The data were cleaned and classified based on the aerosol transmission up to an altitude of 9$\,$km above ground, $T_{9 \, \text{km}}$. The maximum emission of Cherenkov light from atmospheric air showers for vertical 100\,GeV gamma-rays is found at an altitude of close to 9 km above sea level~\citep{bernloehr2000}. Since the majority of Cherenkov light will be produced below 9$\,$km above the ORM, aerosol conditions at higher altitudes will barely affect the recorded gamma-ray events.

\begin{table*}
 \centering
 \caption{Available data (in hours) for all included analysis periods and transmission bins.}
\begin{tabular}{l|l|lllll}
\multicolumn{1}{c|}{Period tag} &  \multicolumn{1}{c|}{Time period} &\multicolumn{1}{c}{$T_{9 \, \text{km}}$ \textgreater 0.95}    &\multicolumn{1}{c}{\textgreater 0.9}                & \multicolumn{1}{c}{0.9 - 0.82} & \multicolumn{1}{c}{0.82 - 0.65} & \multicolumn{1}{c}{0.65 - 0.5} \\ \toprule
ST.03.03   & 27.07.2013 -- 05.08.2014$^*$ &  42.4 & 39.1 & 1.6 & 0.7 & 1.1 \\
ST.03.05   & 31.08.2014 -- 22.11.2014     &  5.7 & 10.1 & 1.8 & 0.3 & 0.5\\
ST.03.06   & 24.11.2014 -- 28.04.2016     &  58.6 & 70.1 & 7.6 & 5.6 & 5.4\\
ST.03.07   & 29.04.2016 -- 02.08.2017     & 21.3 & 31.5 &  3.8 & 2.4 & 1.2\\
ST.03.09   & 10.11.2017 -- 29.06.2018     &  9.3 & 14.3 & 2.6 & 0.2 & 0\\
ST.03.10   & 30.06.2018 -- 30.10.2018     &  6.3 & 7.4 & 0 & 0.1 & 0\\
ST.03.11   & 01.11.2018 -- 15.09.2019     & 19.2 & 20.6 & 1.2 & 1.8 & 1.2\\
ST.03.12   & 16.09.2019 -- 22.02.2020     & 13.4 & 28.6 & 7.7 & 5.2 & 0.9\\ \toprule
Total      & 27.07.2013 -- 22.02.2020     & 176.2 & 221.6 & 26.3 & 16.1 & 10.4\\  \midrule
\end{tabular}
\label{tab:data}
\end{table*}

The Crab Nebula data were then separated into five groups: The first is data with the best possible quality ($T_{9 \,\text{km}}>0.95$), which will be used for the construction of the reference spectra in Sect.~\ref{sec:reference}. The second group contains all data with good quality ($T_{9 \,\text{km}}>0.9$) referred to as the highest transmission region. For the analysis of LIDAR corrections under suboptimal conditions, a high transmission region ($0.82<T_{9 \,\text{km}}<0.9$), a medium transmission region ($0.65<T_{9 \,\text{km}}<0.82$) and a low transmission region ($0.5<T_{9 \,\text{km}}<0.65$) was defined. A detailed overview of all used data for all included analysis periods and transmission bins is shown in Table~\ref{tab:data}. The analysis period~ST.03.03 (marked with~$^*$) is interrupted in parts by ST.03.04. Period~ST.03.04 has been excluded since it only contains stereo data from 27.08.2014 until 30.08.2014 and some mono data from 19.06.2014 until 04.07.2014. Due to its shortness, the remainder does not contain sufficient statistics. The same applies for period~ST.03.08, which covers three months only.

The data were analyzed using the MAGIC Analysis and Reconstruction Software ~\citep[MARS;][]{2013ICRC...33.2937Z}. All spectral fits were obtained by using a forward unfolding approach~\citep{unfolding}. Spectral points were obtained also from forward unfolding except for Sect. \ref{sec:period_averaged}. Here, the Tikhonov method was applied ~\citep{unfolding}.


\subsection{Construction of the reference spectra}
\label{sec:reference}
In order to quantify the improvements in spectra recorded under less than optimal conditions, a reference spectrum must be defined to serve as a benchmark. Due to the large time coverage of the data set, the instrument response of MAGIC has changed over time due to hardware upgrades or environmental impact. Consequently, new analysis periods with their associated sets of simulation data have been introduced, whenever changes were significant enough. Since the accuracy of the simulations can vary from period to period, reconstructed spectra can appear slightly different despite originating from a non-varying source. The intrinsic systematic uncertainty of MAGIC for the absolute energy scale was estimated to lie below 15\% and for the flux normalization at 11\%~\citep[for medium energies at around 300~GeV,][]{aleksic:2016}. 

The reference spectra were built from data with $T_\mathrm{9\,km}>0.95$ only. The resulting spectra were fitted to a log-parabola function for every analysis period:
\begin{equation}
    \frac{\mathrm{d} \phi}{\mathrm{d} E} = f \cdot \left( \frac{E}{275 \, \mathrm{GeV}} \right)^{a - b^2 \cdot \text{log}_{10}\left( \frac{E}{275 \, \mathrm{GeV}} \right) }
.\end{equation}

For the normalization energy, a value of 275$\,$GeV has been chosen. This corresponds to the mean de-correlation energy of the individual reference spectra and aims to minimize the correlation of the flux normalization parameter, $f$, with the index and shape parameter, $a$ and $b$.

\begin{figure}
\centering
  \resizebox{\hsize}{!}{\includegraphics{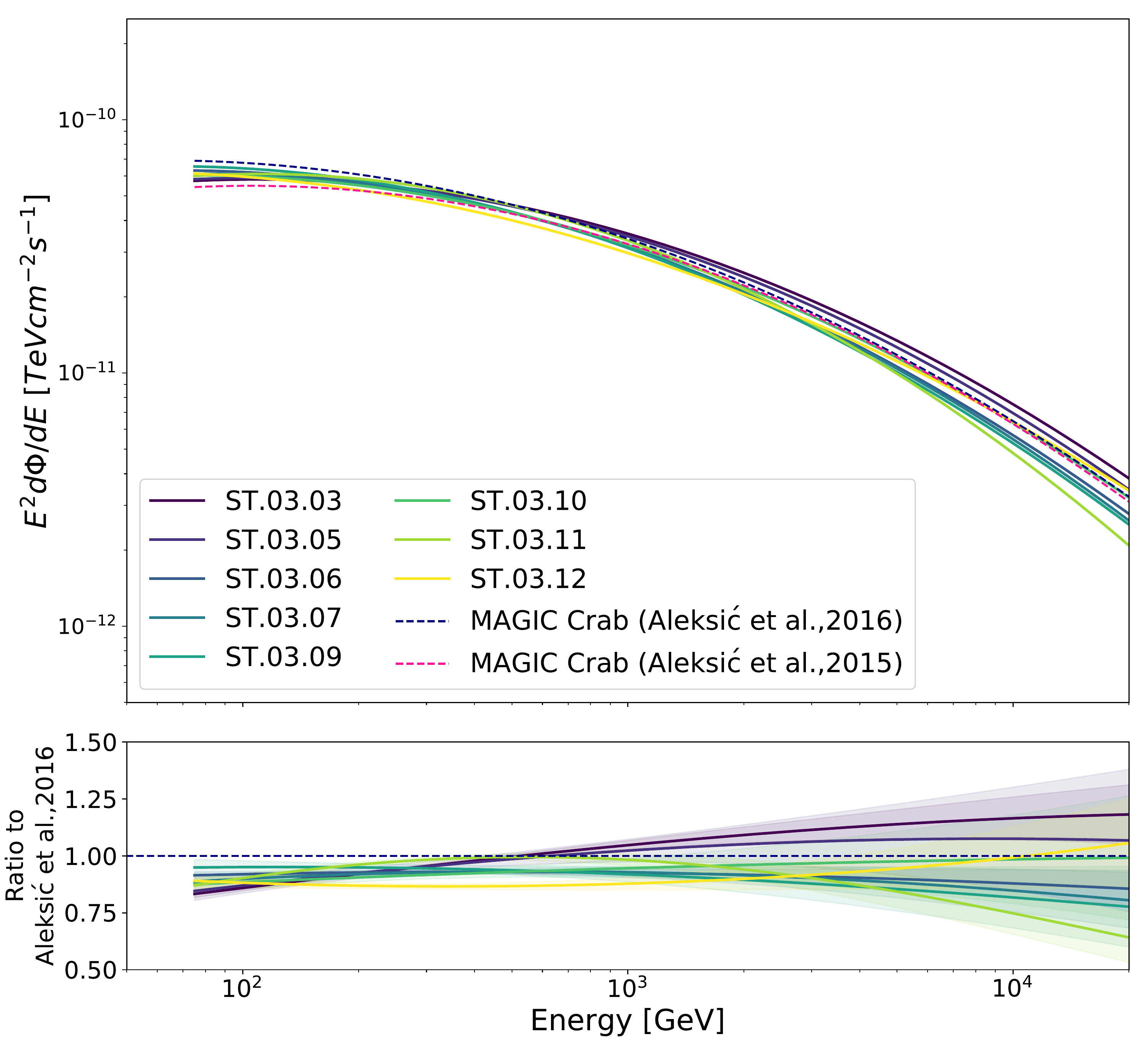}}
  \caption{Reference spectra of all included analysis periods. Top: All spectral fits and, for comparison, spectra published by MAGIC in \protect\citet{aleksic:2016} and~\protect\citet{aleksic:2015}. Bottom: Ratio between the spectrum of a given analysis period and the published spectrum of~\protect\citet{aleksic:2016}. \revtwo{The shaded bands show the one sigma uncertainty band of the spectral fits.}}
  \label{fig:all_ref}
\end{figure}

\revone{Fig.~\ref{fig:all_ref} (top) shows the reconstructed spectral energy distributions (SEDs) together with the two published Crab Nebula SEDs~\citep{aleksic:2015} and~\citep{aleksic:2016}. In order to visualize the relative deviation of the spectra, the ratios of the respective SEDs and the one from~\citet{aleksic:2016} are shown in the lower part of Fig.~\ref{fig:all_ref}. The shaded band in the lower plot shows the one sigma uncertainty band of the spectral fits. Figure~\ref{fig:example_ref} additionally shows two individual example SEDs with their one sigma uncertainty band from the first (ST.03.03) and last (ST.03.12) analysis period used in this study in comparison with two published spectra of the Crab Nebula.}

Table~\ref{tab:paras} shows the analysis period and the fitted spectral parameters for all reference spectra. The reconstructed fluxes show an excess fluctuation of 3.5\%, compatible with the 5\% night-wise excess fluctuations found in~\citet{aleksic:2016}. The spectral index $a$ shows about 0.03 excess fluctuation and the curvature parameter $b$ about 0.02.

\begin{table}
  \centering
 \caption{Fitted spectral parameters of the reference spectra of all periods.}
\begin{tabular}{l|lll}
\multicolumn{1}{c|}{Period tag}  &\multicolumn{1}{c}{f} & \multicolumn{1}{c}{a} & \multicolumn{1}{c}{b} \\ 
\multicolumn{1}{c|}{}  &\multicolumn{2}{l}{\scriptsize{($\times 10^{-10}$TeV cm$^{-2}$ s$^{-1}$)}} &   \\ 
\toprule
ST.03.03 &$7.05 \pm 0.06$ & $-2.19 \pm 0.01$ & $0.48 \pm 0.01$ \\
ST.03.05 & $7.04 \pm 0.16$ & $-2.20 \pm 0.03$ & $0.48 \pm 0.03$ \\
ST.03.06 & $6.90 \pm 0.05$ & $-2.27 \pm 0.01$ & $0.47 \pm 0.01$ \\
ST.03.07 & $6.89 \pm 0.08$ & $-2.26 \pm 0.01$ & $0.49 \pm 0.02$ \\
ST.03.09 & $7.04 \pm 0.13$ & $-2.29 \pm 0.01$ & $0.48 \pm 0.02$ \\
ST.03.10 & $6.80 \pm 0.13$ & $-2.25 \pm 0.03$ & $0.46 \pm 0.03$ \\
ST.03.11 & $7.29 \pm 0.08$& $-2.23\pm 0.02$  & $0.54 \pm 0.02$ \\
ST.03.12 & $6.45 \pm 0.09$ & $-2.28 \pm 0.02$ & $0.43 \pm 0.02$ \\ \midrule
Weighted means &  $6.94 \pm 0.22$ & $-2.24 \pm 0.04$ & $0.48 \pm 0.03$ \\ \bottomrule   
\end{tabular}
\label{tab:paras}
\end{table}


These resulting spectra will later on serve as a reference for a comparison of uncorrected and corrected data taken under nonoptimal atmospheric conditions. The latter can then be compared with the corresponding reference spectrum constructed by data taken in the same period. This approach aims to minimize the systematic uncertainties of the analysis. 

\begin{figure*}
  \centering
  \includegraphics[width=.32\textwidth]{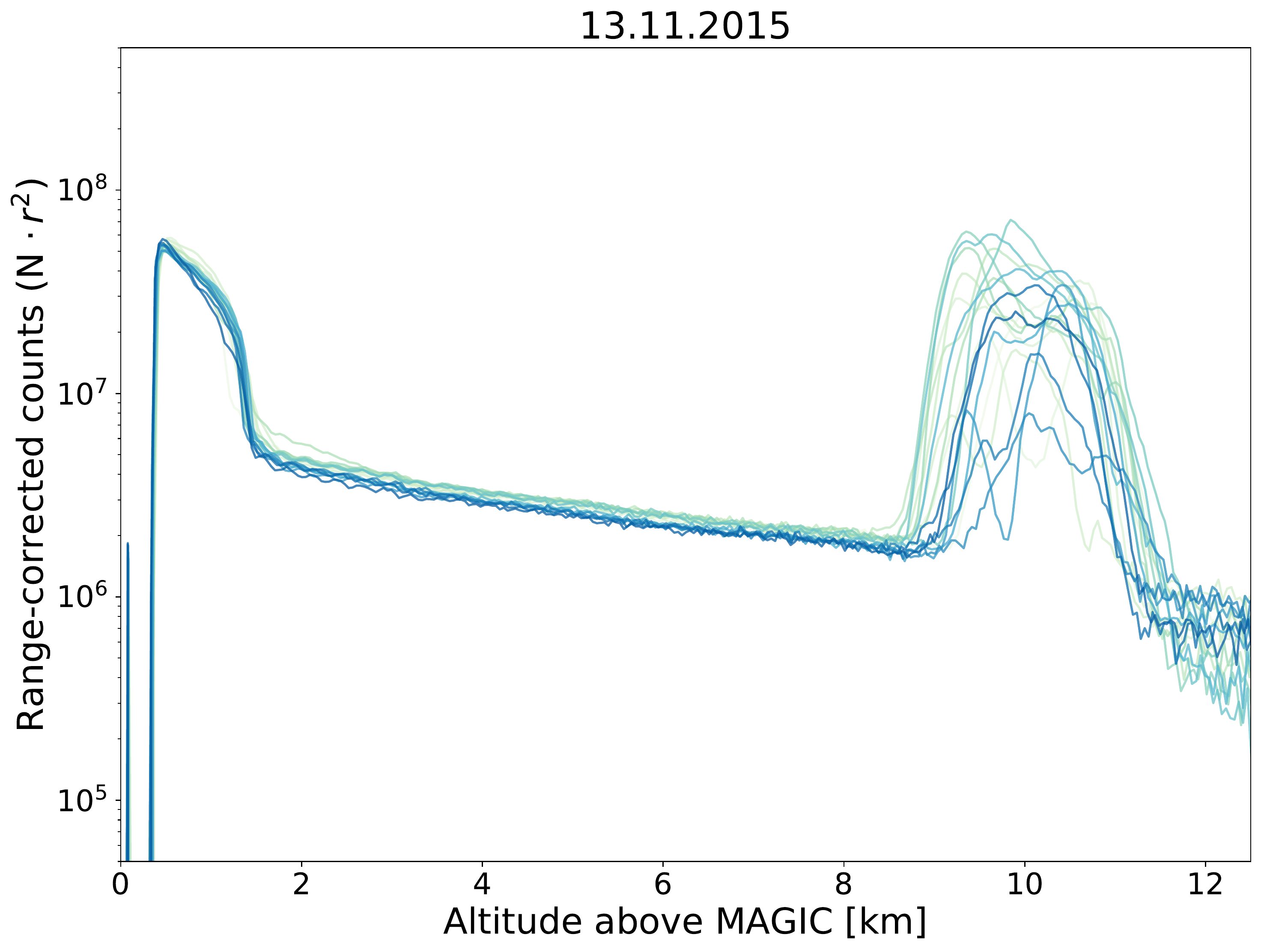}
  \hspace{0.15cm}
  \includegraphics[width=.32\textwidth]{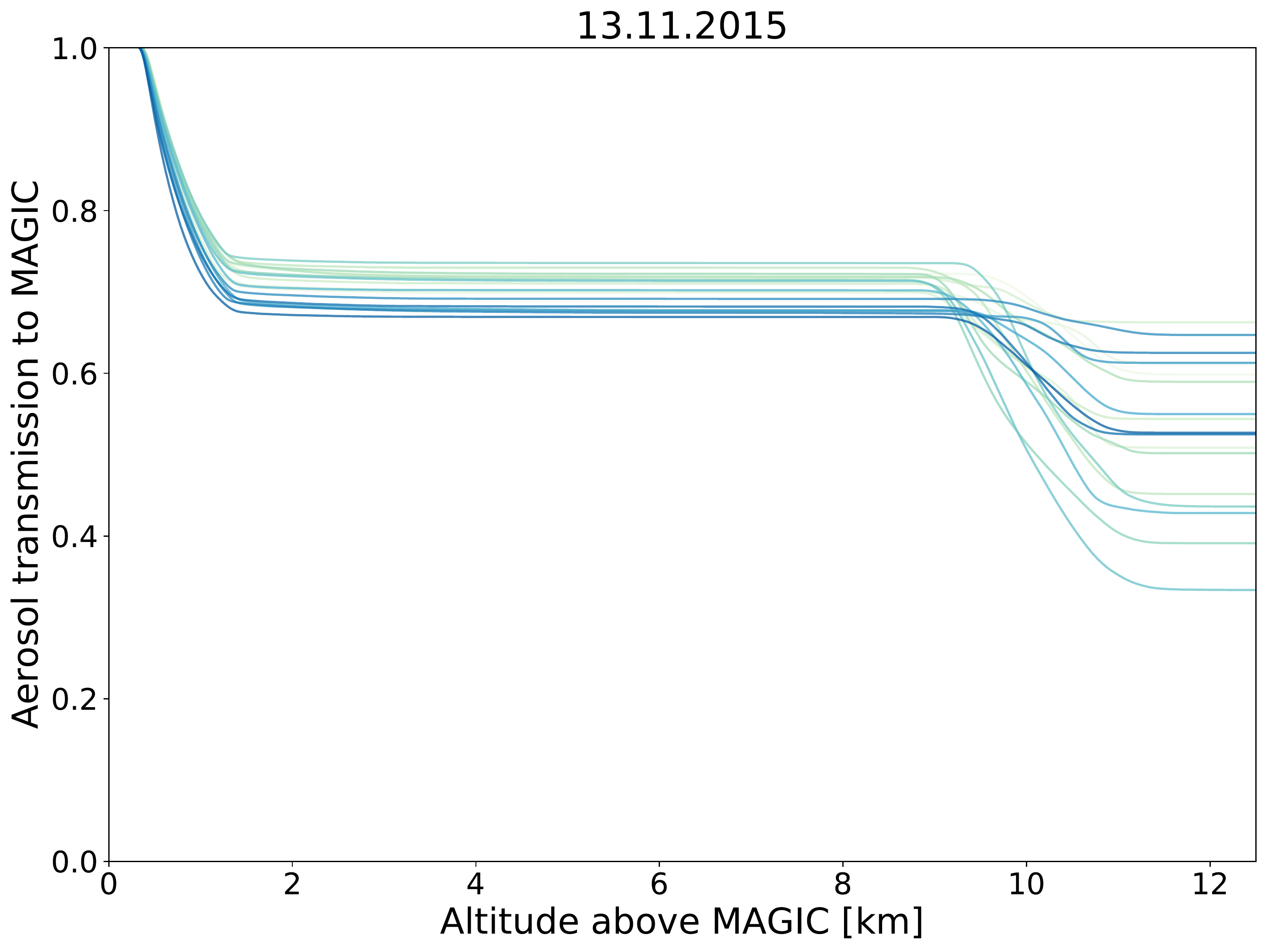}
  \hspace{0.15cm}
  \includegraphics[width=.32\textwidth]{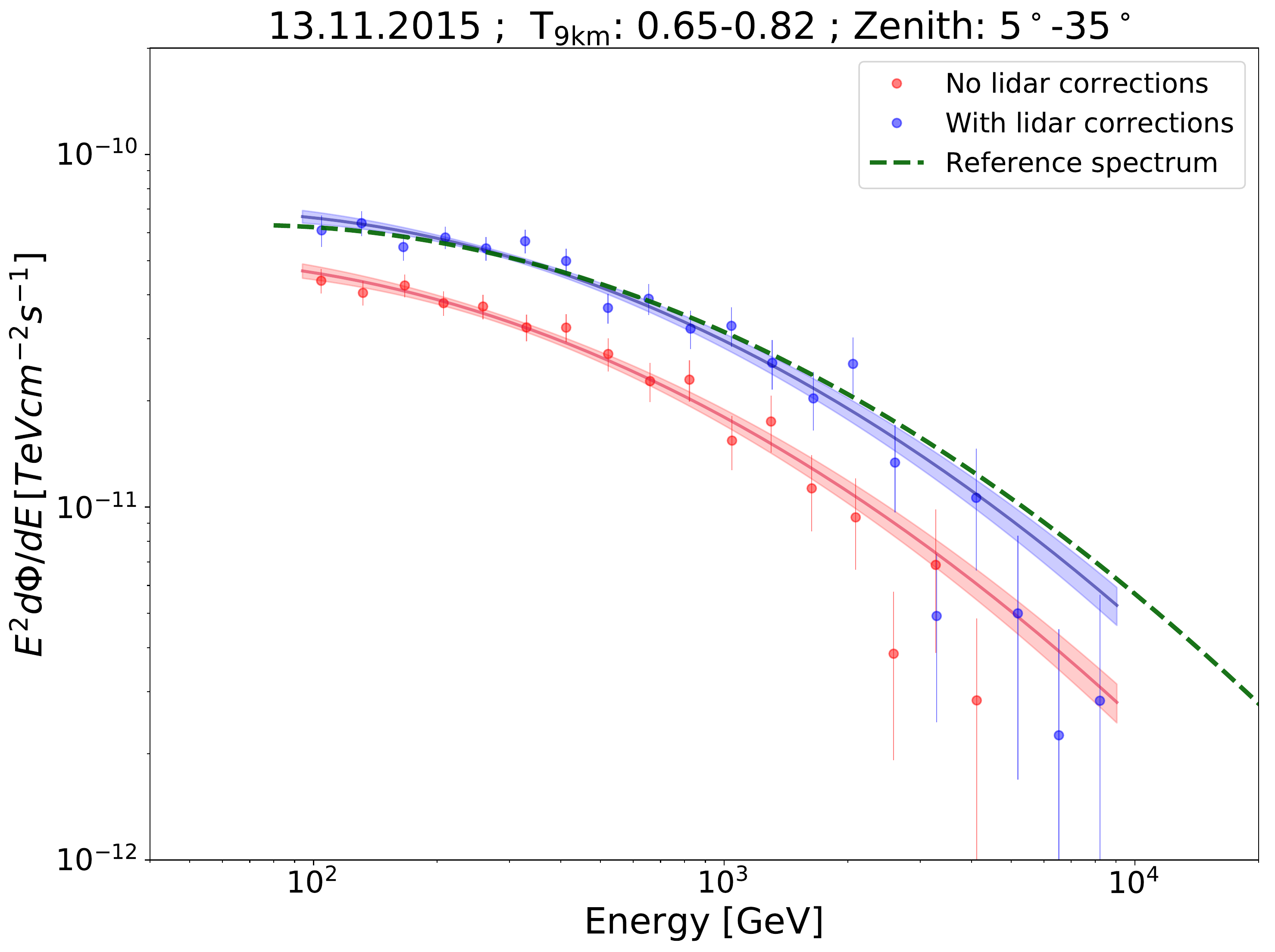}

  \vspace{0.5cm}
  
  \includegraphics[width=.32\textwidth]{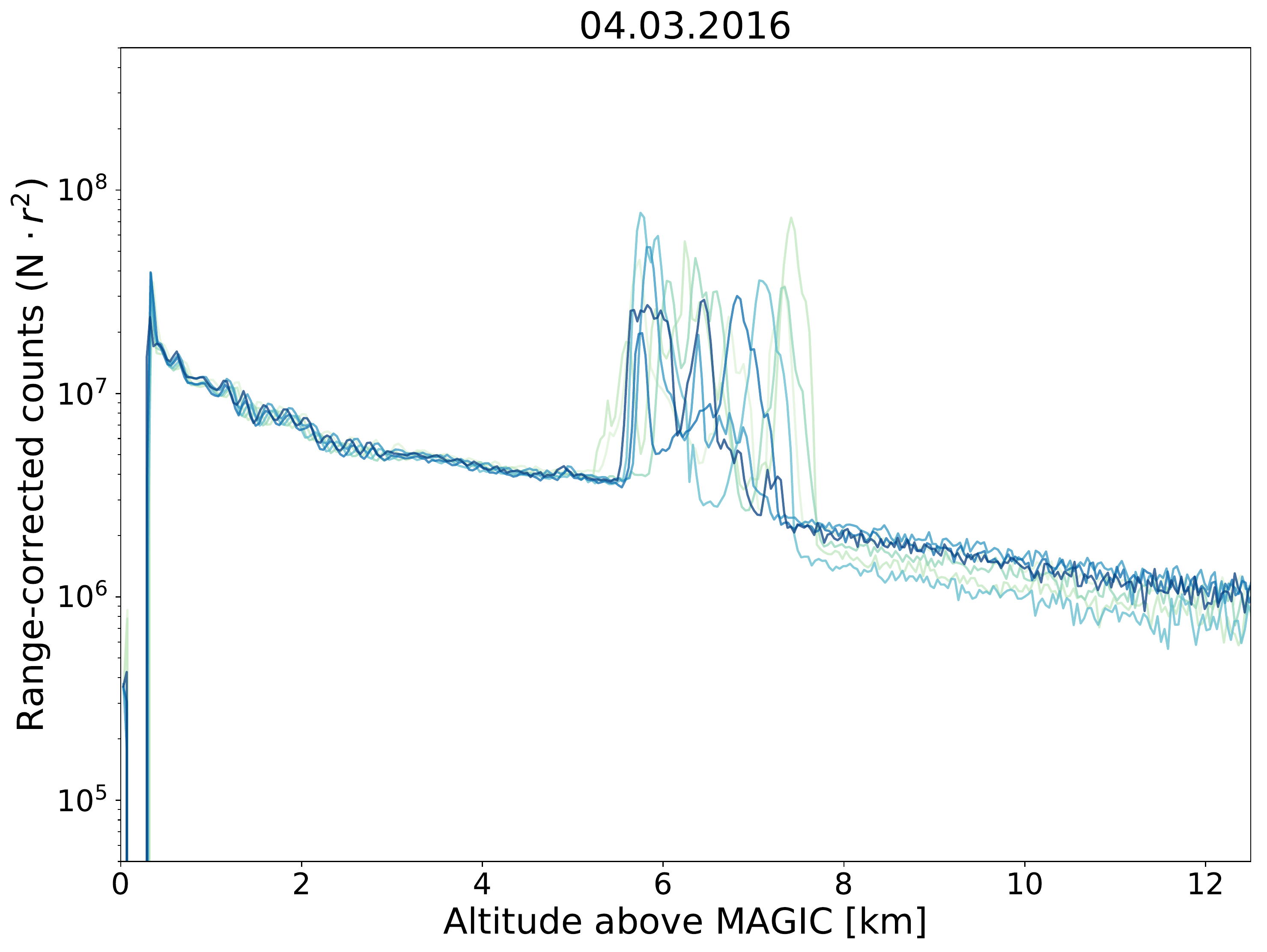}
  \hspace{0.15cm}
  \includegraphics[width=.32\textwidth]{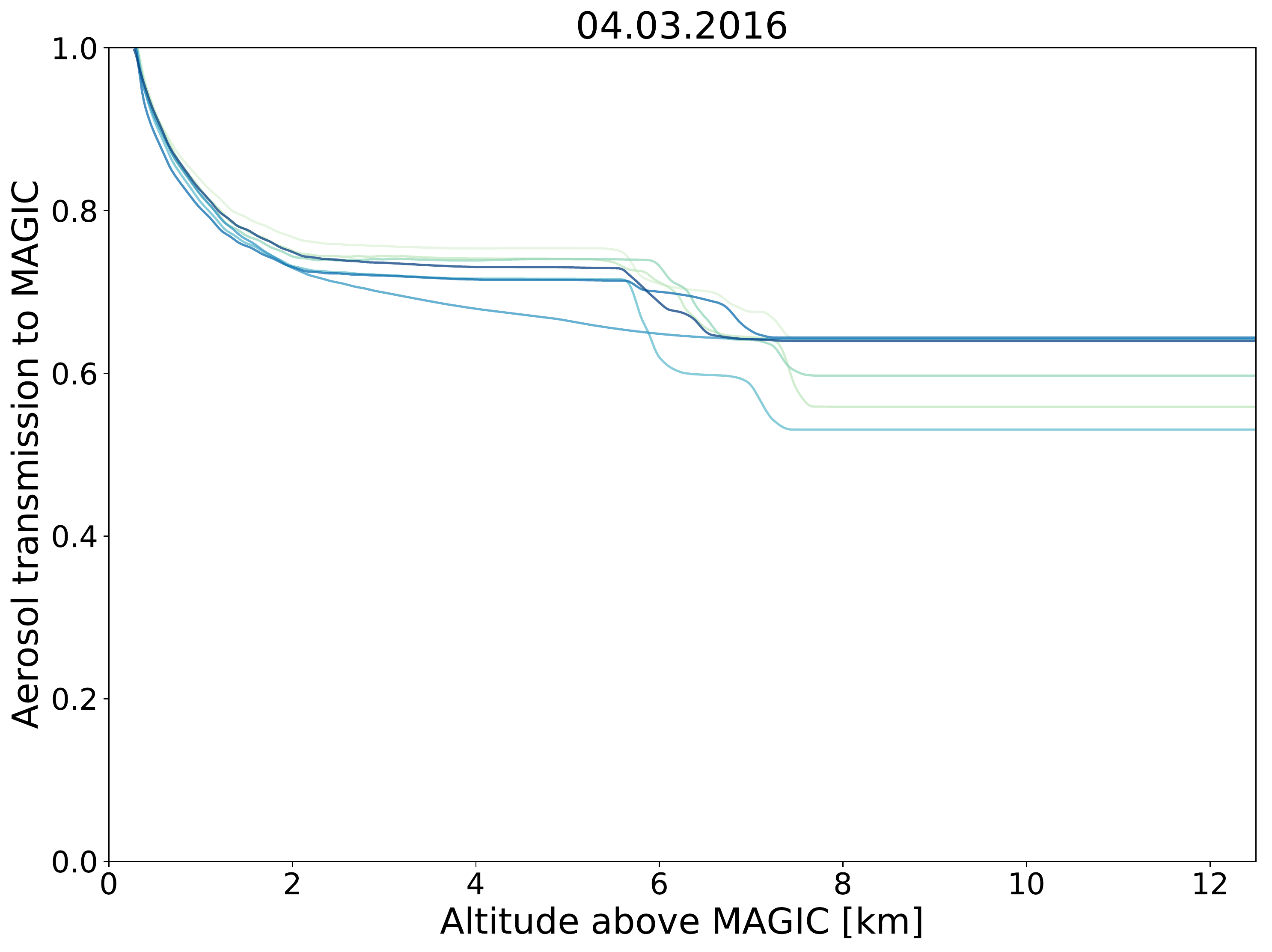}
  \hspace{0.15cm}
  \includegraphics[width=.32\textwidth]{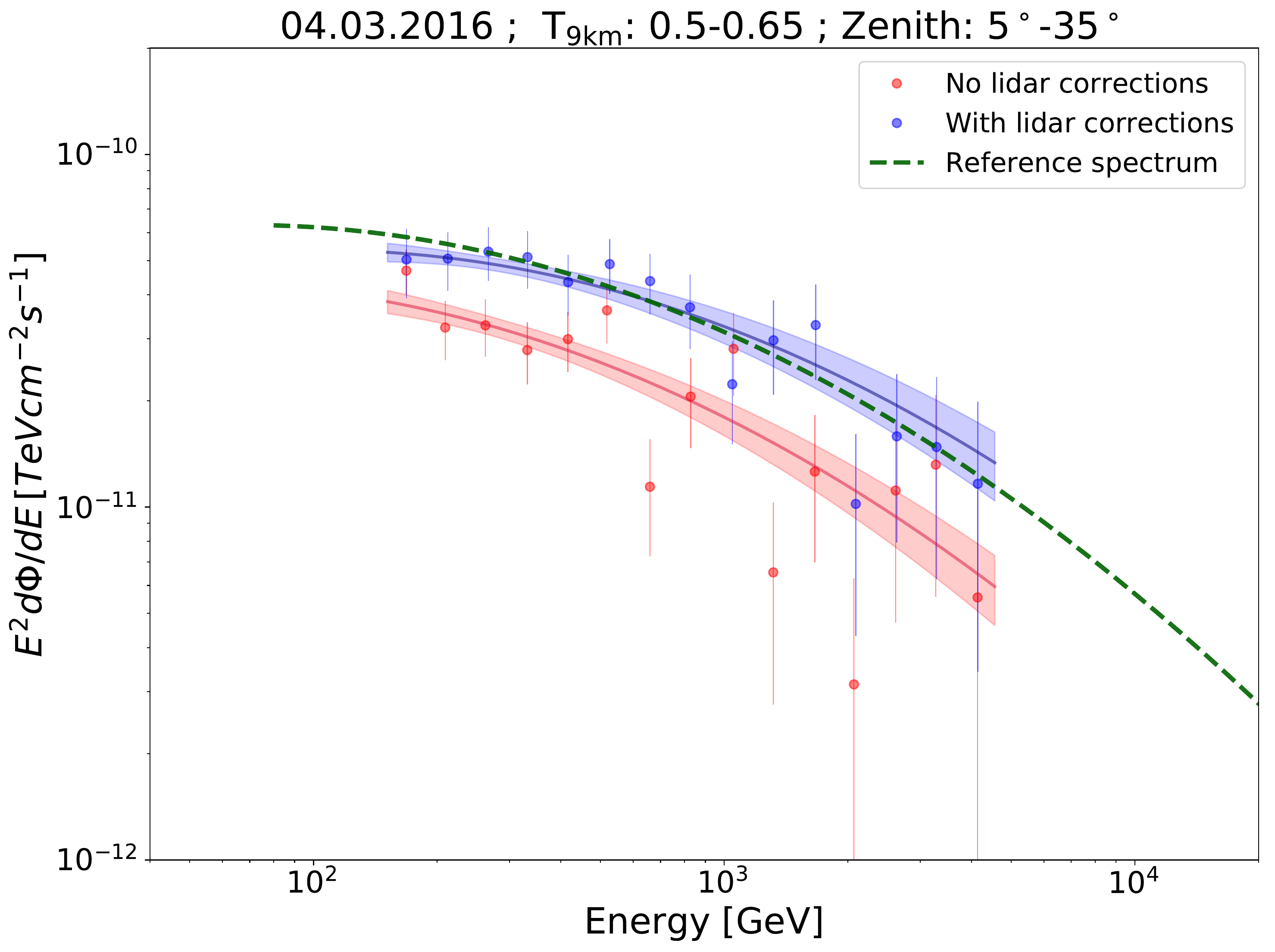}
  
  \vspace{0.5cm}

  \includegraphics[width=.32\textwidth]{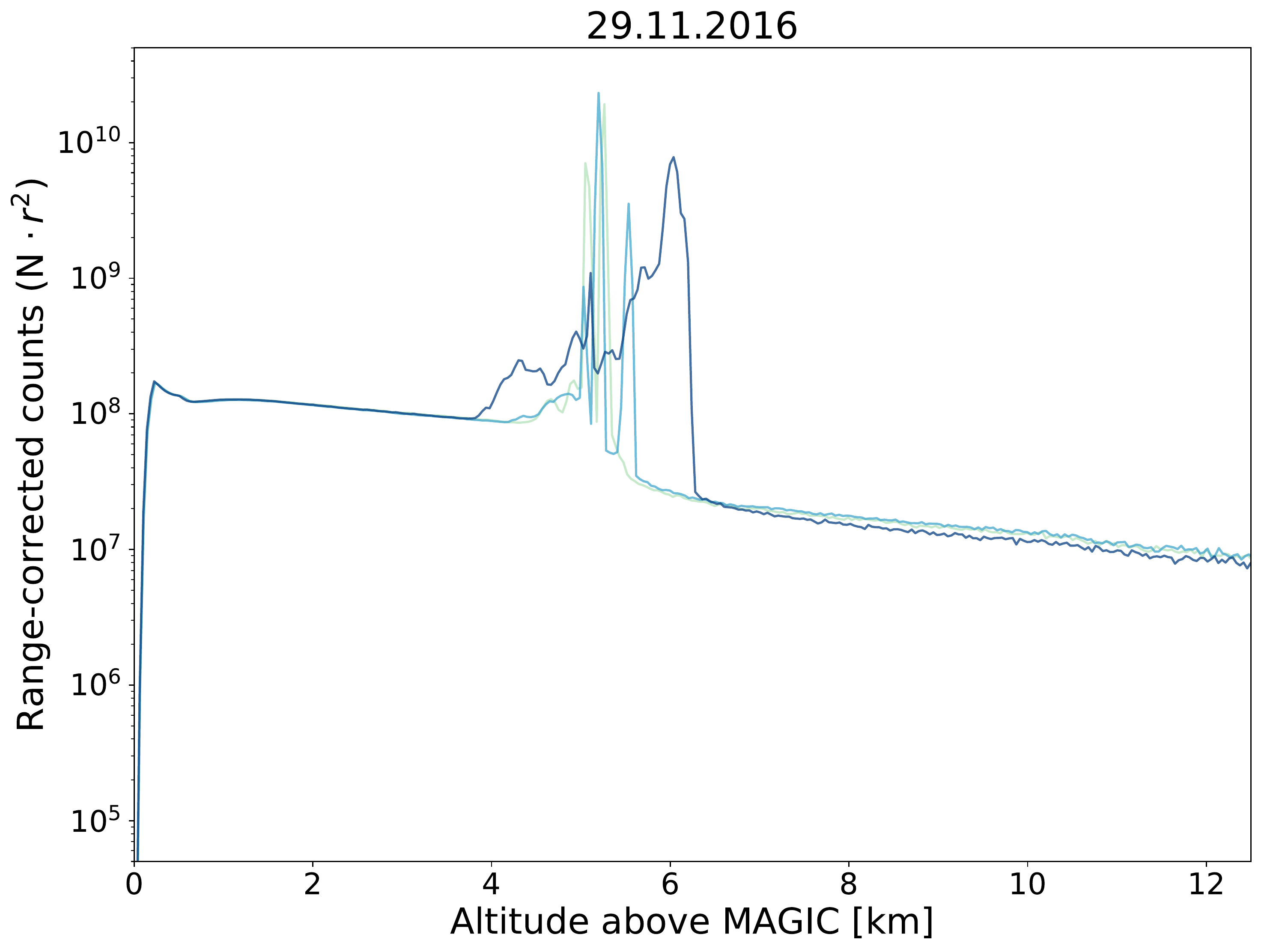}
  \hspace{0.15cm}
  \includegraphics[width=.32\textwidth]{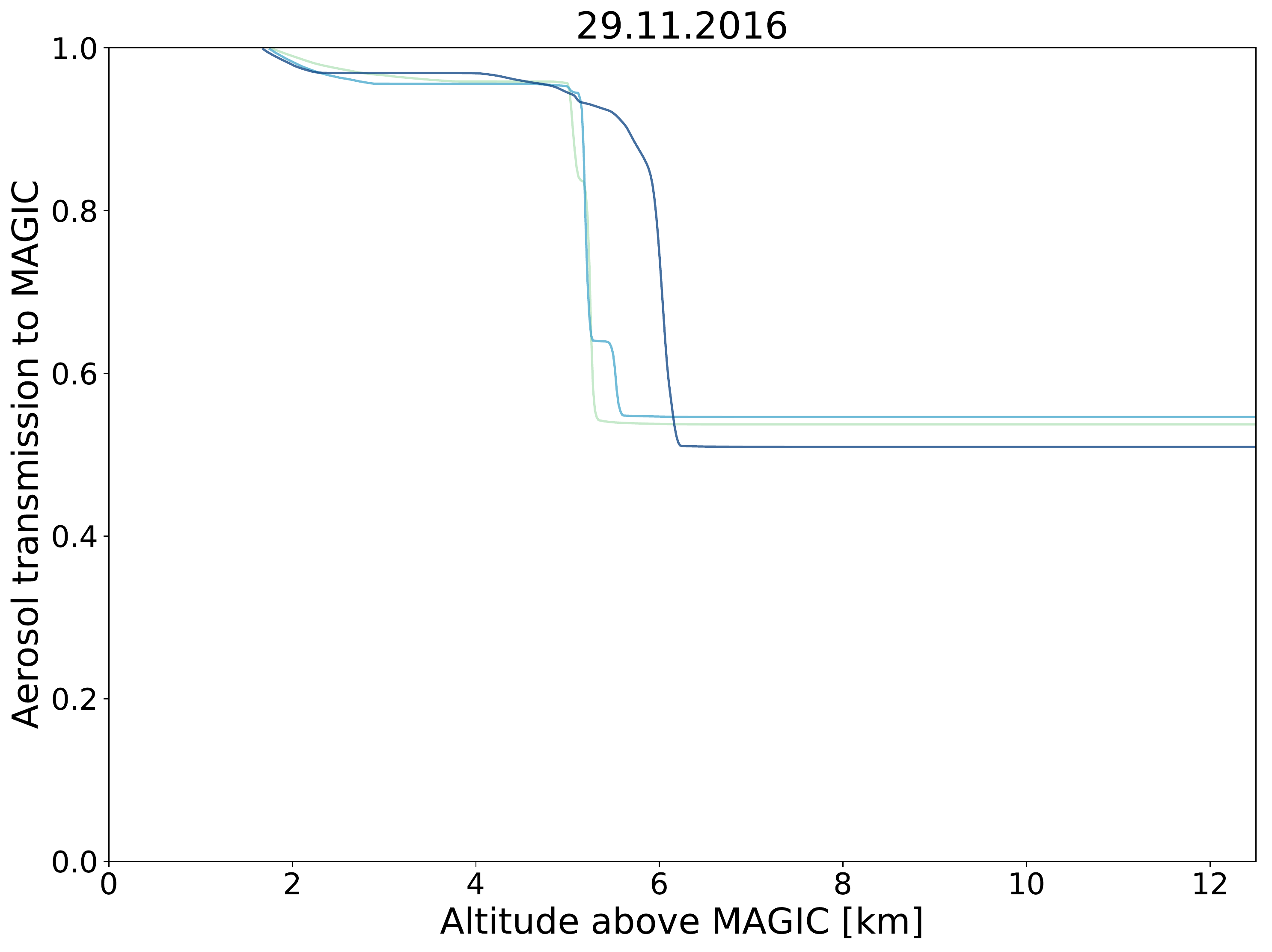}
  \hspace{0.15cm}
  \includegraphics[width=.32\textwidth]{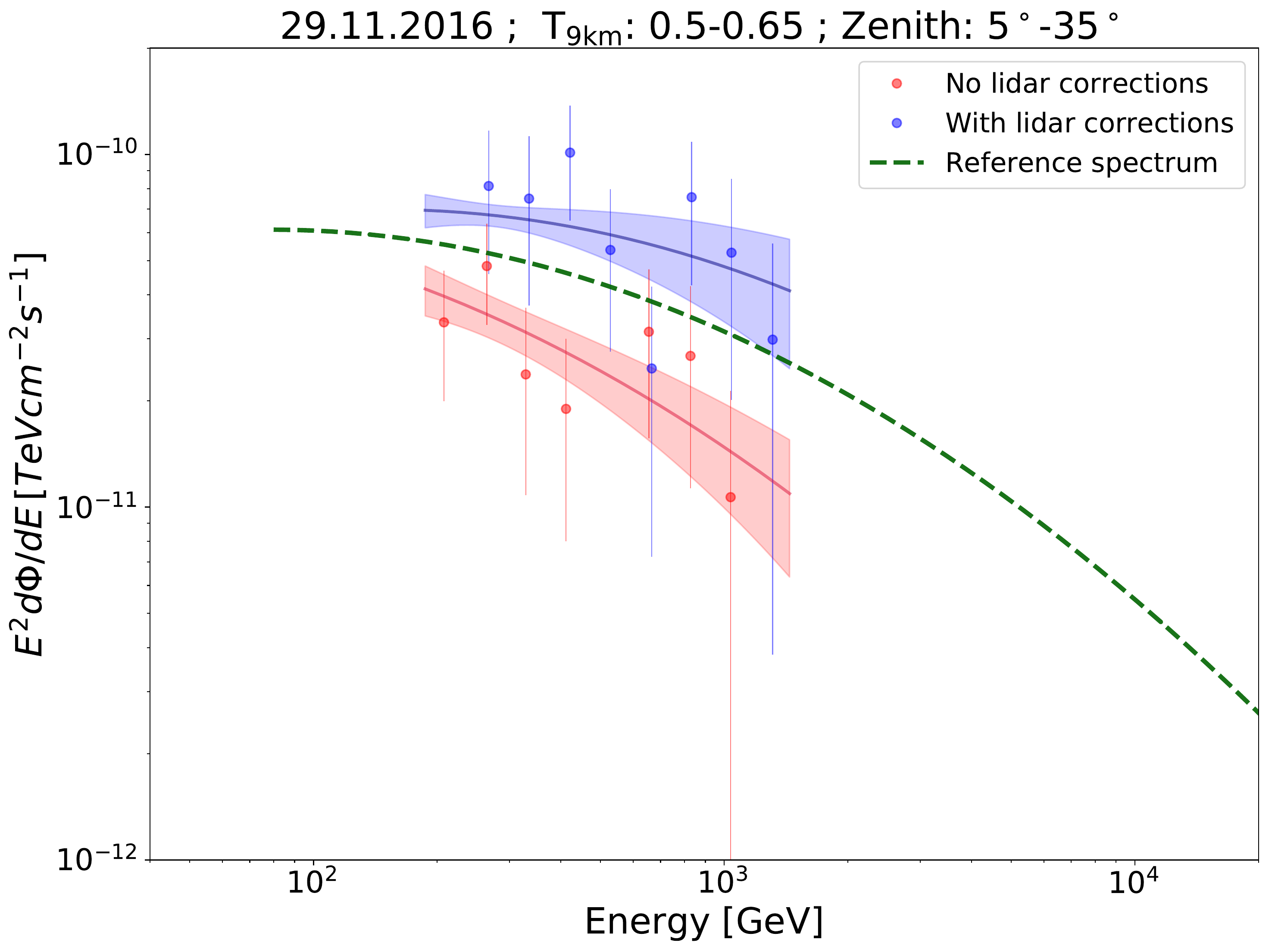}

  \vspace{0.5cm}

  \includegraphics[width=.32\textwidth]{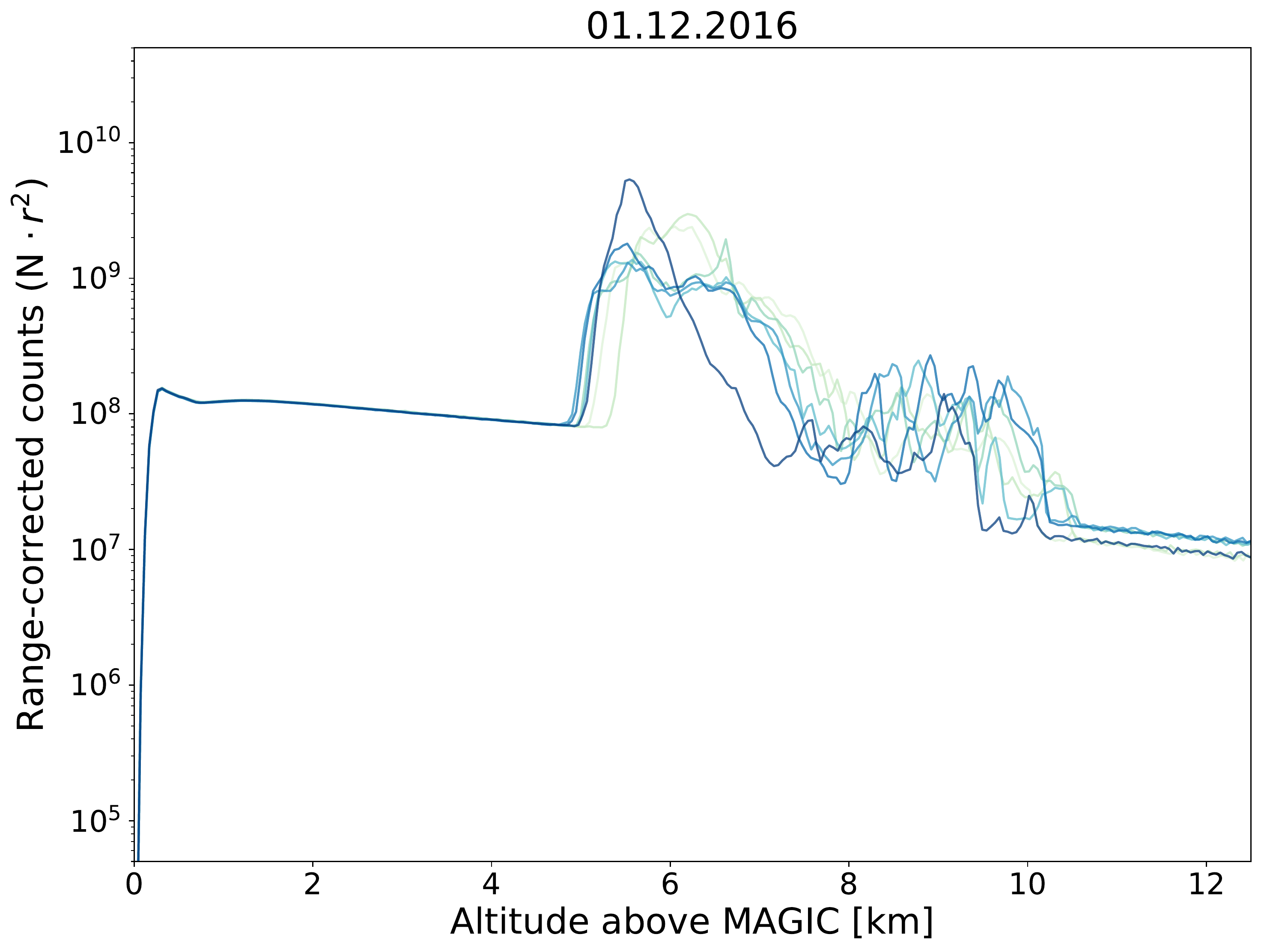}
  \hspace{0.15cm}
  \includegraphics[width=.32\textwidth]{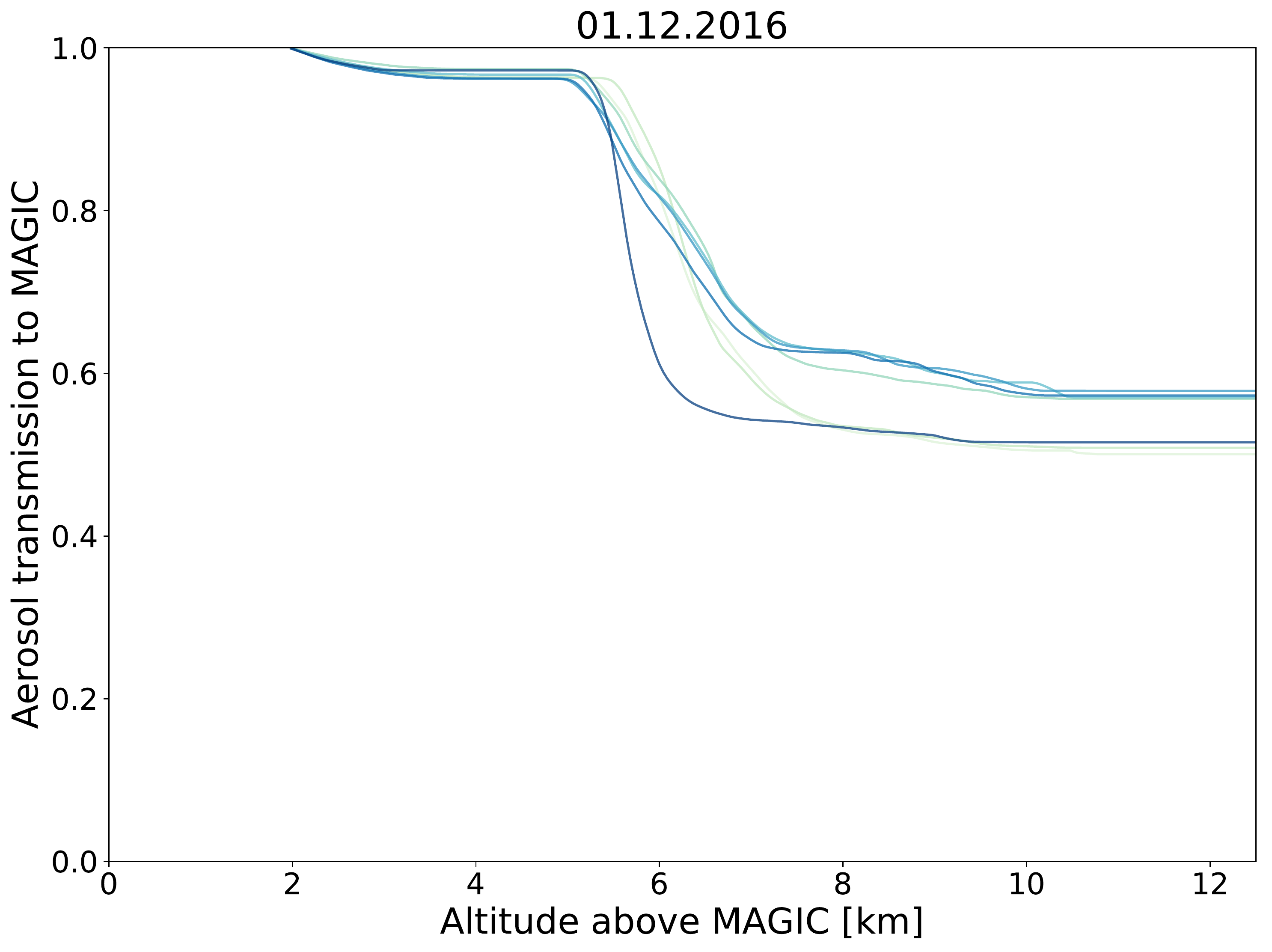}
  \hspace{0.15cm}
  \includegraphics[width=.32\textwidth]{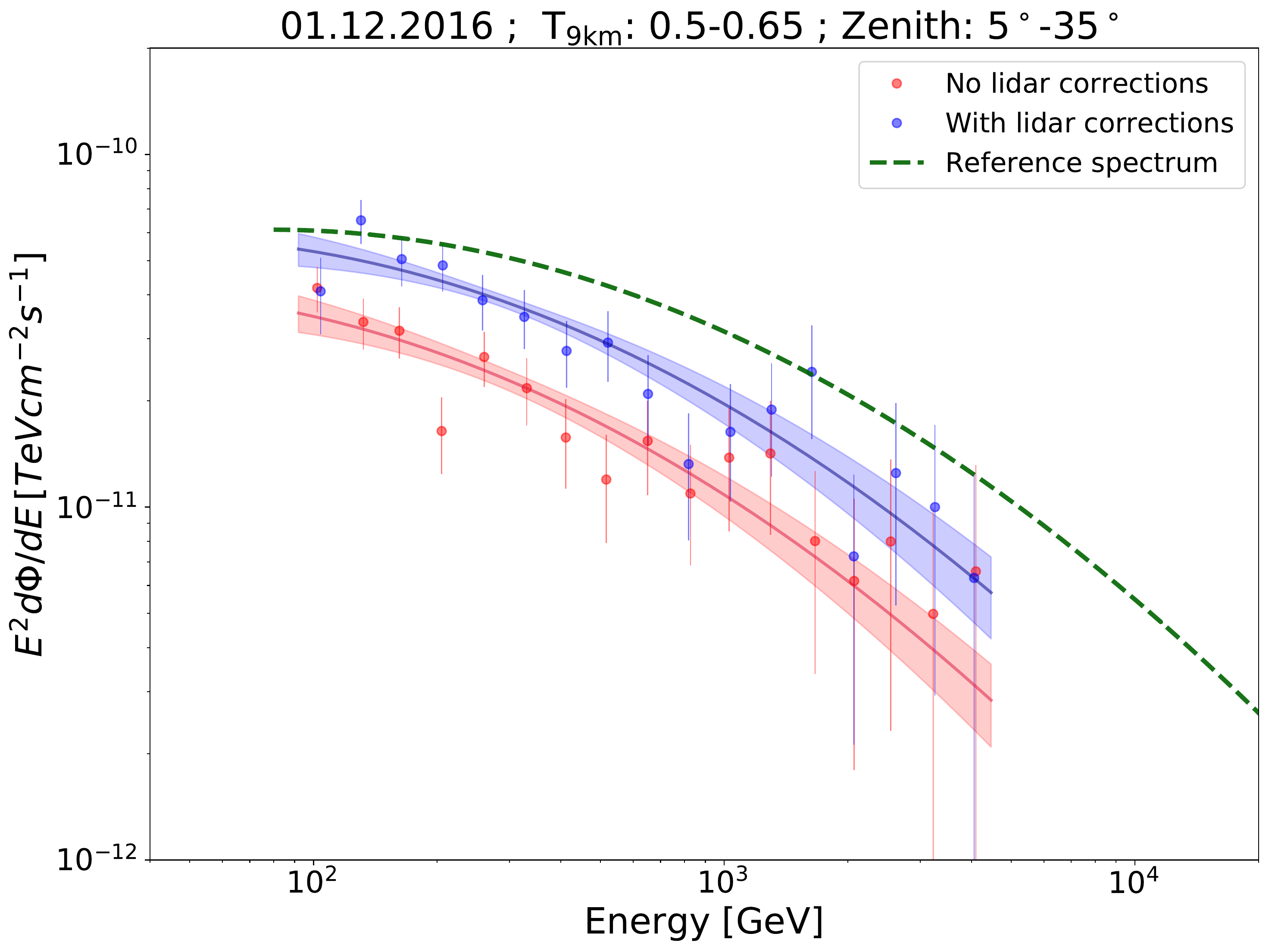}

  \caption{Four example nights that show the effect of the LIDAR corrections. Left column: LIDAR profiles in the form of range-corrected photo-electron counts for four example nights. The consecutive profiles are marked by different color shadings and provide a sample every 4 minutes. Center column: Resulting integral aerosol transmission curves, $T_\mathrm{aer}(h)$. Right column: SEDs of concurrent observations of the Crab Nebula shown without (red) and with (blue) LIDAR correction. For comparison, the reference spectrum of the corresponding period is shown as a dashed green line.}
\label{fig:example_nights}
\end{figure*}
\subsection{Evaluation of the LIDAR performance on a nightly basis}

In the first part of this analysis, the data were processed and analyzed into spectra on a nightly basis. In order to illustrate the method, Fig.~\ref{fig:example_nights} shows four example nights with unfavorable aerosol conditions. In the top-left part, the range-corrected photo-electron counts of the LIDAR, taken on November 13, 2015, are shown. They reveal a high aerosol content from ground up to  $\sim$1.5$\,$km, as well as a layer of clouds between $\sim$8.5$\,$km and $\sim$11$\,$km. In the center, the integrated atmospheric transmission curves $T_\mathrm{aer}(h)$ are shown, indicating $T_{9 \,\text{km}} \approx  0.7$. On the right side, the resulting SED of that night is shown and highlights the impact of the LIDAR corrections on data in the 0.65 to 0.82 transmission bin. The spectrum has been successfully recovered under these conditions. In the second row from top, the same plots are shown for March 4, 2016. This night shows strong calima and  few clouds starting above $\sim$6$\,$km. The resulting SED can be properly corrected as well. On November 29, 2016, lots of fast varying clouds around 5$\,$km make the correction challenging and result in a slight over-correction of the spectrum. The SED was only produced with data in the 0.5 to 0.65 transmission bin and therefore shows limited statistics. On December 1, 2016, a thick layer of clouds with a varying substructure between 5$\,$km and 11$\,$km results in  $T_{9 \,\text{km}} < 0.6$ for all profiles. The resulting correction does not fully reconstruct the SED.
In general, one can see an acceptable reconstruction of the corrected spectra in the two upper cases, where calima is the predominant atmospheric effect. In the two cloud-dominated cases, the reconstruction appears to be more challenging. A detailed description of the effects of clouds on the LIDAR corrections and its systematic uncertainties is discussed in Sect.~\ref{sec:discussion}.

The correction shown in Fig.~\ref{fig:example_nights}, as well as all the following LIDAR corrections until Sect.~\ref{sec:period_averaged}, was performed by using the second correction method, described in Sect.~\ref{sec:method2}. Due to the unfavorable aerosol conditions and sometimes low effective observation times, individual nights can show very limited event statistics. To assist \revone{the convergence of} the fitting procedure, the curvature parameter, $b$, in the log-parabola function was fixed to the obtained value from the respective reference spectrum. Only the amplitude, $f$, which parameterizes the overall scale of the spectrum, and the index parameter, $a$, which parameterizes the spectral tilt, were left free for the fit. \revone{The curved nature of the Crab Nebula spectrum is well established, and the spectral shape can be well determined by high quality data. However, the curvature, is more strongly affected by the high end of the spectrum, where the statistics become additionally more limited and the potential for the curvature to take on very different values is greater for small data sets. The amplitude and the index are primarily dominated by the low energy end of the spectrum, where statistics are large, making these the most important parameters for quantification under suboptimal aerosol conditions.}

The resulting set of parameters was then reconstructed and compared to the reference spectrum. To compare the fluxes, integral fluxes
for three separate energy regions were considered. The bins cover the regions from 200$\,$GeV -- 400$\,$GeV (low energy), from 400$\,$GeV -- 1$\,$TeV (medium energy), and above 1$\,$TeV (high energy).

The comparison of the resulting parameters and fluxes, using $q$ as a substitute variable for both, to the reference spectrum was then performed in two ways. First, the percental deviation from the reference value was determined to get the relative deviation:
\begin{equation}
\label{eq:perc}
D_{\%} = \left( \frac{q_i}{q_\mathrm{ref}} -1 \right) \cdot 100    
.\end{equation}
Second, to estimate the statistical significance of a given deviation, the deviation in units of standard deviations was also determined:
\begin{equation}
D_{\sigma} = \frac{q_i-q_\mathrm{ref}}{\Delta q} 
.\end{equation}
Here, $\Delta q$ contains the resulting statistical uncertainty obtained from the individual spectrum and the reference spectrum:
\begin{equation}
\Delta q = \sqrt{ \Delta q_i^2 + \Delta q_\mathrm{ref}^2 }
.\end{equation}
Together, the two deviations provide a reasonable quantification of how substantial a given quantity deviates from the expected value. A large percental deviation in combination with a small statistical deviation can be attributed to low statistics, whereas a large statistical deviation together with a low percental deviation may be caused by the intrinsic systematic uncertainties of the spectral analysis of MAGIC data. 

\subsubsection{Influence of the LIDAR corrections on the parameter reconstruction}
\label{sec:parflux}

\begin{figure}[h!]
\centering
\resizebox{\hsize}{!}{\includegraphics{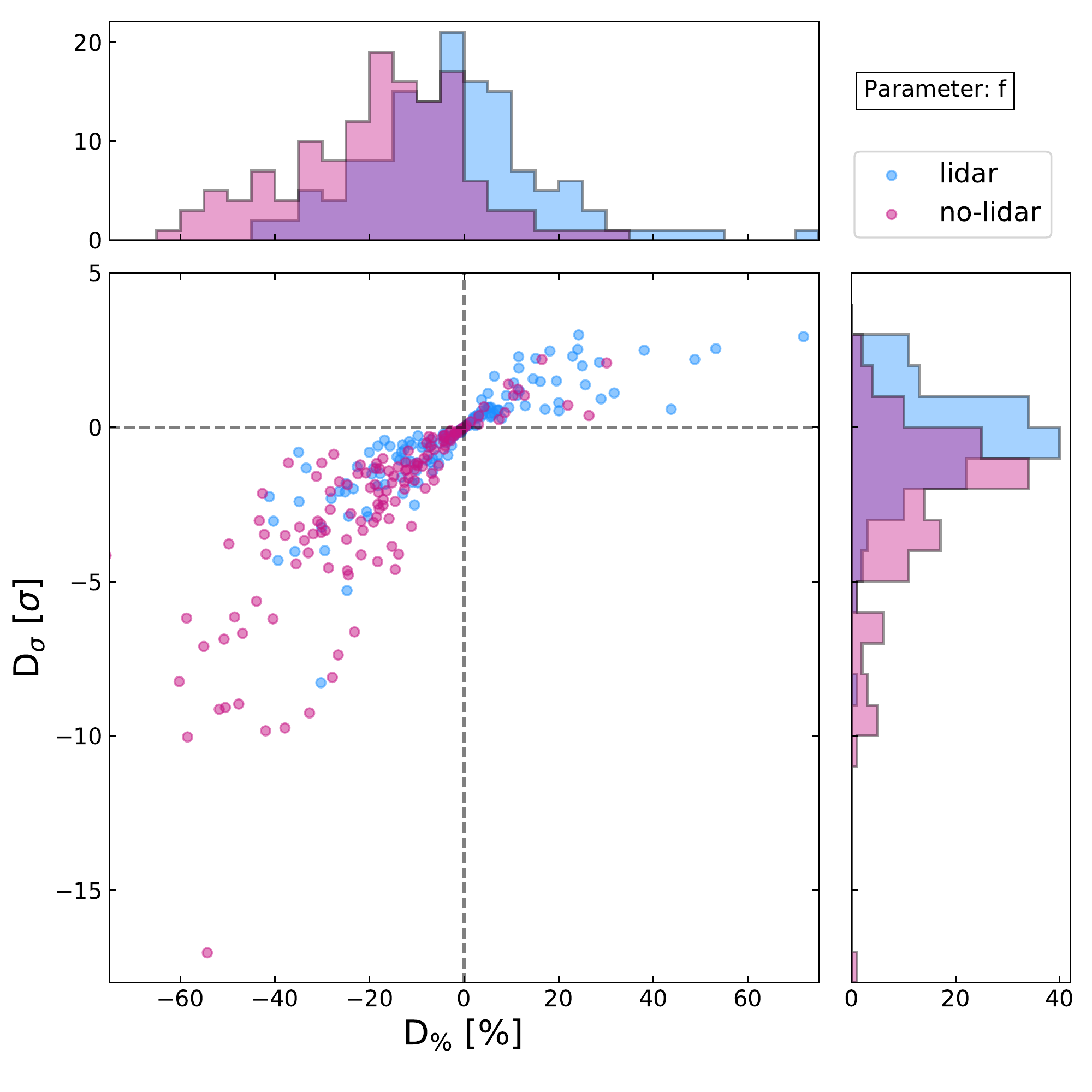}}
\resizebox{\hsize}{!}{\includegraphics{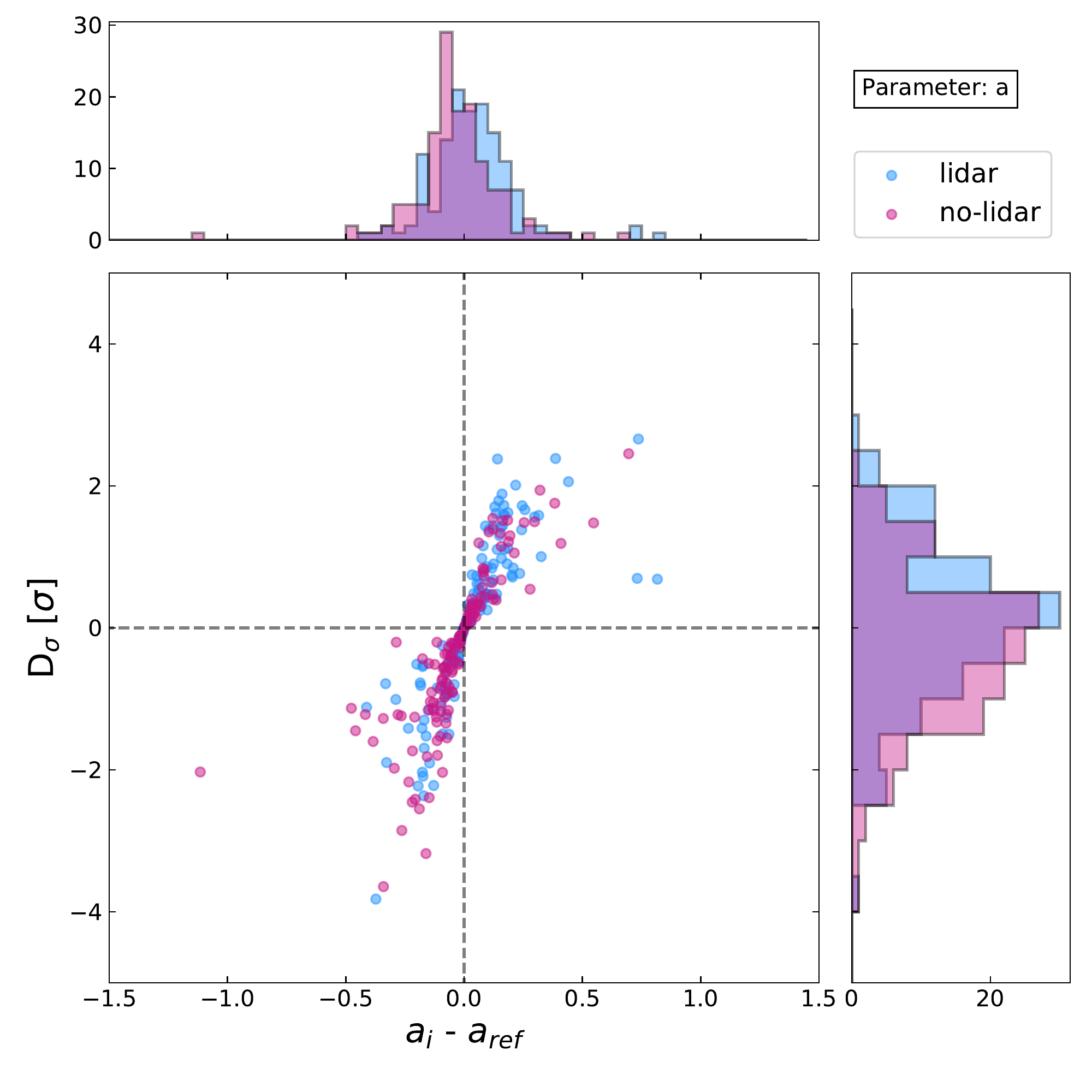}}
\caption{Scatter plots with projected histograms, showing deviations of the spectral parameters of individual nights. Top: Distribution of the percental and statistical deviation of the reconstructed flux normalization obtained from nights with an aerosol transmission between 0.5 and 0.9 from the reference spectrum for individual nights with projected histograms. Bottom: Same for the index parameter but using the difference instead of the percental deviation in the x-axis.}
\label{fig:scatter_plots_paras}
\vspace{10pt}
\end{figure}

In the following section, the impact of nonoptimal atmospheric conditions and LIDAR corrections on the overall spectral shape of the Crab Nebula is investigated on a nightly basis, quantified by the parameters of the log-parabola fits with fixed shape parameter, $b$. The energy threshold of the fit was set to 100$\,$GeV for low zenith data taken between 5$^\circ$ and 35$^\circ$, 200$\,$GeV for medium zenith between 35$^\circ$ to 50$^\circ$ and 400$\,$GeV for low zenith data between 50$^\circ$ to 62$^\circ$. The results are portrayed in the form of scatter plots showing both types of deviations from the reference values for individual nights. The images show the deviation of the flux of a single night as a scatter point and the overall distribution of nights as projected histograms. Positive values express an over-estimation of the parameter compared to the reference, whereas negative values correspond to an under-estimation. 

Figure~\ref{fig:scatter_plots_paras} shows the results for the flux normalization, $f$, and the spectral index, $a$, from all individual nights. The plots show data from the three transmission bins, taken under suboptimal conditions with $T_{9 \, \text{km}}$ between 0.5 and 0.9, and all zenith ranges combined. On the upper part, the distribution of deviations of individual nights are presented for the flux normalization. The histograms of both the percental and the statistical deviations show a strong improvement produced by the LIDAR corrections: the error weighted mean of the percental deviations moves from $-24.2\pm0.3$\% to $-4.3\pm0.5$\% and the mean of the deviation significances from -2.7$\,\sigma$ to -0.3$\,\sigma$. \revone{Also, the number of nights with statistical deviation greater than 3$\,\sigma$ has been considerably reduced from 48 (out of 137 tested samples) down to 7 due to the LIDAR corrections}. 

In the lower plot of Fig.~\ref{fig:scatter_plots_paras}, the deviations of the spectral index parameter $a$ are shown. Here, the differences 
$a_i-a_{ref}$ (instead of the percental deviation Eq.~\ref{eq:perc}) are shown on the horizontal axis. Both uncorrected and LIDAR-corrected distributions show mean values very close to the origin: \revone{the weighted mean of the differences shifts from a slight negative bias of $-0.05\pm0.01$ to $0.00\pm0.01$ after corrections. However, the clear-night case, Table~\ref{tab:paras}, already shows an error of 0.04 on the mean, making it compatible with both values and no significant biases can be claimed.} All in all, suboptimal atmospheric conditions do not seem to strongly distort the spectral tilt.

In order to investigate the influence of the zenith angle and transmission range on the LIDAR corrections, the data have been further classified into the previously mentioned transmission bins. Data with $T_{9 \, \text{km}}$>0.9 are now also included.
Additionally, the data were split into three telescope pointing zenith angle bins: a low zenith bin from 5$^\circ$ to 35$^\circ$, a mid zenith bin from 35$^\circ$ to 50$^\circ$ and a high zenith bin from 50$^\circ$ to 62$^\circ$. After having split the data in this manner, the number of nights in a given realm has further decreased and the available statistics reduced. Therefore, we condensed the information in the following manner: 
for the percental deviation, an error-weighted mean was calculated 
from individual nights and for the statistical deviation, a standard mean over individual nights. Additionally, covariance error ellipses have been constructed to visualize the distribution of the underlying data. The ellipses were derived from the standard mean, standard deviation and the Pearson coefficient. 
For normally distributed data, the 1$\,\sigma$ ellipse is expected to enclose roughly 68\% of the data, whereas the 2$\,\sigma$ encloses around 95\%.

\begin{figure*}[!ht]
  \centering
  \includegraphics[width=17cm]{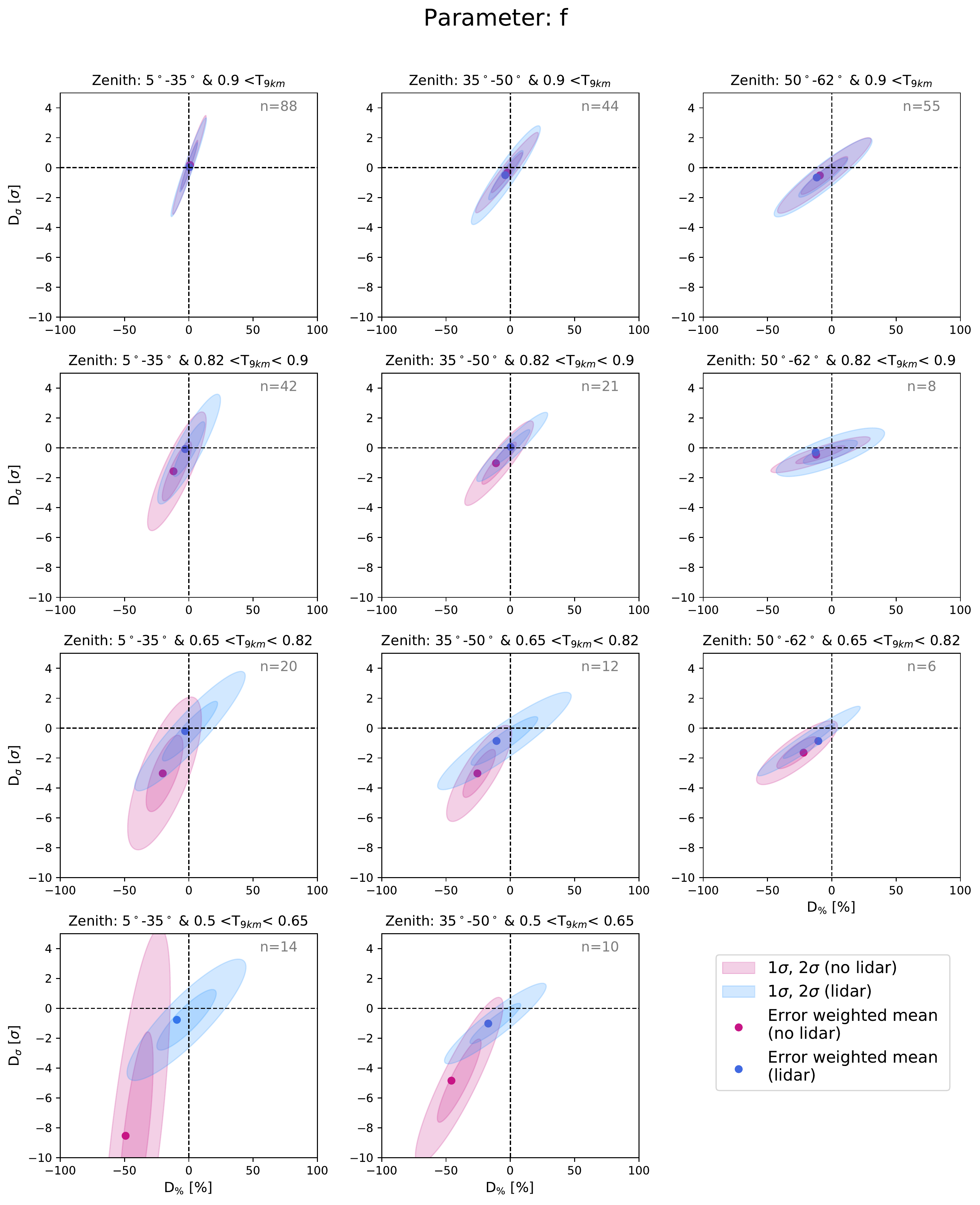}
 \caption{Error-weighted mean percental and mean statistical deviation of the flux normalization parameter, $f$, without (red dot) and with (blue dot) LIDAR corrections for eight zenith and aerosol transmission  $T_{9 \, \text{km}}$ bin combinations. The error ellipses are given by the shaded area. \revone{Note that, by construction, the uncertainty ellipses may not pass through the coordinate system origin, reflecting reconstruction biases}. The number of averaged nights, $n$, is provided in the top-right corner of each plot. In the low aerosol transmission and high zenith bin, sufficient data are not available.}
 \label{fig:mean_par_f}
\end{figure*}

Figure~\ref{fig:mean_par_f} depicts the results for the flux normalization parameter for eight zenith and $T_{9 \, \text{km}}$  combinations. An individual plot shows 
results obtained before and after applying the LIDAR corrections, marked with red and blue color, respectively. The shaded areas indicate the 1$\,\sigma$ and 2$\,\sigma$ covariance error ellipses. We note that since the ellipses have been constructed using the geometric mean, the error-weighted means do not always lie exactly on the center of the ellipses in the horizontal axis.

In almost all cases in the low zenith region, the flux normalization is reconstructed without bias, as illustrated by the good agreement of the means with the origin. Only in the lowest $T_{9 \, \text{km}}$ bin, the error-weighted mean shows a remaining bias of $(-9\pm2)$\% after using corrections. Under medium zenith angles, the corrections produce a bias-free 
reconstruction in the two higher transmission bins. Only below $T_{9\,\textrm{km}}<0.82$, a residual bias of $(-11\pm3)$\% is found, worsening to $(-17\pm4)$\% and -1.0$\,\sigma$ in the lowest transmission bin. Nevertheless, 
significant improvement can be seen here as well. 
Finally, the high zenith bin is shown in the right column. In the highest $T_{9 \, \text{km}}$ bin, the LIDAR corrections produce a mild worsening indicated by the slightly wider ellipse and the mean value moving away from the origin. A downward flux correction of up to 4\% is actually possible if the LIDAR reconstructs better aerosol conditions than those employed for the standard Elterman model used in the MC simulations~\citep{garrido2013}. 
In the transmission bin $0.82<T_{9 \, \text{km}}<0.9$, the corrected weighted mean shows almost no improvement toward the origin, but the error ellipse  widens slightly after applying corrections. Both distributions are, however, compatible with the origin and with only eight nights, the statistics is rather limited. Below $T_{9\,\textrm{km}}\approx 0.82$ 
the corrected mean shows an offset of $-10\pm6$\%.
As previously mentioned, the spectral amplitude is normalized at the overall sample wide de-correlation energy of 275$\,$GeV. This is very close to the energy threshold under clear conditions. Under suboptimal conditions, the threshold is even higher, which requires an extrapolation of the spectrum to lower energies in order to determine the flux normalization. This effect might additionally impair the reconstruction of the amplitude in the high zenith region. 
In the lowest $T_{9 \, \text{km}}$ bin, there are not sufficient data available. 

\begin{figure*}[!ht]
  \centering
  \includegraphics[width=17cm]{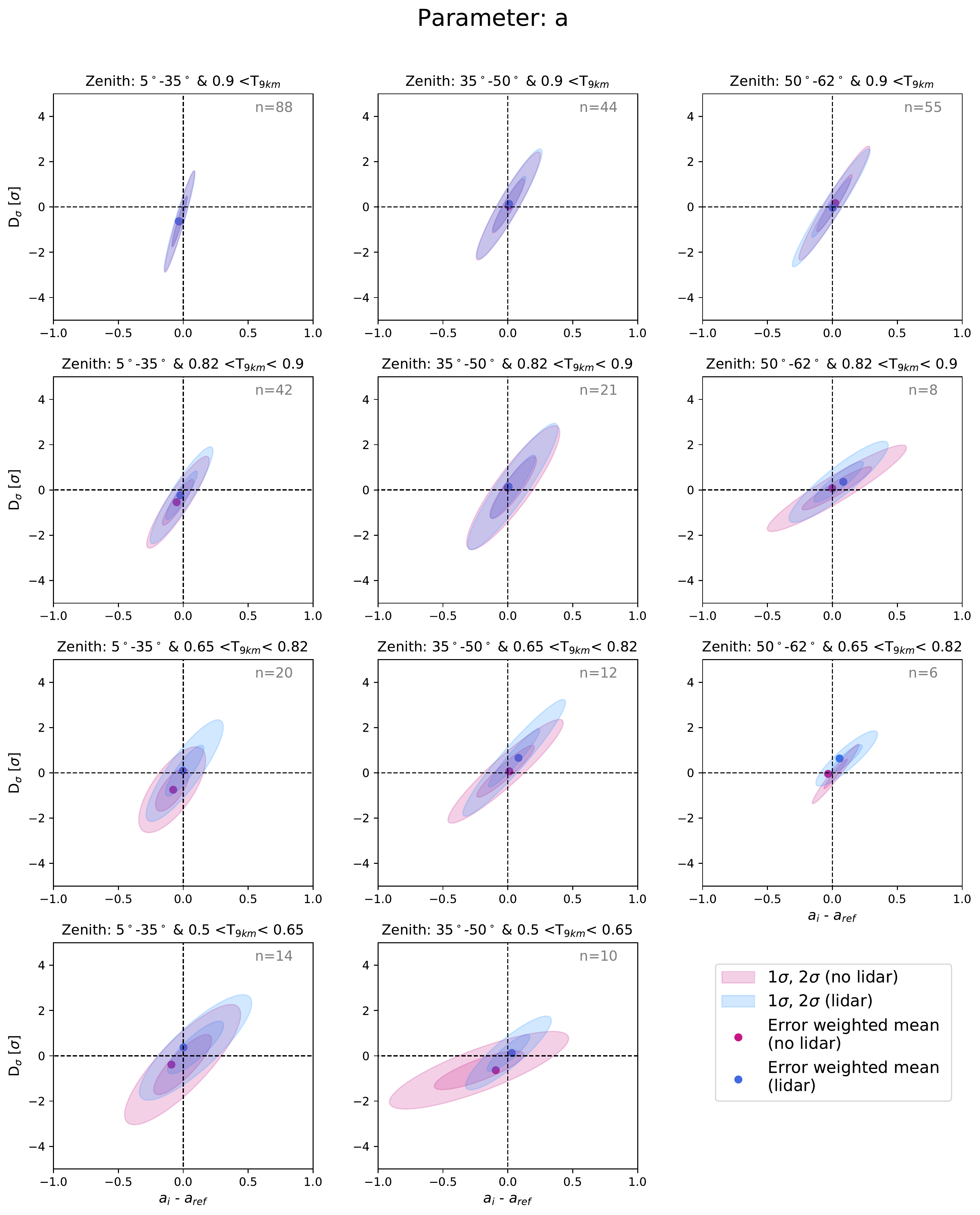}
 \caption{Same as Fig.~\ref{fig:mean_par_f}, but for the index parameter, $a$. In the low aerosol transmission and high zenith bin, sufficient data are not available.}
 \label{fig:mean_par_a}
\end{figure*}

 An analogous analysis for the index parameter, $a$, can be seen in Fig.~\ref{fig:mean_par_a}. As  already observed earlier, in the majority of cases the distributions of the index parameter $a$ do not change significantly before and after applying LIDAR corrections. At $T_{9 \, \text{km}} > 0.82$, uncorrected and corrected data produce almost identical results, except for the largest zenith angles, where a reduction of the spread can be observed. In the medium transmission range, $0.65 < T_{9 \, \text{km}} < 0.82$, improvements for low zenith angles are observed, but also slight worsening in the form of over-corrections at larger zeniths.
 Only under low aerosol transmissions, $0.5 < T_{9 \, \text{km}} < 0.65$, significant improvement is achieved for the shape parameter, in the form of narrower distributions and mean values close to the origin .   

\subsubsection{Influence of the LIDAR corrections on the integral flux reconstruction}
\label{sec:integflux}

\begin{figure}[!ht]
  \centering
  \resizebox{\hsize}{!}{\includegraphics{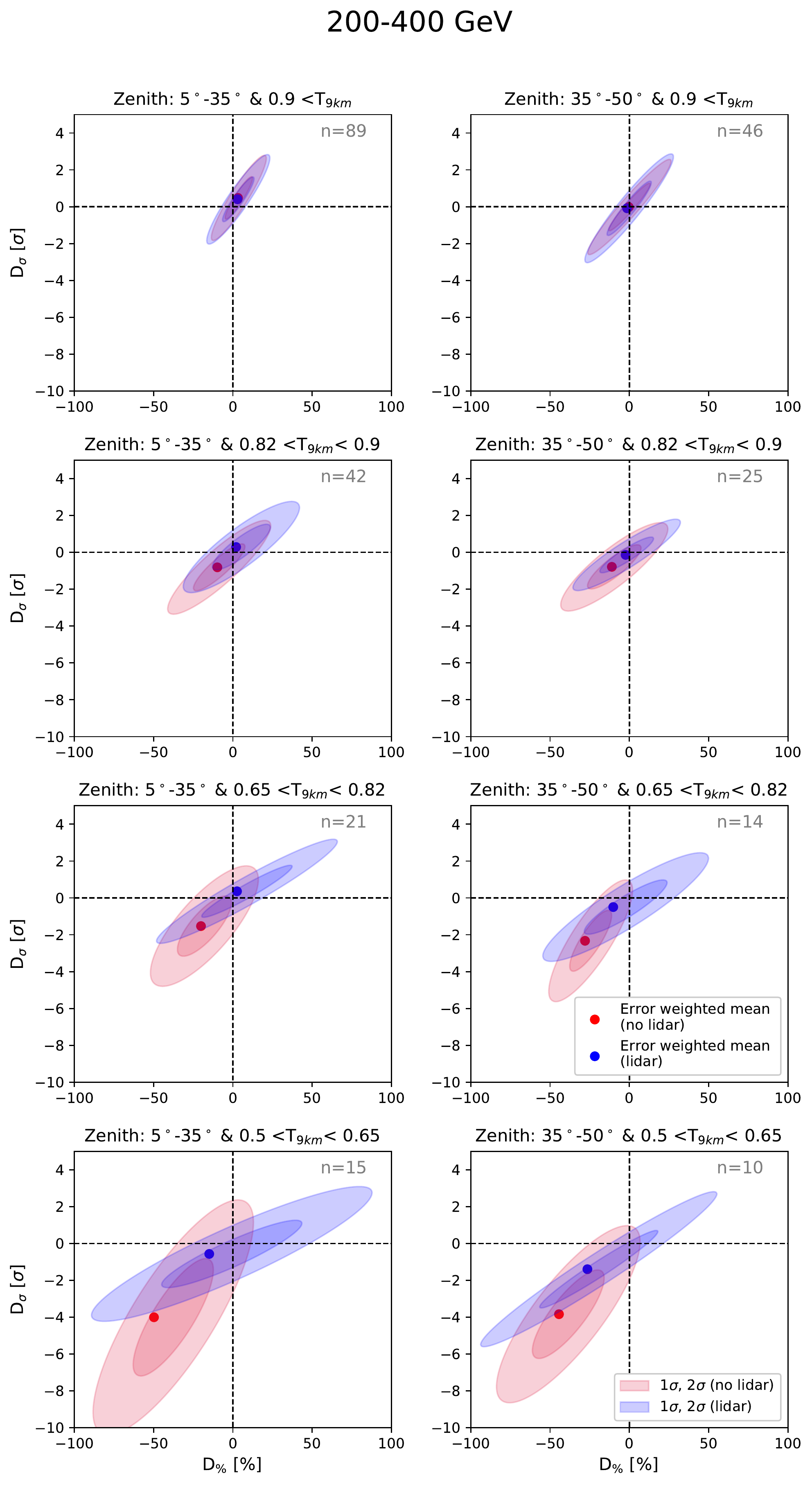}}
 \caption{Error-weighted mean percental and mean statistical deviation of the flux between 200$\,$GeV and 400$\,$GeV from the reference spectrum shown without (red dot) and with (blue dot) LIDAR corrections for six zenith and aerosol transmission  $T_{9 \, \text{km}}$ bin combinations. The error ellipses are given by the shaded area. The number of averaged nights, $n$, is provided in the top-right corner of each plot.}
 \label{fig:mean_200-400}
\end{figure}

\begin{figure*}[!ht]
  \centering
  \includegraphics[width=17cm]{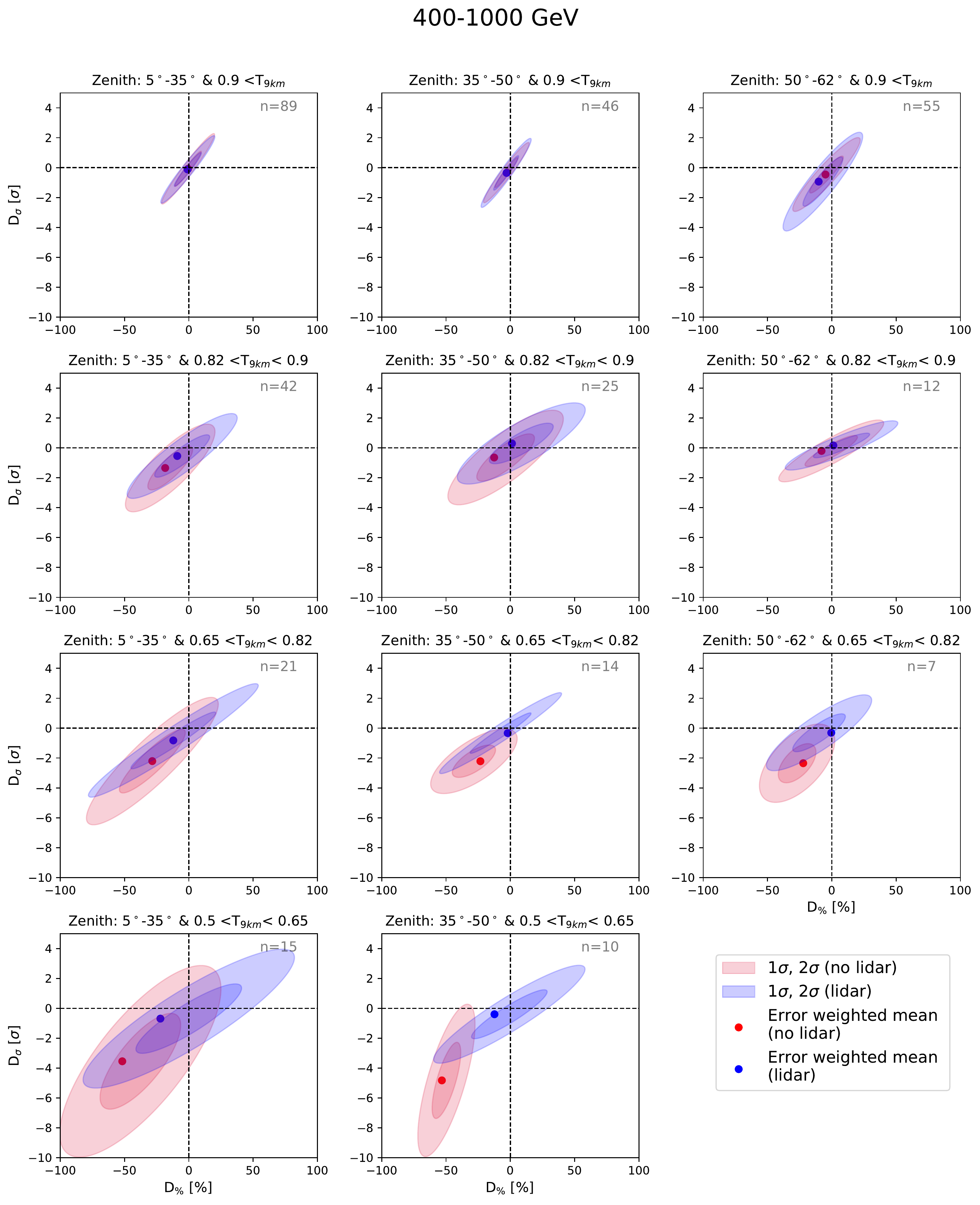}
 \caption{Same as Fig.~\protect\ref{fig:mean_200-400}, but for the energy region between 400$\,$GeV and 1$\,$TeV. In the low aerosol transmission under high zenith bin, sufficient data are not available.}
 \label{fig:mean_400-1000}
\end{figure*}

In the following, we investigate deviations of {integrated fluxes} from clear-night expectation of single nights for the three
energy bins: 200--400~GeV, 400--1000~GeV and $>$1000~GeV. 
As before, the error-weighted means and distributions of both types of deviations were computed for different zenith and $T_{9 \, \text{km}}$ ranges. 

Figure~\ref{fig:mean_200-400} shows the results derived for the flux in the 200--400~GeV energy range. 
The highest zenith bin has been excluded here, since the energy threshold, for optimal atmospheric conditions, lies already at $\sim$200~GeV (at 50$^\circ$) and $\sim$420~GeV (at 62$^\circ$)~\citep{aleksic:2016}. Lower aerosol transmission will even further increase the energy threshold and make this energy range susceptible to complicated threshold effects.

The results shown in Fig.~\ref{fig:mean_200-400} confirm the previously discussed improvements: 
In the highest transmission bin, almost no difference can be observed between uncorrected and corrected data. 
Both means are very close to the origin, indicating no significant offset of the reconstructed fluxes from the reference on average. 
At  $0.82<T_{9 \, \text{km}}<0.9$, 
further improvements can be obtained by applying corrections, resulting in both distributions moving very close to the origin. 
The data taken under $0.65<T_{9 \, \text{km}}<0.82$ are corrected satisfactorily for low zeniths. At medium zeniths, a small bias of $(-10\pm4)\%$ remains. For the lowest aerosol transmission case, an improvement of the distribution means from $(-50\pm3)$\%  to $(-15\pm4)$\% and from -4.0\,$\sigma$ to -0.6\,$\sigma$, respectively, can be observed at low zenith angles. The LIDAR corrections provide an almost full recovery of the flux on average. For medium zenith ranges, improvements from $(-44\pm3)$\%  to $(-26\pm5)$\% and from -3.8\,$\sigma$ to -1.4\,$\sigma$ are observed. The rather low transmission under an increased zenith angle may make
threshold effects  start to play a role here. 

The results of the corrected data obtained for the medium energy range 400-1000~GeV are shown in Fig.~\ref{fig:mean_400-1000}. For low and medium zenith angles and aerosol transmission $T_{9 \, \text{km}}>0.9$, the data show an almost identical distribution before and after correction.
At high zeniths, differences become apparent: mean values of both uncorrected and corrected data show a small offset from the origin. Data taken close to the highest zenith angles of 62$^\circ$ may suffer from threshold effects here, leading to a slight under-reconstruction of the flux. The LIDAR corrections seem to worsen the flux reconstruction even further: 
the standard deviation increases from 12\% to 15\% and 1.2\,$\sigma$ to 1.6\,$\sigma$. As previously observed, the LIDAR corrections may cause a downward correction of the flux in some cases. 
The results for $0.82<T_{9 \, \text{km}}<0.9$ without corrections show clear offsets of up to around -18\% for all three zenith ranges. After applying corrections, the distributions become well-centered, except in one case, where -10\% is achieved. Similarly, for $0.65<T_{9 \, \text{km}}<0.82$, the corrections achieve similar improvements. In the lowest aerosol transmission region $0.5<T_{9 \, \text{km}}<0.65$, an improvement from $(-52\pm3)$\% to $(-22\pm4)$\% and from -3.5\,$\sigma$ to -0.7\,$\sigma$, respectively, can be observed at low zeniths. After correction, a deviation of the mean from the origin remains, but the distribution of individual nights is found close to the origin. A few nights with better statistics seem to show some strong remaining offset. The larger spread of percental deviations can be attributed to the worse statistics in this low aerosol transmission bin. At medium zeniths, a significant improvement is achieved, with a remaining systematic offset of $(-12\pm5)$\% from the origin. In the zenith angle region above 50$^\circ$ there are not sufficient data available to draw conclusions.

\begin{figure*}[!ht]
  \centering
  \includegraphics[width=17cm]{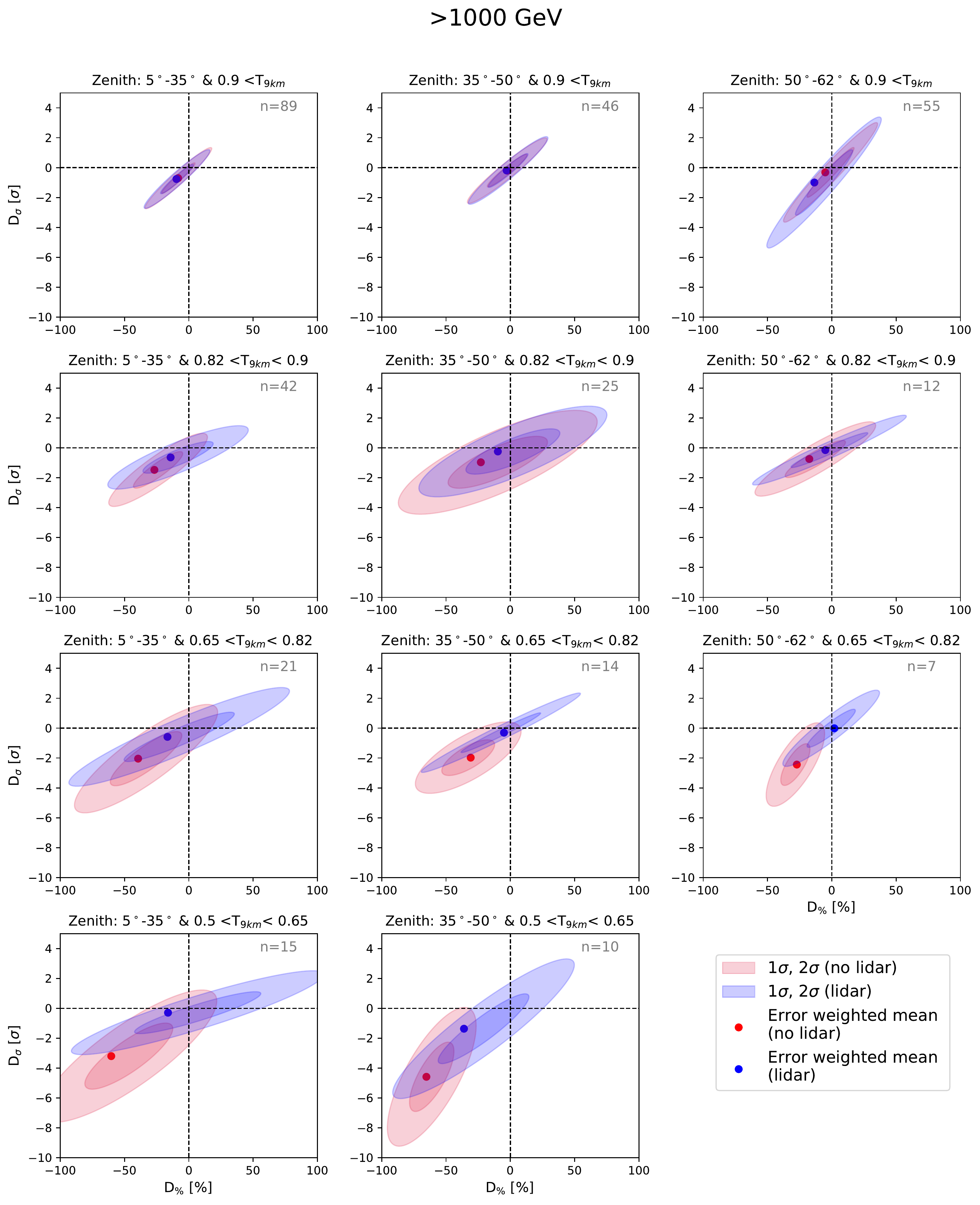}
 \caption{Same as Fig.~\protect\ref{fig:mean_200-400}, but for the energy region above 1$\,$TeV. In the low aerosol transmission under high zenith bin, sufficient data are not available.}
 \label{fig:mean_1000}
\end{figure*}

Finally, Fig.~\ref{fig:mean_1000} shows the results obtained for the high-energy region above 1$\,$TeV. In the top row, there is almost no difference between uncorrected and corrected data for low and medium zenith angles. At high zeniths, the LIDAR corrections lead again to a wider distribution of deviations, as well as less centered mean values. For high aerosol transmission, $0.82<T_{9 \, \text{km}}<0.9$, the results are also similar to the previous energy ranges: small offsets from the origin without corrections get significantly reduced to below -15\% on average after  applying the LIDAR corrections. In the medium aerosol transmission bin, $0.65<T_{9 \, \text{km}}<0.82$, the corrections work well for the higher zenith bins. At low zeniths, an offset of $(-16\pm4)$\% persists after corrections. 
The worse performance in this instance might be due to a few outliers resulting from the worse event statistics at such high energies. At low aerosol transmissions, $0.5<T_{9 \, \text{km}}<0.65$, conclusions are also similar to those obtained from previous energy ranges: at low zeniths, data are strongly corrected, but show a remaining bias of around $(-15\pm6)$\% from the origin after correction. At medium zeniths, a significant discrepancy of about $(-36\pm6)$\% remains. Again, there are not sufficient data available in the zenith angle region above 50$^\circ$. 
In general, the offsets observed for all uncorrected data 
are larger compared to those obtained at lower energies, reaching  $\lesssim$-60\% in the lowest aerosol transmission bin. This can be interpreted as a direct result of the spectral softening at higher energies: the steeper the spectrum, the larger the effect of a wrongly reconstructed energy on the resulting flux. Also, the ellipses are much wider along the horizontal axes, a direct result of the lower event statistics above 1$\,$TeV. 
The width of the distributions in terms of statistical significance, however, remains very similar to the previous energy ranges. 

\subsection{Comparing two methods of LIDAR corrections with period-averaged data}
\label{sec:period_averaged}

\begin{figure*}
\centering
  \includegraphics[width=17cm]{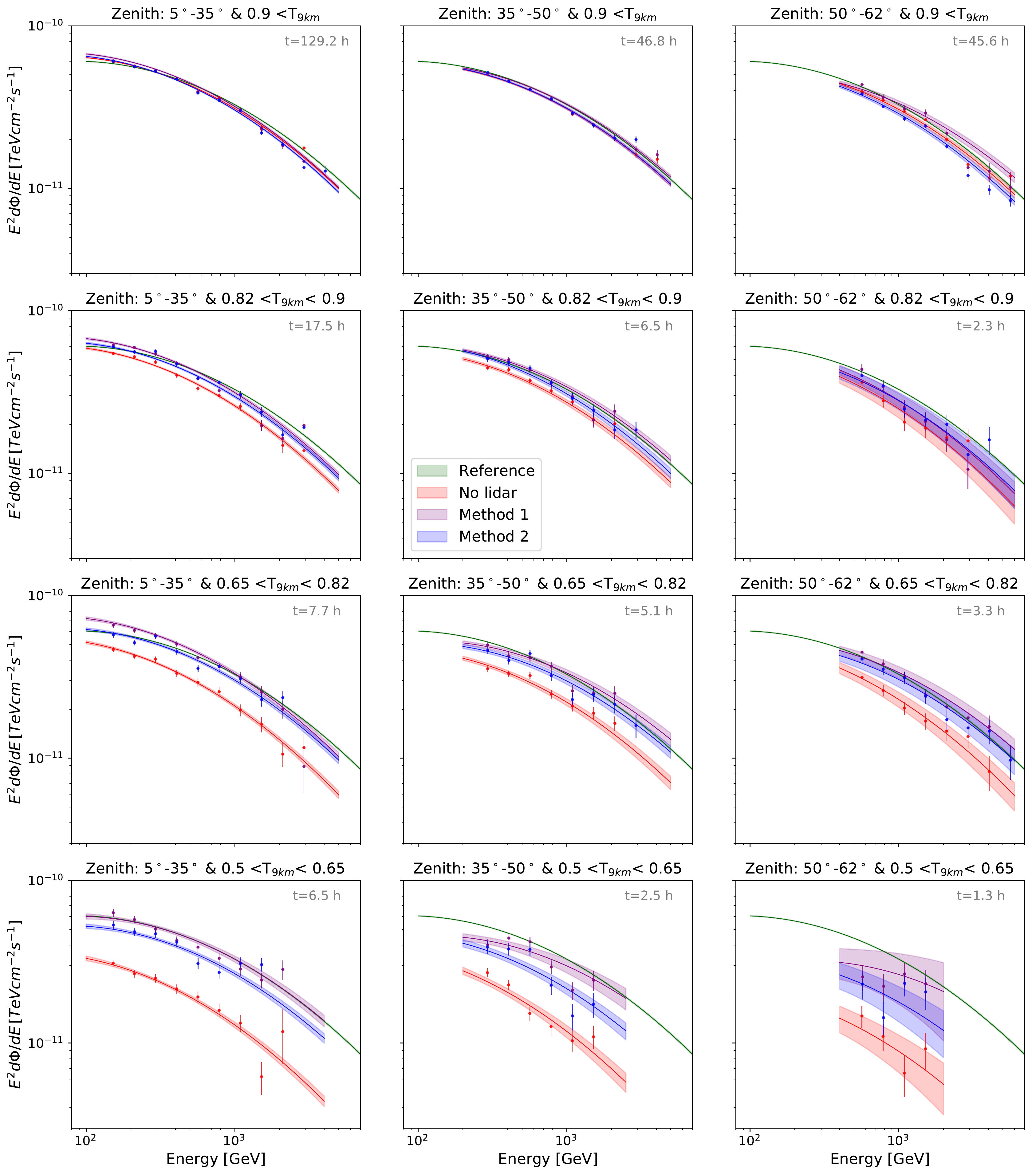}
  \caption{Period-averaged SEDs and corresponding flux points for nine different zenith angle and atmospheric transmission bins without (red) and with either type (purple and blue) of LIDAR corrections. The reference spectrum is given in green. The effective time of the data used for the individual spectra is displayed in the top-right corner of each subplot.}
  \label{fig:spectra}
  \vspace{10pt}
\end{figure*}

\begin{figure*}
\centering
  \includegraphics[width=17cm]{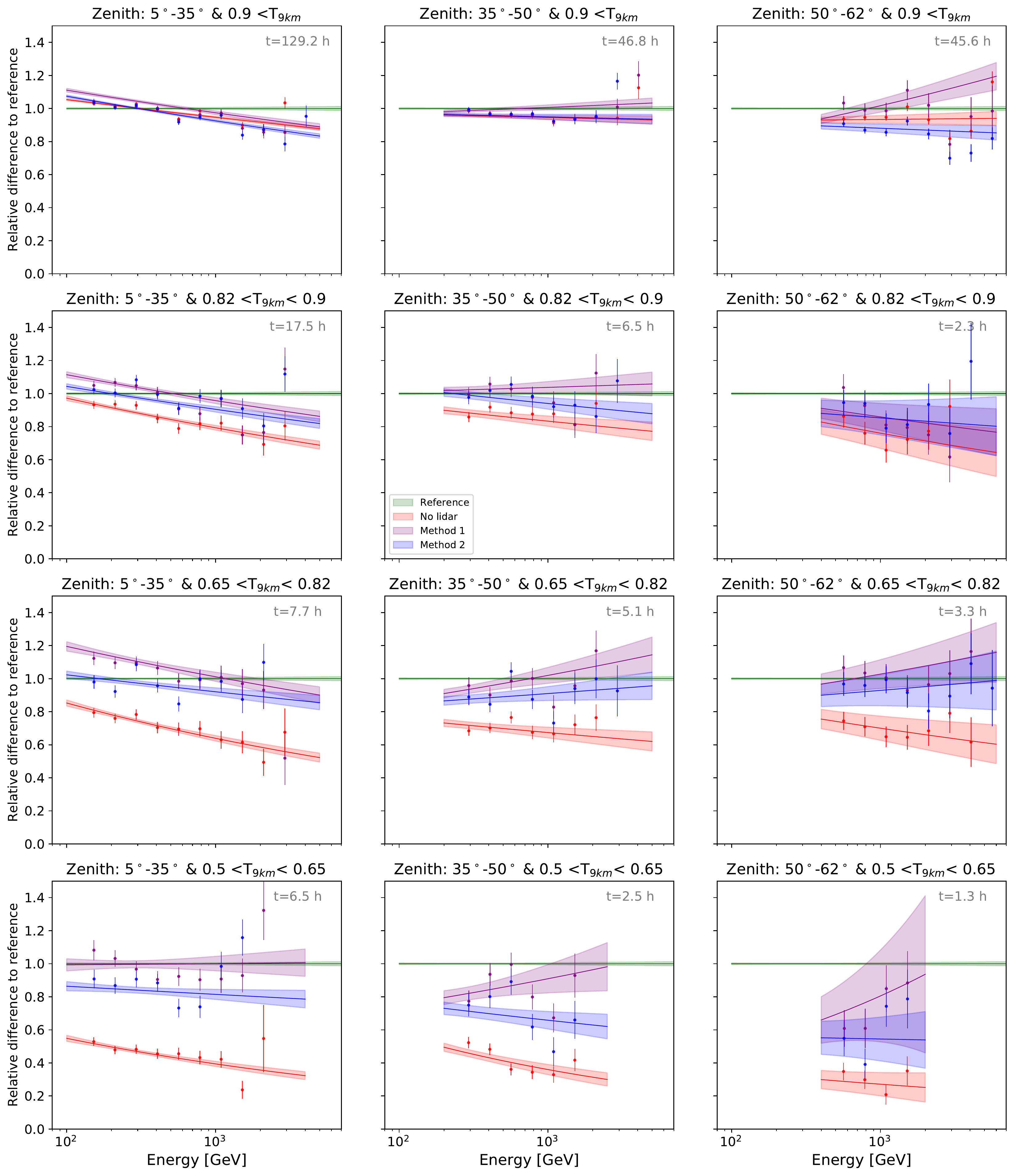}
  \caption{Relative difference between the period-averaged spectral fits and flux points to the reference spectrum for nine different zenith angle and atmospheric transmission bins without (red) and with either type (purple and blue) of LIDAR corrections. The reference spectrum is given in green. The effective time of the data used for the individual spectra is given in the top-right corner of each subplot.}
  \label{fig:rel_diff}
\end{figure*}

So far, all reported results have been obtained \revone{from averaging over multiple single-night spectra.} Only after reconstruction, the statistical behavior of individual spectra was investigated. In this section, data spanning over all nights and across different analysis periods have been combined into single spectra, only separated by zenith angle and aerosol transmission. This strategy leads 
the smallest statistical uncertainties possible for a given aerosol parameter region. Figure~\ref{fig:spectra} shows
the estimated SED without applying LIDAR corrections and those obtained after applying both types of 
correction algorithms, described in Sects.~\ref{sec:method1} and~\ref{sec:method2}. The blue SEDs 
correspond to  Method~II, used to obtain all previously shown results. 
SEDs obtained with  Method~I are shown in purple. The effective observation time corresponding to the used data is provided in the top-right corner of each individual plot. The reference spectrum has been obtained here by combining all data (used for the respective period-wise references) into a single spectrum, which was subsequently fitted to a log-parabola (\revone{resulting fit: $f$=(6.96$\pm$0.03)$\cdot10^{-10}$~TeV~cm$^{-2}$s$^{-1}$, $a$=-2.238$\pm$0.006, $b$=0.485$\pm$0.007}). This period-averaged reference spectrum also provides the shape parameter $b$, which had been fixed for all individual spectral fits of the previous sections. In order to make uncertainties better comparable, the reference spectrum of Fig.~\ref{fig:spectra} 
was then also fitted with the shape parameter fixed, yielding the shown statistical uncertainty band. 
For a more detailed representation, Fig.~\ref{fig:rel_diff} contains the residuals in terms of the relative difference with respect to the reference SED. 

The results from the combined spectral analysis support the previously drawn conclusions: the spectra taken under good aerosol conditions, $T_{9 \, \text{km}}>0.9$ and low zeniths show that the flux normalization is fitted correctly with both methods with a slight divergence for the spectral tilt. A similar behavior had already been observed in the top-left part of Fig.~\ref{fig:mean_par_a}. At medium zeniths, all three spectra show deviations below 10\%, and the corrections barely alter the spectra. In the high zenith bin,  Method I corrects the spectrum harder than the reference. Method II causes a slight downward flux correction and with that does not improve the uncorrected SED.

In the second row, the results for high aerosol transmissions ($0.82 < T_{9 \, \text{km}} < 0.9$) are shown. The spectrum without corrections at zeniths below 35$^\circ$ results to be too soft, which leads to a discrepancy from the reference of up to 30\% at a few TeV. Both correction methods upscale the flux, but fail to recover the correct spectral tilt. 
On the contrary, Method I corrects the flux normalization stronger than Method II and crosses the reference spectrum 
at around 500\,GeV, whereas the normalization for Method II coincides with the reference at 200\,GeV. 
In the medium zenith range, 
Method I reconstructs compatible with the reference, whereas Method II produces a slightly too soft 
spectrum. At high zenith angles, both LIDAR correction methods produce very similar results, 
achieving corrections very close to the reference spectrum, with only a slight under-reconstruction.

In the aerosol transmission bin ($0.65 < T_{9 \, \text{km}} < 0.82$), the LIDAR corrections are able to substantially improve the reconstructed spectra. Before corrections, deviations range from 20\% for low energies up to 50\% for high energies. Again, all three spectra show a stronger spectral tilt compared with the reference. At low zeniths, the flux normalization of the corrected spectrum of Method I crosses the reference spectrum at about 1$\,$TeV, leading to a light over-correction for lower energies. Method II normalizes the flux correctly at around 150$\,$GeV, resulting in a good overall agreement, but slight under-correction for higher energies. In the medium zenith bin, 
both methods slightly over-correct the spectral tilt, 
with flux crossings at 700$\,$GeV (Method I) and 7$\,$TeV (Method II). At high zeniths, the uncorrected spectra again shows an excessively steep spectral tilt, which gets correctly reconstructed with 
both methods; however, the statistics are quite low here. 
 
 For the lowest aerosol transmission bin, ($0.5 < T_{9 \, \text{km}} < 0.65$),  at low zeniths Method I 
 is able to fully correct the original deviations from 40\% up to 70\%. 
 Method II shows a rather constant deviation of $\sim$15\%, after correction, across the entire energy range. At medium  zeniths 
 Method I reconstructs a slightly too flat spectrum crossing the reference 
 at 2.5$\,$TeV. Method II under-reconstructs the spectrum by 25-35\%. For the highest zenith bin, only 1.3$\,$h of data are available in total, resulting in poorly constrained spectra. Method I reconstructs the spectrum compatible with the reference, whereas Method II results in too low reconstructed fluxes.

Altogether, both methods of using the LIDAR data to correct spectra obtained with IACTs work well across most domains. Method I almost always produces corrected spectra compatible with the reference, whereas Method II consistently seems to under-correct.

\section{Discussion}
\label{sec:discussion}
\subsection{Systematic uncertainties}
The LIDAR correction method has been used in the past to increase the amount of scientifically usable data impaired by nonoptimal atmospheric conditions \citep[e.g.,][]{2021ApJ...908...90A,MAGIC_M15,2018MNRAS.480..879M,2017MNRAS.470.4608A}. The method is, however, affected and limited by systematic uncertainties, which can be classified into the following categories: uncertainties stemming from the LIDAR hardware and data analysis itself, those related with the combined LIDAR and IACT observation modes, and finally those arising from IACT data correction.

\subsubsection{LIDAR hardware and data analysis}

The first source of uncertainties stems from the LIDAR itself and the estimation of the aerosol transmission profile. The LIDAR hardware, the aerosol extinction profile reconstruction method, and its associated  systematic uncertainties have been extensively described in~\citet{LIDAR1}. 
The LIDAR is absolutely calibrated with a correlated uncertainty of 0.01 for the aerosol optical depth of complete layers to ground and a night-wise uncorrelated uncertainty of 0.015.  This leads to a combined accuracy of $\lesssim$2\% for the aerosol transmission of a full layer, independent of assumptions on the LR. 

For unfavorable Cherenkov light emission heights, however, like those located inside an extended aerosol or cloud layer, assumptions on the LR do affect the 
accuracy of reconstructing the aerosol transmission from that point to its base height. In that case, the accuracy is about 0.1~times  the optical depth of that layer. This uncertainty may be reduced in the future through the use of Raman LIDARs~\citep{Gaug:RamanLIDAR:2019}.

The correction method presented in this work 
introduces further systematic uncertainties:  
the wavelength correction of the aerosol extinction coefficients, described in Sect.~\ref{sec:data_correction}, 
introduces a  systematic uncertainty of 0.5\% for the ground layer transmission during clear nights and up to 5\% for the case of nights affected by strong calima, apart from a probably small bias due to the use of diurnal \AA ngstr\"om coefficients for nocturnal observations. That uncertainty can be easily remedied through the use of a laser operating closer to the average Cherenkov wavelength or a system adding an additional UV line to the currently used green one~\citep{Gaug:RamanLIDAR:2019}.

\begin{figure*}
    \centering
    \includegraphics[width=17.5cm]{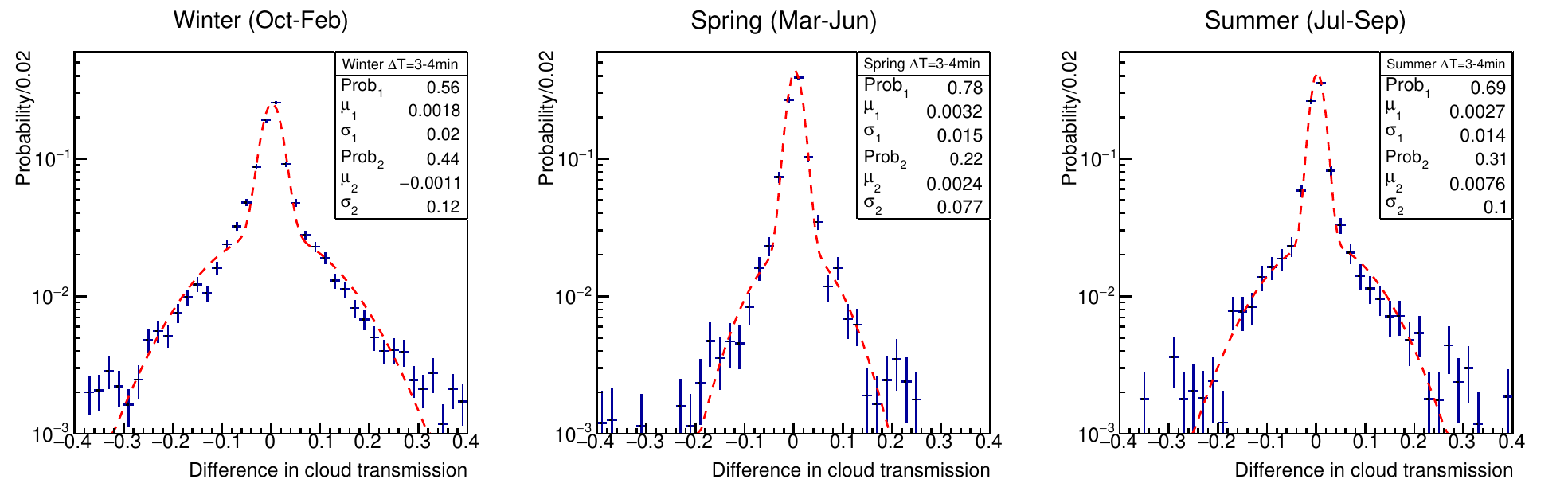}
    \includegraphics[width=17.5cm]{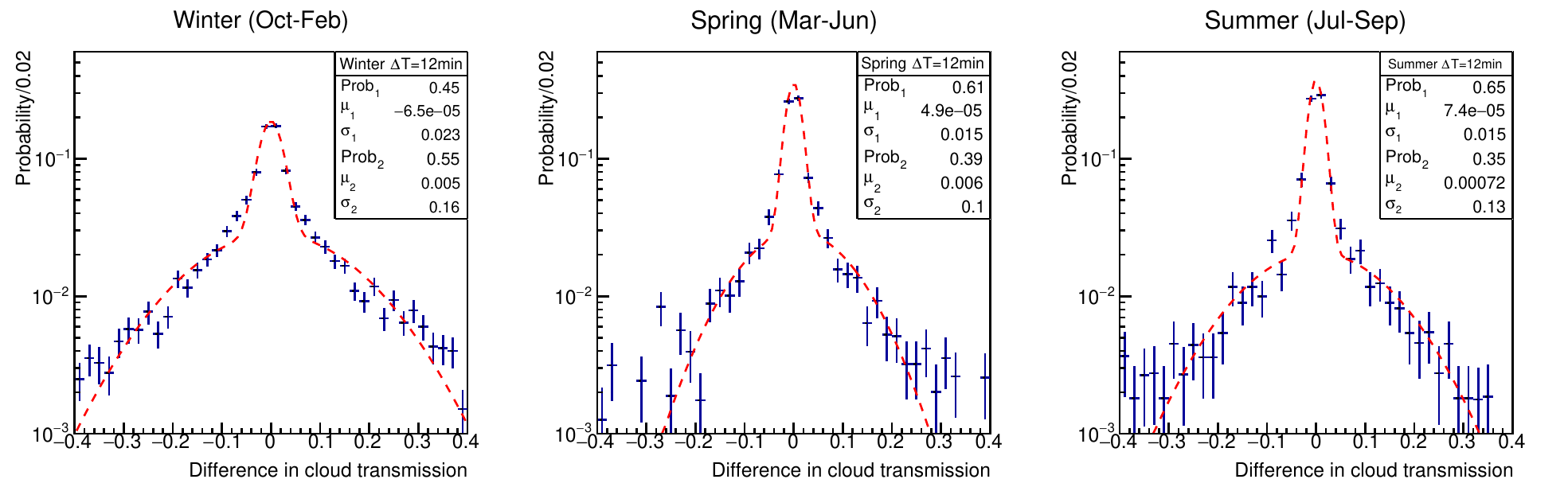}
    \caption{Differences in cloud transmission for LIDAR data sets separated by \revone{five} or four minutes (top, according to the LIDAR operation schemes before or after 2016; see~\protect\citealt{LIDAR1}), or by 12~minutes (bottom).  The distributions have been separated by seasons: the winter months,
    where the highest cloud fractions are observed (left); the spring months,
    with the lowest cloud fraction (center); and the summer months, 
    affected by a subtropical climate and frequent breakdown of the temperature-inversion layer below the observatory (right). All data have been selected such that at least one of the two data sets separated by the given time span contains at least one cloud and both belong to the same tracked sky coordinates. Total cloud transmissions from 12~km above ground down to the end of the ground layer     have been compared. The distributions are normalized, meaning no information about the relative occurrence frequency of clouds between seasons can be read from the plots. All distributions have been fit to a double-Gaussian distribution. The narrower Gaussian of the two is interpreted as the LIDAR continuing to observe the same cloud, whereas the wider one may represent cases where one of the two data sets has already missed the cloud. The scales \textit{Prob}$_1$ and \textit{Prob}$_2$ show the relative frequency of one or the other case occurring. \revtwo{The bin width of the difference in transmission is 0.02.}}
    \label{fig:dts}
\end{figure*}

\revone{Figure~\ref{fig:dts} (top row) shows differences in cloud transmission for subsequent LIDAR data sets, affected by clouds (separately for each  season). All distributions show a double-Gaussian structure that suggests two types of cases: case~1 represents both data sets observing the same cloud and produces the narrow part of the distribution, whereas case~2 shows the wider distribution and corresponds to cases, where only one out of the two subsequent LIDAR data sets observes the cloud.}
When different parts of a cloud move through the FoV of the LIDAR during its 100-second-long data acquisition, or when the LIDAR FoV covers several substructures of a cloud at a same time, uncertainties $\lesssim$2\% are obtained (see the values obtained for $\sigma_1$ and \textit{Prob}$_1$ in Fig.~\ref{fig:dts}, top), affecting about 55\%-80\% of the cloud data. Furthermore, it may be possible that the cloud moves through the LIDAR FoV during its exposure, and a part of the laser returns are already characterized by the absence of cloud. We estimate that $<$44\% of cloud observations during winter, $<$22\% during spring, and $<$31\% during summer are hampered by rapid cloud movement (see the values obtained for $\sigma_2$ and \textit{Prob}$_2$ in Fig.~\ref{fig:dts}, top, \revtwo{representing width and relative area of the wider of the two Gaussians fitted}) with corresponding uncertainties ranging from 8\% to 12\%.  We note that our Crab Nebula test sample shows high prevalence of winter data and may hence be overly affected by this uncertainty. 
Uncertainties of this type can only get reduced through faster LIDARs, operating with shorter exposure times and higher sampling rates.


\subsubsection{Combined LIDAR and IACT observation modes}

In order to avoid pointing the laser into the FoV of the MAGIC telescopes, the LIDAR tracks the telescopes with a $5^{\circ}$ offset with respect to their optical axes. In the case of stratified aerosol layers, for example calima or extended clouds, this mismatch between observed and characterized FoVs produces a negligible error. Only in the case of very localized layers, for example fine cirrus, can this lead to a stronger discrepancy between the estimated transmission profile
and the real aerosol transmission  that has affected the Cherenkov light observed by the IACT. In the worst case, science data affected by such a cloud may remain completely uncorrected, whereas other data in temporal proximity become over-corrected, because the cloud has moved out of the FoV of the IACT, but into the one of the LIDAR. The cloud base heights of such events lie above 8~km a.s.l. at La~Palma~\citep{Gaug:Atmohead2022}, which makes particularly low-energy gamma-ray events prone to this effect. Instead, Cherenkov light of higher energy showers gets emitted largely below the cloud. 
\revone{Figure~\ref{fig:dts} (bottom row) shows differences in cloud transmission for LIDAR data sets separated by 12~min, which is the typical time that the MAGIC telescopes' FoV will enter the line of sight characterized by the LIDAR during normal tracking. As before, all distributions show a double-Gaussian structure, which suggests two types of cases: a cloud observed in both data sets or a cloud observed in only one of the two. From the statistics of these distributions, }
we estimate that up to 55\% of the data (particularly during winter) affected by cirrus suffer from such FoV mismatches, leading to \revone{an uncertainty of the aerosol transmission estimate} proportional to half the cloud's optical depth at low energies (i.e., up to 8\% in transmission correction; see values for \textit{Prob}$_2$ and  $\sigma_2$ shown  in Fig~\ref{fig:dts}, bottom), and about 20\% to 30\% of it for the high energies, resulting in a $\sim$6\% error on the transmission correction. This class of uncertainties can only be remedied with a passive device following the exact FoV of an IACT~\citep{Ebr:2021}.

\begin{figure}
    \centering
    \resizebox{0.495\hsize}{!}{\includegraphics{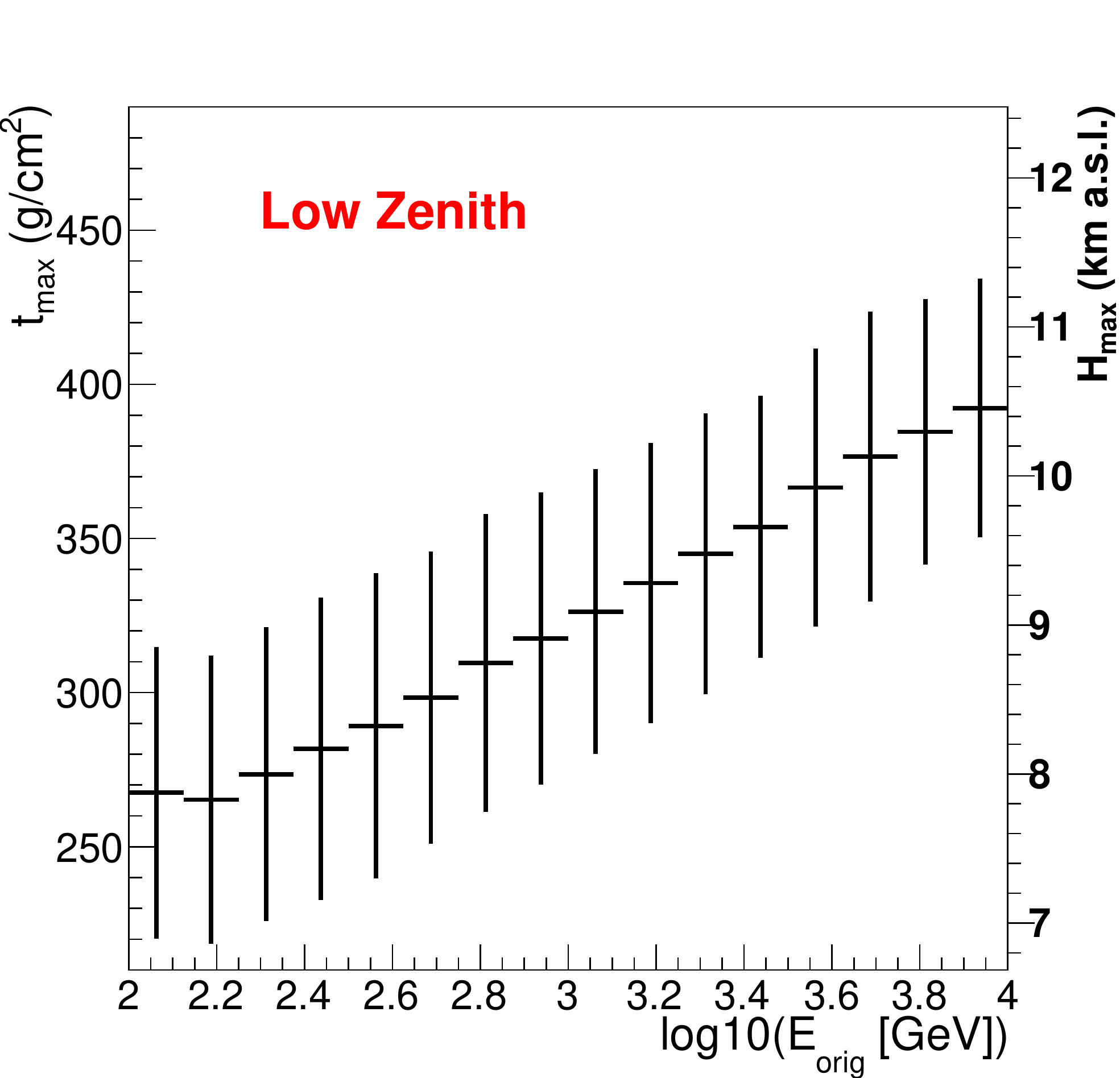}}
    \resizebox{0.495\hsize}{!}{\includegraphics{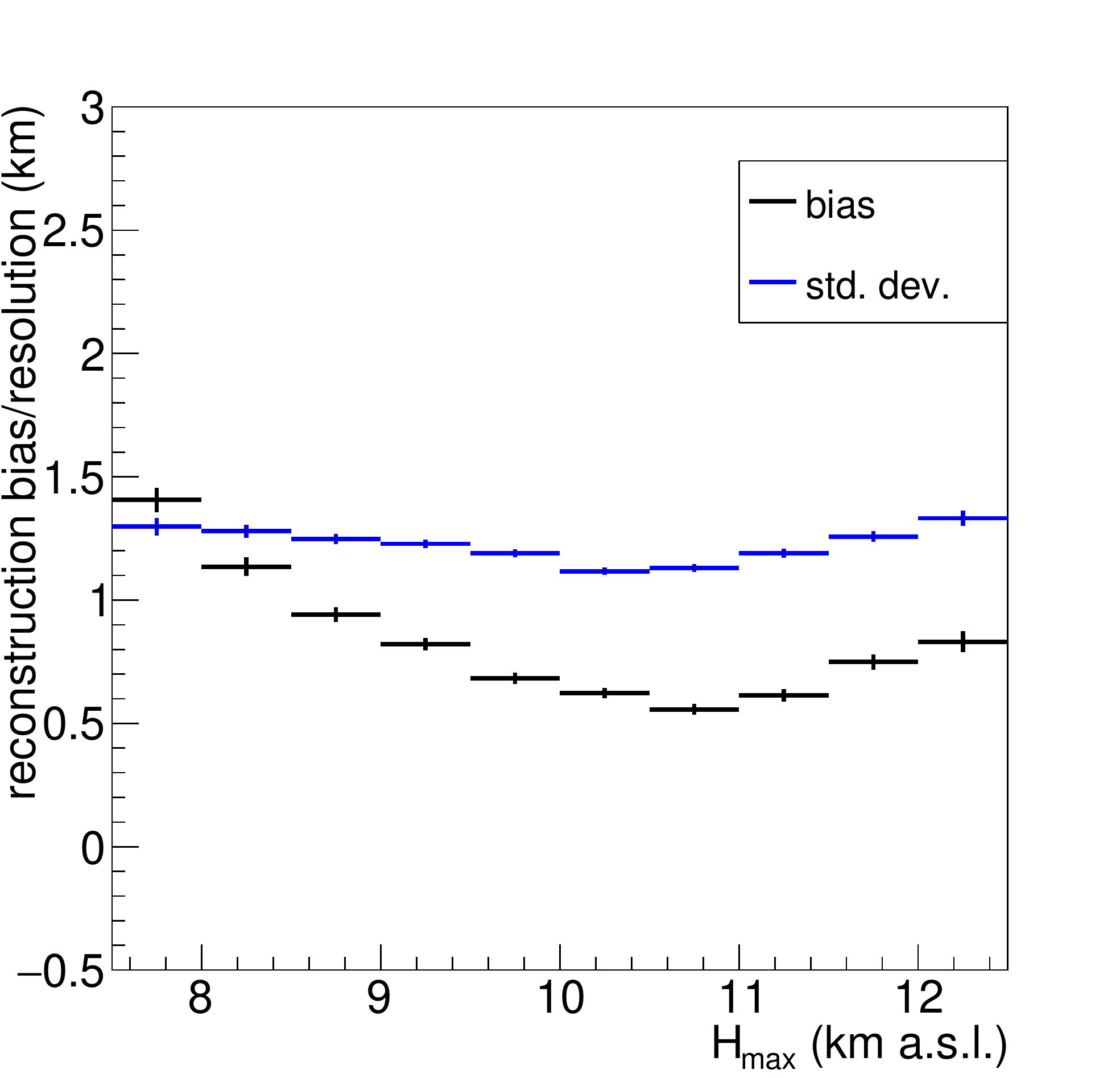}}
   \resizebox{0.495\hsize}{!}{\includegraphics{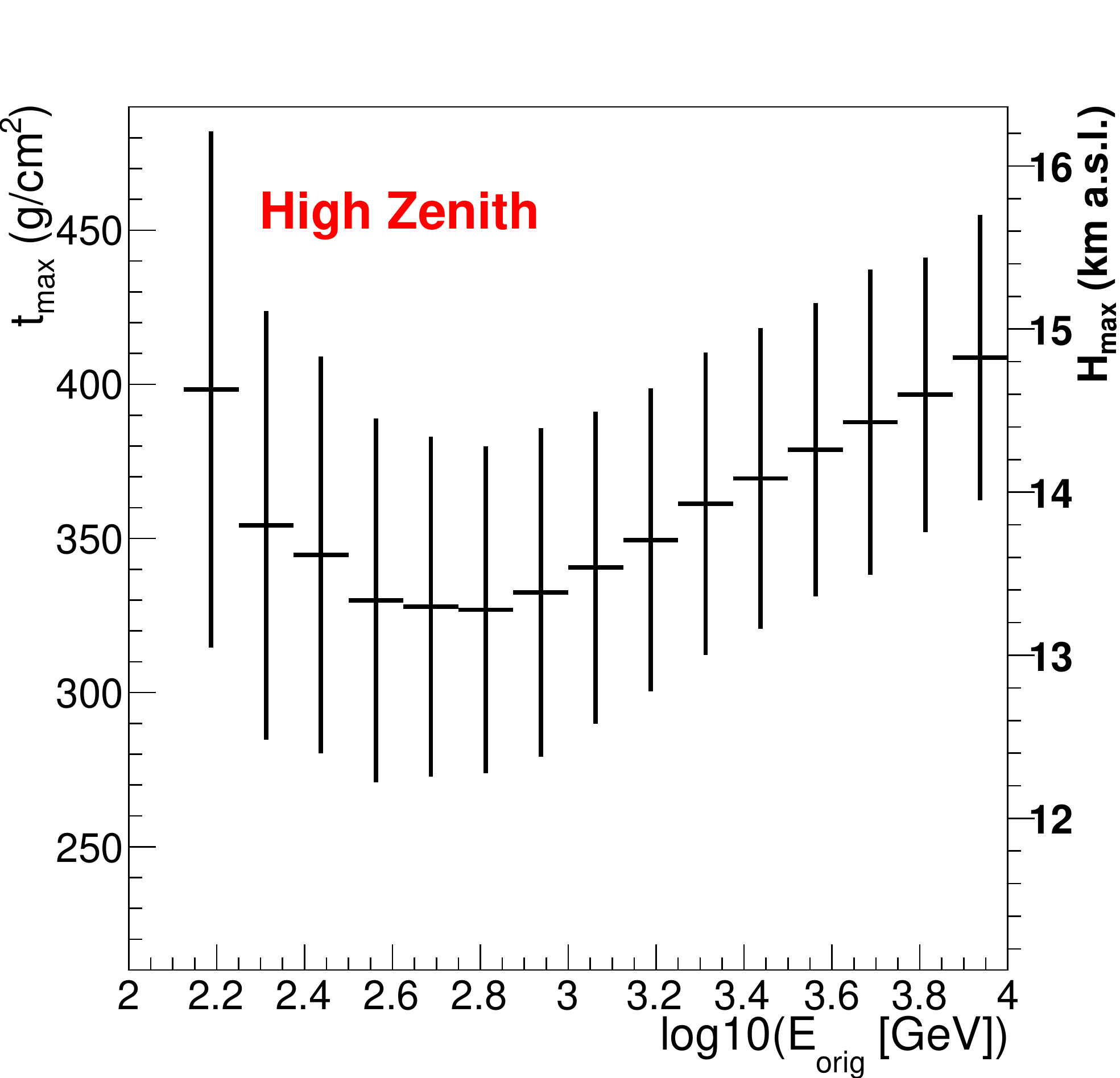}}
    \resizebox{0.495\hsize}{!}{\includegraphics{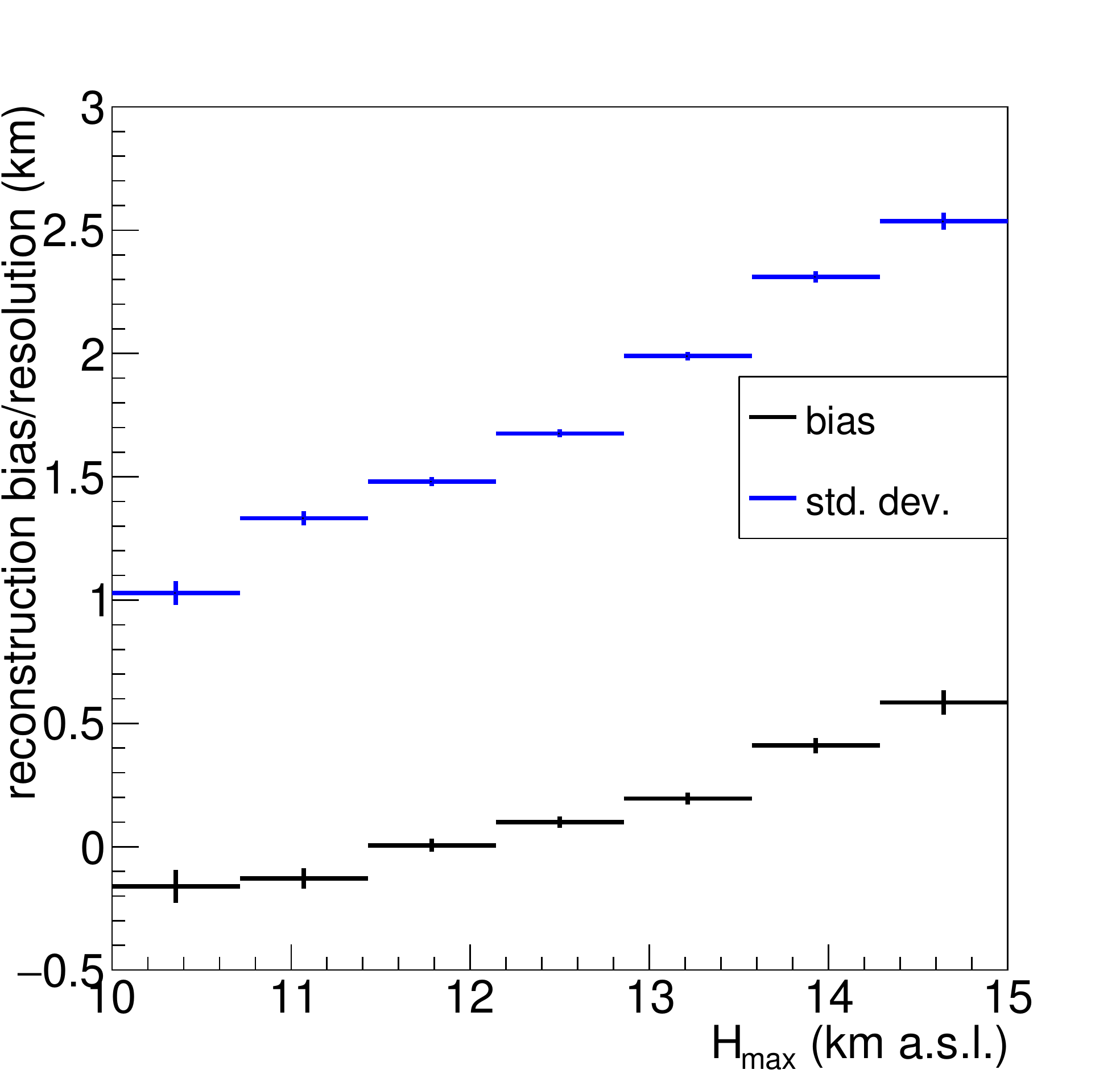}}
    \caption{Shower maxima obtained from gamma-ray simulations. 
    Left: Profile of shower maxima as a function of the logarithm of energy. The left axis shows the slant depth traversed and the right axis the corresponding altitude above sea level. Right: Bias and resolution of the {reconstructed} height of shower maximum as a function of the {simulated} shower maximum. The top plots show fully detected and reconstructed showers at low zenith angles, below 35$^{\circ}$, and the bottom row at high zenith angles, 60$^{\circ}$.}
    \label{fig:hmaxs}
\end{figure}

\subsubsection{Gamma-ray data correction method}

The correction of the gamma-ray event energy described in Sect.~\ref{data_energy} requires an estimated emission profile of the Cherenkov light. Monte Carlo simulations suggest that the longitudinal shower evolution can be described fairly well by a Gaussian profile while the  shower maximum has to be determined from stereo reconstruction.

Figure~\ref{fig:hmaxs} (left) shows the expected particle shower maxima, both in terms of slant depth (left axis) and atmospheric altitude (right axis), of gamma-ray showers detected by \revone{the MAGIC telescopes} and fully analyzed with the MAGIC \revone{analysis chain}. The well-known increase of penetration depth with energy can be observed, combined with energy-dependent threshold and selection effects. One can see that the average shower maximum height decreases by $\sim$2.5~km from the lowest to the highest observable gamma-ray induced shower energies. On the other hand, shower-to-shower fluctuations are on the order of $\pm$(1--1.5)~km, almost as large as the energy dependence of the average heights. It is therefore important to correctly reconstruct the height of the shower maximum~\citep{Chadwick:1996a} on an event-by-event basis. 
On the right-hand side of Fig.~\ref{fig:hmaxs}, these altitudes have been compared with the reconstructed height of the maximum of Cherenkov light emission and characterized in terms of bias and resolution. We note that this is not a true statistical bias, because two slightly different quantities are compared: on one hand the (simulated) maximum of particles in the shower and on the other hand the (reconstructed) maximum of {detected} Cherenkov light emission. The latter is affected by atmospheric extinction and image leakage outside the camera. For this reason, a strong dependence of both quantities on impact distance is also present (not shown in Fig.~\ref{fig:hmaxs}). We used the findings of Fig.~\ref{fig:hmaxs} to derive a conservative estimate of the uncertainty of the shower maximum reconstruction in terms of combined bias and resolution. 
The reconstruction uncertainty affects actually rather energy resolution (and through the energy-bin spill-over in spectrum unfolding also the reconstructed spectral index), whereas the reconstruction bias increases the systematic uncertainty of the full spectrum reconstruction. 
We assume an average reconstruction bias of $\lesssim$0.8~km in the dominant range (e.g., from 9~km to 12~km a.s.l. of shower maxima at vertical incidence). For a typical geometrical width of cirrus of $\sim$2.5~km~\citep{LIDAR1}, the effect roughly scales with $\sim$0.3 times the aerosol transmission of cirrus (see Eq.~\ref{tau_aer}). In our Crab Nebula test sample, about half the data are affected by cirrus ranging above $\sim$9~km a.s.l.
Finally, the limited resolution of this method adds to the energy resolution for data affected by cirrus. Since cirrus are mostly found below 14~km a.s.l.~\citep{LIDAR1}, we assume an average uncertainty of $\sim$1.5~km on the reconstructed height of shower maximum, leading to a resolution of the correction of about 
\revone{$0.6\, \cdot \, (1-\overline{T}_\mathrm{aer})$}, 
which needs to be added quadratically to the energy resolution of $\sim$16\%~\citep{aleksic:2016}. 
 This  uncertainty is intrinsic to the reconstruction method proposed and cannot be remedied  by better auxiliary hardware. Instead, time-interval-wise tailored simulations are required to improve on this part~\citep{Gaug:2017Atmo}.

The width of the longitudinal emission profile is also energy-dependent and fluctuates around the most probable value for a given energy with a standard deviation of 10-15\%. 
Fluctuations of the shower width, in combination with a cloud layer crossing the air shower, primarily lead to a degradation of the energy resolution, as in the case of the air shower maximum reconstruction. Assuming a linear growth of extinction across the layer leads to a contribution of $0.3\cdot\overline{T}_\mathrm{aer}$ to the energy resolution.

Finally, the assumptions made in Method~I and Method~II 
lead to systematic uncertainties. We cannot derive a simple estimate for these errors without extensive simulations.\ Instead, we used the excess fluctuations, observed in the previous sections, to characterize the overall {observed} uncertainty. We attributed the remainder of these fluctuations to the sum of the previously estimated uncertainties to derive a number for the methodological assumptions.

\begin{table*}
    \centering
    {\small
    \begin{tabular}{ccccl}
    \toprule
    Type  
          &   clouds     &   calima    &    this sample  & comments\\
         &  (\%)        &  (\%)       & (\%) \\
    \midrule
   \multicolumn{5}{c}{Uncertainties on optical depth of layer - LIDAR only}   \\
   \midrule
Absolute calibration of LIDAR                & --  & 2    &  1  & \\
Light emission inside layer / assumptions LR & 
 10
& 
\llap{$\ll$} 10      
&
\llap{$\lesssim$} 6
&
\\
Color correction                             & \llap{$<$} 1  & \llap{$<$} 5 & \llap{$\lesssim$} 1 
& \\
   Larger LIDAR FoV than cloud substructures   & 2     & --    & 1.5 \\
   Rapid cloud movement during LIDAR exposure   &  12 & --    &  6 
&
\\
\midrule
Quadratic sum for optical depth  & 
 16
&  
\llap{<} 8
& 
\llap{$\lesssim$} 9  &    \\
    \bottomrule  \addlinespace[0.1cm]
   \multicolumn{5}{c}{Systematic errors on $\overline{T}_\mathrm{aer}$ - science data correction}   \\
   \midrule
Rapid cloud movement / mismatch obs. vs. char. FoVs &  8 & -- &  4 \\
Air shower maximum reconstruction &
\llap{$\lesssim$} 30
& --    
& 
\llap{$\lesssim$} 15 
& affect mainly low and 
\\
Assumptions mean width air shower long. profile    
& 
\llap{$\lesssim$} 5    
 & -- 
 & 
\llap{$\lesssim$} 2.5 
& medium zenith angles
 \\
\midrule
Quadratic sum for $1-\overline{T}_\mathrm{aer}$  & 
\llap{$\lesssim$} 31 
&  
--
& 
\llap{$\lesssim$} 16  & $\lesssim$ 11\% for high zenith angles
\\ 
   \bottomrule  \addlinespace[0.1cm]
   \multicolumn{5}{c}{Resolution on $\overline{T}_\mathrm{aer}$ - energy resolution}   \\
   \midrule
   Air shower maximum reconstruction &
\llap{$\lesssim$} 60
& --    
& 
\llap{$\lesssim$} 33 
& 
\\
Assumptions mean width air shower long. profile    
& 
\llap{$\lesssim$} 30   
 & -- 
 & 
\llap{$\lesssim$} 17 
& 
 \\
\midrule
Quadratic sum for resolution of $1-\overline{T}_\mathrm{aer}$  & 
\llap{$\lesssim$} 70 
&  
--
& 
\llap{$\lesssim$} 37  & 
\\ 
\bottomrule 
    \end{tabular}
    }
    \caption{Systematic uncertainties for the aerosol optical depths that affect the observed Cherenkov light of the MAGIC telescopes.   
    }
    \label{tab:uncertainties}
\end{table*}

\begin{table*}
\newcolumntype{C}{ @{}>{${}}c<{{}$}@{} }
    \centering
    {\small
    \begin{tabular}{r>{$\!\!\!\!\!\!<~~$}c>{$\!\!\!\!\!\!<~~$}lc *3{rCl} c *3{rCl}} 
    \toprule
    \multicolumn{3}{c}{$T_{9\,\text{km}}$} & Prediction  & \multicolumn{3}{c}{\footnotesize Observed} &  \multicolumn{3}{c}{\footnotesize Observed} & \multicolumn{3}{c}{\footnotesize Reduced atm.} &
      Prediction &
    \multicolumn{3}{c}{\footnotesize Observed} & \multicolumn{3}{c}{\footnotesize Observed} &
    \multicolumn{3}{c}{\footnotesize Reduced atm.} \\
    \multicolumn{3}{c}{bin} & uncertainty &
    \multicolumn{3}{c}{\footnotesize bias} &
    \multicolumn{3}{c}{\footnotesize excess fluct.}  &  
       \multicolumn{3}{c}{\footnotesize excess fluct.}  &  uncertainty &
       \multicolumn{3}{c}{\footnotesize bias} & \multicolumn{3}{c}{\footnotesize excess fluct.} &
          \multicolumn{3}{c}{\footnotesize excess fluct.}  \\
         \multicolumn{3}{c}{} & {\footnotesize $\Delta\overline{T}_\mathrm{aer} / \overline{T}_\mathrm{aer}$} & \multicolumn{3}{c}{\footnotesize for $f$} & \multicolumn{3}{c}{\footnotesize for $f$} & 
         \multicolumn{3}{c}{\footnotesize for $f$} & 
        {\footnotesize for $a$} & 
         \multicolumn{3}{c}{\footnotesize for $a$} & 
         \multicolumn{3}{c}{\footnotesize for $a$} & 
         \multicolumn{3}{c}{\footnotesize for $a$} \\
    \midrule  
   0.9~ &  $T_{9\,\text{km}}$ & 1.0 
   &  1\%  
   & (0&\pm&2)\% &  (6&\pm&3)\% &   (4&\pm&5)\% &  
   0.004 & 
   -0.03&\pm&0.02  &  0.04&\pm&0.03 & 0.00&\pm&0.04\\
   0.82 &  $T_{9\,\text{km}}$ & 0.9 
   &  3\%  
   & (-2&\pm&2)\% & (\revone{8}&\pm&1)\% &  (\revone{7}&\pm&2)\%  & 
   0.01 &
   -0.01&\pm&0.02  &  \revone{0.10}&\pm&0.05 & \revone{0.09}&\pm&0.06\\ 
   0.65 &  $T_{9\,\text{km}}$ & 0.82 
   & 7\%  
   & (-5&\pm&4)\% & (\revone{18}&\pm&4)\% & (18&\pm&5)\% &
   0.03 &
   0.03&\pm&0.05  &  0.09&\pm&0.06 & 0.08&\pm&0.08   \\
   0.5~ &  $T_{9\,\text{km}}$ & 0.65 
   & 15\% 
   &  (-11&\pm&4)\% & (\revone{16}&\pm&8)\% & (\revone{16}&\pm&9)\% & 
   0.05 &
   0.02&\pm&0.03  &  \revone{0.06}&\pm&0.11 & \revone{0.05}&\pm&0.15    \\ 
      \bottomrule 
    \end{tabular}
    }
    \caption{Comparison of estimated vs. observed  systematic uncertainties after the LIDAR correction Method~II. {Predictions} have been calculated from the total displayed in Table~\protect\ref{tab:uncertainties}. {Observed} quantities refer to the average of the observations from three zenith-angle bins (e.g., displayed in Fig.~\protect\ref{fig:mean_par_f}). {Reduced} quantities have been obtained from "observed," after quadratic subtraction of the excess uncertainty observed with the clear-nights reference spectrum: 4\% for {flux}  and 0.04 for {spectral index} $a$. }
    \label{tab:uncTbins}
\end{table*}

\subsubsection{Comparison with excess fluctuations}

Table~\ref{tab:uncertainties} summarizes the systematic uncertainties described so far, relative to $\overline{T}_\mathrm{aer}$. We note that many of the listed uncertainties may be correlated.\ On the other hand, the total uncertainty cannot exceed the correction itself. It has been shown in the past~\citep{garrido2013} that the relative energy reconstruction bias $(E_\mathrm{corr}-E_\mathrm{est})/E$ (above threshold) scales roughly as $1/\overline{T}_\mathrm{aer}$, and hence additional systematic uncertainties on the energy scale are expected to roughly scale $\propto\Delta \overline{T}_\mathrm{aer} / \overline{T}_\mathrm{aer}$. Similarly, flux reconstruction biases $(F_\mathrm{corr}-F_\mathrm{est})/F$ at a given energy \revone{may scale, at first order,} as $\overline{T}_\mathrm{aer}$, 
\revone{since the reconstructed flux is inversely proportional to the effective area, which itself scales, at first order and above the energy threshold, as $\overline{T}_\mathrm{aer}$~\citep{garrido-master,garrido2013}}. The combination of both scalings sometimes gives the wrong impression that aerosols do not alter much at all a pure power-law spectrum. 
In this framework, the LIDAR correction-induced systematic uncertainty on an SED normalization is expected to scale roughly $\propto\Delta \overline{T}_\mathrm{aer} / \overline{T}_\mathrm{aer}$.

\begin{figure}
    \centering
    \resizebox{0.495\hsize}{!}{\includegraphics{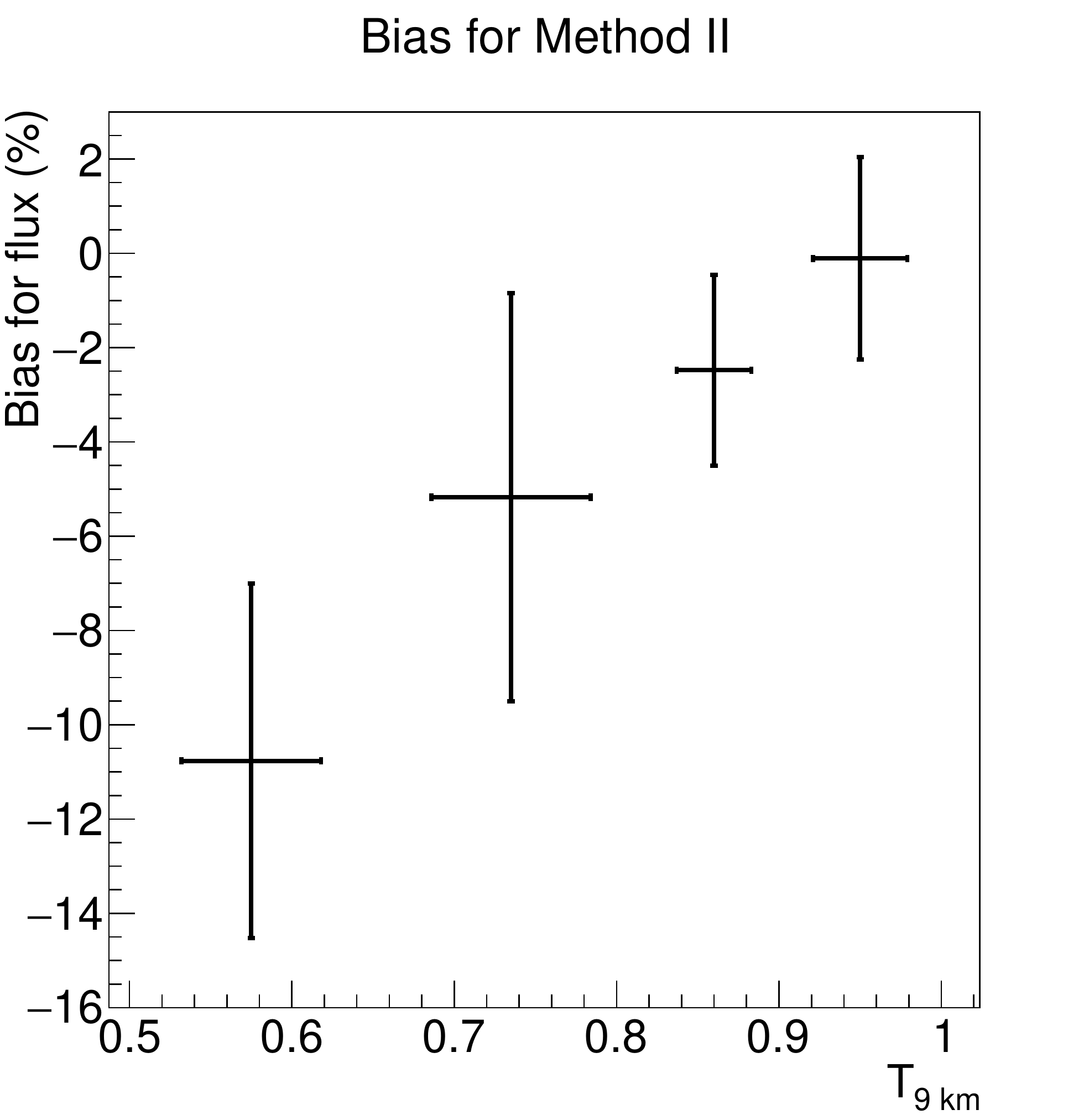}}
    \resizebox{0.495\hsize}{!}{\includegraphics{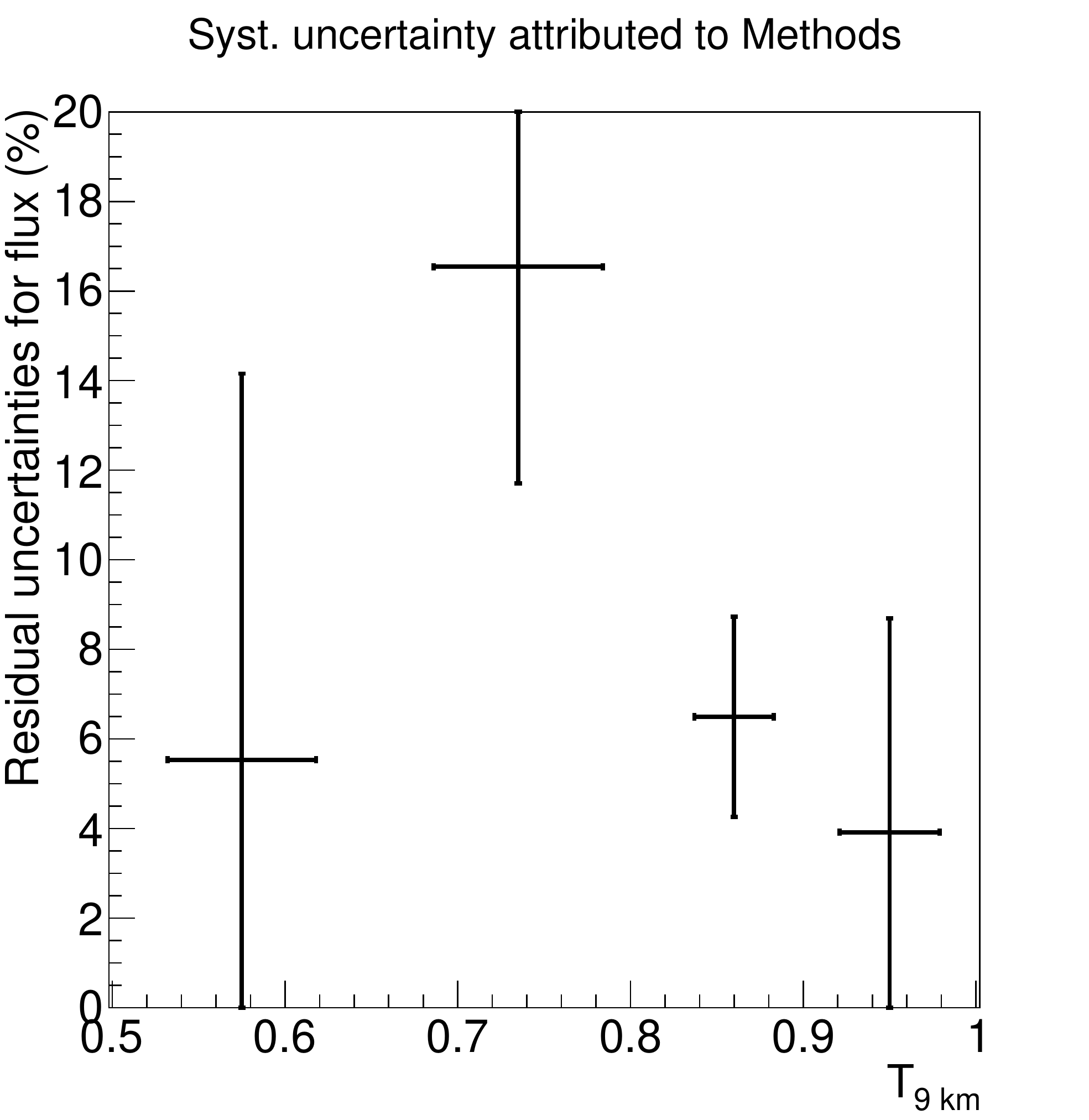}}
    \resizebox{0.495\hsize}{!}{\includegraphics{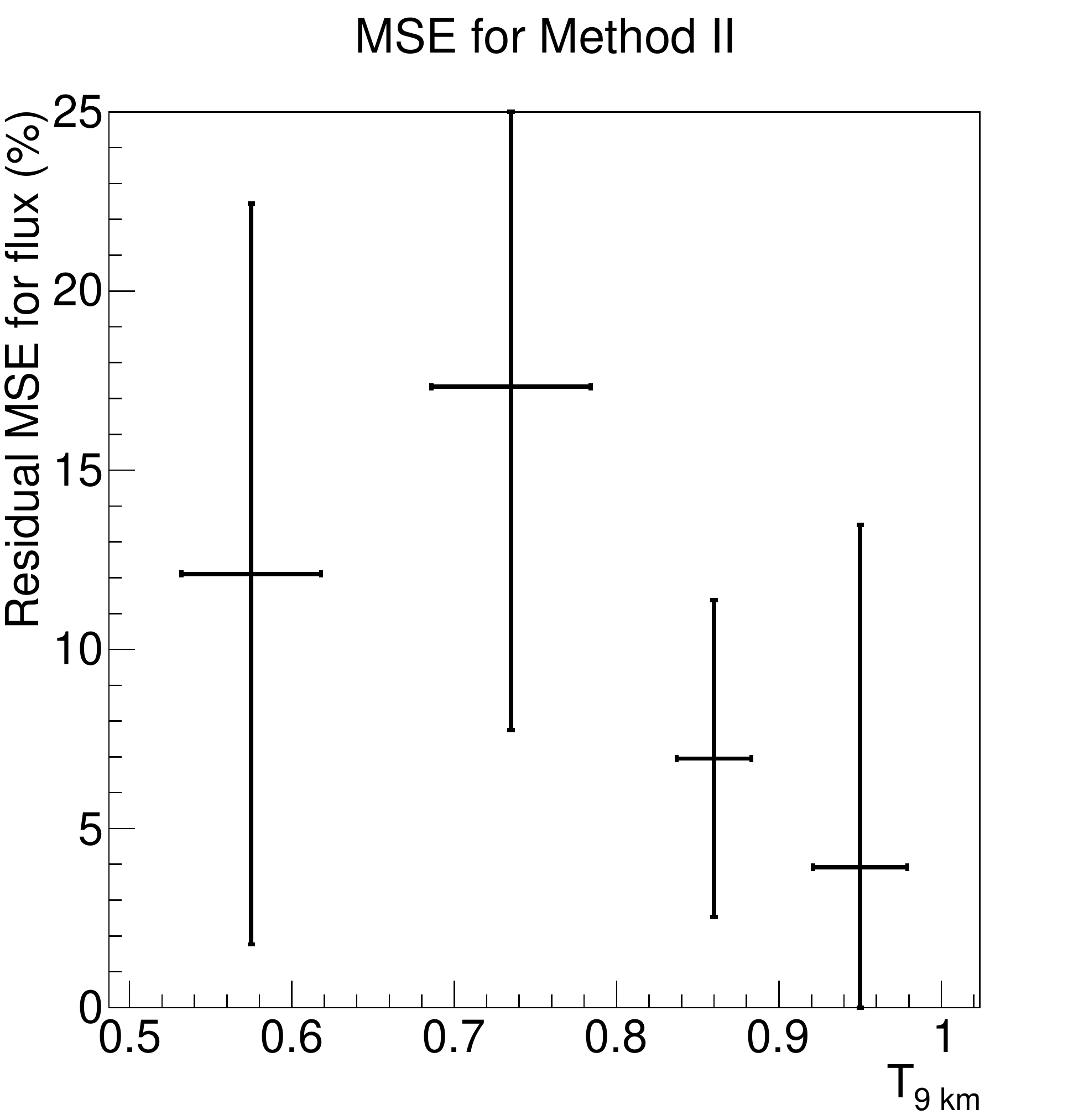}}
    \caption{ 
    Residual flux reconstruction bias after corrections with Method~II.\ Excess uncertainties are attributed to the method (i.e., after quadratically subtracting the second from the fifth column in Table~\protect\ref{tab:uncTbins}) and the mean squared error, as a function of aerosol transmission at 9~km above ground.
    }
    \label{fig:biassyst}
\end{figure}

In an attempt to  experimentally test these uncertainties and quantify their impact, we calculated \revone{the square root of variances in excess of expected purely statistical fluctuations 
so-called} excess fluctuations 
\revone{$\sqrt{\mathrm{Var}(D_{\%})-100/(N-1)\sum_{i=1}^N (\Delta q_i)^2/q_\mathrm{ref}^2}$}
using the data shown in Figs.~\ref{fig:mean_par_f} and~\ref{fig:mean_par_a}, and quadratically subtract from these the 4\% period-wise excess fluctuations on the flux observed in the absence of aerosol-induced atmospheric changes, obtained in Sect.~\ref{sec:reference}. 
The results are presented in Table~\ref{tab:uncTbins}, together with residual biases for Method~II (Method~I can be considered basically bias-free). 
The second column of Table~\ref{tab:uncTbins} presents a prediction for the uncertainty of $\Delta \overline{T}_\mathrm{aer} / \overline{T}_\mathrm{aer}$ (obtained from Table~\ref{tab:uncertainties}), which can be compared to the excess fluctuations 
in the fourth column, observed for the flux reconstruction.  These numbers have been processed into Fig.~\ref{fig:biassyst}, where also the mean squared error is shown.  
We find that, within the available precision, we must attribute $\lesssim$17\% residual mean squared error to the assumptions made in Method~II. For Method~I,  we assume about $\lesssim$15\%, since the bias is much smaller here. 
Unfortunately, the MAGIC collaboration decided to move from Method~I to Method~II as the officially supported LIDAR correction method in 2018. 

Finally, the 6$^\mathrm{th}$ column of Table~\ref{tab:uncTbins} shows predictions for excess uncertainties on the spectral index, obtained from the additional fluctuations on $\overline{T}_\mathrm{aer}$ (see the last rows of Table~\ref{tab:uncertainties}), which increase energy resolution. Observed excess fluctuations for the estimated spectral index $\alpha$ are shown in the last column of Table~\ref{tab:uncTbins}. Here, we observe marginal excess fluctuations at the level of $\sim$1\,$\sigma$ only, comptable with predictions. 


The residual uncertainties on the flux reconstruction may be attributed to changes of the \revone{Hillas parameters of events taken under suboptimal conditions, which do not necessarily resemble parameters of events of accordingly down-scaled energy and hence influence effective areas in a different way than the assumed scaling of energy at which the effective area is evaluated}. \citet{sobczynska_influence_2014} found that the shapes of the (size-independent) {scaled} width and length \revone{parameter distributions} are only marginally distorted by clouds at different altitudes, even if these are totally opaque (see their Fig.~5).
This finding supports our assumptions of effective areas \revone{as a function of energy} from cloud-affected data \revone{resembling} those from clear nights after rescaling the energy accordingly. On the other hand, \citet[][see their Fig.~6]{garrido2013} demonstrated that this assumption breaks down below the energy threshold of an IACT. We expect hence a stronger effect on the \revone{spectral index} at high zenith angles, where the energy threshold moves up.  Such an effect can be seen, at least for Method~I, in  Fig.~\ref{fig:rel_diff}, where increasing observation zenith angles seem to "flip" the relative difference of the reconstructed spectral index from positive to negative values. On the contrary, Method~II seems to be unaffected by this effect, however shows anyhow also much larger discrepancies in the reconstructed flux. 

On the other hand, the assumptions made for each method -- the scaling of energy with inverse average transmission of emitted Cherenkov light (for both methods) and  the scaling of event rate with effective area in Method~I and the common shift of effective area for all events ending up in a same $\overline{T}_\mathrm{aer}$ bin -- may have introduced parts of the additional systematics. 
Although the assumption of energy scaling seems to be well supported by simulations~\citep[see, e.g.,~Fig.~2 of][]{doro2014}, the other assumptions inherent to the two different methods have not been tested by simulations so far, at least to the knowledge of the authors. This entails that any future improvement of the atmospheric correction method shall pay particular attention to improving the model for the modified IRFs, above all the effective area~\citep[one possible alternative has been proposed in][]{Gaug:2017Atmo}.

\subsection{Possible improvements of the method}
\revone{One of the key aspects of reaching smaller systematic uncertainties of future IACTs, in particular the CTA~\citep{cta,ctaconcept}, consists of improving corrections of IACT data taken under nonoptimal atmospheric conditions~\citep{Gaug:2017Atmo}.}

If LIDARs are to be used for such an endeavor, we have shown that the main uncertainty comes from the data correction method itself: namely the air-shower maximum reconstruction and the correction model used for rescaling of energy and effective area. 
This entails the need for time-interval-wise simulations matching the observed aerosol extinction profile with that encountered by the LIDAR and simulating the full interaction of all gamma-ray showers with the observed clouds.  The correction of data taken during dust intrusions into the  ground-layer, on the contrary, are much less affected by systematic and can be corrected in an easier and more reliable way ~\citep[as already observed by][]{dorner2009}. 

\subsection{Comparison with other methods from literature}

An earlier method for correcting atmospherically impaired data taken by the MAGIC telescopes was presented by~\citet{dorner2009}. Their method relies on an estimate of the integrated transmission on a daily basis after comparing distributions of the Hillas parameter "size" (i.e., the total amount of light in a shower image) of clear nights with those from nights affected by {calima}. The integrated transmission is used to upscale photo-electron counts obtained during the affected night, and as such an energy correction is achieved. The method is hence equivalent to using Eq.~\ref{eq:Ebiascorr} for air showers located always entirely above the \revone{scattering} ground layer. The method can only be applied for ground-layer aerosols stable on longer timescales. 

\citet{nolan2010} used a ceilometer, operated at 905~nm at the H.E.S.S. \revone{telescopes} site, to derive the distribution of ground-layer dust and particularly its maximum height. 
Aerosol conditions were simulated for three standard desert-like aerosol models (of different height profiles, but matching \revone{aerosol optical depths}) \revone{using the MODTRAN software}~\citep{modtran4:1996} and gamma-rays incident at eight zenith angles leading to a set of look-up tables (LUTs)        for the H.E.S.S. IRFs. Data affected by cirrus had been previously removed from the data, instead of getting corrected. Finally, the cosmic-ray background rates were used to pick a corresponding LUT entry. The accuracy of the method was tested with a sample of Crab Nebula data leading to a difference of $(14 \pm 5)$\% for the flux normalization at 1~TeV,  probably due to the  important systematic uncertainty on the color correction for their case. This method correctly takes into account the response of Cherenkov light to ground layers located entirely below its emission height, but cannot be used for higher layer heights and particularly not for data affected by clouds.  
A similar approach has been pursued by~\citet{connolly:phd} for changes of the ground layer at the VERITAS array \revone{site}, claiming 6-7\% systematic uncertainty on the energy correction, dominated by large color correction from infrared to UV wavelengths. 

Since 2006, the H.E.S.S. collaboration is operating an elastic LIDAR located 850~m from the telescopes measuring at 355~nm and 532~nm~\citep{Bregeon:2016}. 
 The LIDAR samples the atmosphere with a fixed vertical direction from 0.5~km above ground before each observation run,  the obtained extinction profiles are used to create individual {IRFs on a run-wise basis}~\citep{Devin:2019}. As in~\citet{nolan2010}, data affected by cirrus had been previously excluded from the test data sample, and only the temporal evolution of the ground-layer extinction profiles are getting corrected.
Excess  fluctuations ranging from 10\% to 18\% have been found (e.g., Fig.~5.14 of \citealt{Devin:2018a} and Fig. 5 of \citealt{Devin:2019}), which is compatible with uncertainties due to the overlap function correction and the assumed LRs. 


\citet{Sobczynska:2020} use MC simulations to study the impact of 0.5~km thick clouds at various altitudes on the energy bias and effective area of very high energy gamma rays observed by the small-sized telescopes of the CTA. The presented energy and effective area correction algorithms resemble Eqs.~\ref{eq:Ebiascorr} and~\ref{eq:aeffcorr}, though no time variability was simulated.   
Nevertheless, effective collection areas get systematically under-estimated by $\sim$15\% for $T_\mathrm{cloud} \sim 0.6$, compatible with the flux reconstruction bias observed in Table~\ref{tab:uncTbins} and Fig.~\ref{fig:biassyst}.

\subsection{Application to sources with different spectra or varying fluxes} 

\revone{This study has been carried out uniquely with Crab Nebula data, and one may ask to what extent the obtained results on correction accuracy are applicable to less hard spectra or varying sources, such as gamma-ray bursts~\citep{2019Natur.575..455M}. Approaching this question is complex and depends on the nature of the  atmospheric phenomena affecting IACT data: systematic uncertainties in the case of ground-layer aerosol enhancements only can be considered, in first order, independent of the respective spectral shape parameters, since for low zenith angle observations, the scattering layer is normally found below the entire emission region of Cherenkov light. In that case, aerosol scattering probabilities are the same for all Cherenkov photons irrespective of the shower height, and $\overline{T}_\mathrm{aer}$ becomes a constant independent of shower energy~\citep[see also Fig.~7 of~][ $\vartheta=20^\circ$ case]{Holch:2022}. Only at large observation zenith angles and aerosol optical depths $\gtrsim 0.3$, the TeV gamma-ray energy range gets affected differently from the GeV range~\citep[$\vartheta=50^\circ$ case of][]{Holch:2022}, and hence hard sources affected differently by corrections than soft ones. Nevertheless, $\overline{T}_\mathrm{aer}$ should, by construction, correct for this dependence with the corresponding systematic uncertainties described in Table~\ref{tab:uncertainties}. In this case, no difference is expected in systematic uncertainties between stable and fast varying sources since the aerosol conditions are themselves stable. }

\revone{The situation is different for the case of cirrus: here the dominating systematics (mismatches between characterized and affected FoVs, rapid cloud movement during exposure) can strongly affect varying sources, if the timescale of gamma-ray flux variability is shorter than the variability timescale of the cloud cover. In that case, it becomes impossible to disentangle atmospheric effects from intrinsic fluctuations of the source emission and residual systematics. Their effect on reconstructed light curves and spectra have to be studied then on a case-by-case basis, involving detailed simulations.  On the contrary, if the gamma-ray sources are stable, we can assume that the reconstruction of the low energy part of the spectrum gets corrected through larger values of $\overline{T}_\mathrm{aer}$ and consequently affected by larger systematics (as can already be discerned, e.g., in the difference between Method~I and~II close to the energy thresholds in Fig.~\ref{fig:rel_diff}). Such a scenario will lead to larger systematic uncertainties on the reconstructed spectral index of soft sources than hard ones, though a quantitative estimate cannot be given here if the exact height of the cloud layer is variable or unknown a priori.  }

\section{\revone{Summary}}
An elastic LIDAR serves as the main atmospheric monitoring instrument for the MAGIC telescopes. In addition to classifying the respective nocturnal aerosol conditions (and applying corresponding data quality cuts), we have shown that the obtained aerosol extinction profiles can be better used to correct IACT science data affected by enhancements in the aerosol ground layer and by clouds. 

 Sufficient coverage with LIDAR profiles is achieved by probing the atmosphere every four minutes, in a direction offset by 5$^\circ$  from the pointing of the MAGIC telescopes. The derived extinction profiles at 532~nm are then corrected upward to their equivalent at
 the average wavelength of detected Cherenkov light -- 400~nm in our case. Corresponding  mean \AA ngstr\"om exponents were derived from Sun-photometer data located at a distance of $\sim$400~m separately for calima and non-dusty periods. Clouds have been assumed to exhibit gray extinction always. 
 The aerosol extinction profiles were converted into aerosol transmission profiles and interpolated in time.
 
Our method for the energy reconstruction correction of individual gamma-ray showers calculates an aerosol transmission, $\overline{T}_\mathrm{aer}$, encountered by the observable part of  emitted Cherenkov light of each individual gamma-ray shower. The height of the shower maximum, obtained from standard IACT stereo parameter reconstruction, is used to center a Gaussian light emission profile \revone{model} of energy- and zenith-angle-dependent width. That emission profile is then folded into the wavelength-corrected LIDAR aerosol transmission profile on an event-by-event basis. 

For the correction of effective collection areas and energy-redistribution functions, we have assumed that gamma-ray shower images, observed under a given aerosol transmission, resemble those from showers of correspondingly lower energy observed under optimum  conditions. Since the aerosol transmission, which scales the energy correction, has become event-wise, the two IRFs are now calculated on an event-by-event basis as well. 
Two methods for recombining them again into a global average have been presented. The average depends only on the energy and zenith angle and combines the individual time spans spent under a given variable atmosphere.

Method~I, implemented in 2013, assumes that the effective area at a given energy can be assumed to be proportional to the rate of events of that energy.  This  correctly takes the relative occurrence of energy-wise corrections into account but leads mathematically to averaging inverse effective areas. Method~II was introduced in 2018 and has become the standard method supported by the MAGIC collaboration. It calculates average transmission corrections, weighted by the elapsed time spent under a particular observation condition. This method does not depend on any assumption about the scaling of event rates with effective area. It neglects, however, the possible distribution of effective area shifts under a given atmospheric condition, which is relevant for the case of clouds crossing the FoV of the telescopes. 

Almost seven years of Crab Nebula observations with the MAGIC telescopes have been used to assess the performance of the proposed LIDAR corrections for different zenith angles and measured aerosol transmission at 9\,km ($T_{9 \, \text{km}})$ above 0.5. The sample consists of data affected by dust intrusions (calima) and clouds; about half the sample is affected by the latter. We compared unfolded gamma-ray fluxes in several energy ranges with those obtained from data taken under optimum atmospheric conditions, both for uncorrected data and data corrected with the now standard Method~II.
Furthermore, the detailed spectral shapes of the reconstructed SEDs were compared for both correction Methods~I and~II. 

We find that for $T_{9 \, \text{km}}\!>\,$0.9, LIDAR corrections do not improve the reconstructed fluxes. 
Under moderately degraded conditions, $0.65<T_{9 \, \text{km}}<0.9$, both correction methods work very well, independently of the zenith angle.
The usage of LIDAR corrections is therefore recommended, though a systematic uncertainty of 10\%--18\% needs to be attributed to the correction \revone{of the flux}. Interestingly, this is in the same ballpark as the systematic uncertainties for the flux normalization claimed by the MAGIC Collaboration for optimum atmospheric conditions~\citep{aleksic:2016}. 

Data taken under highly degraded conditions ($0.5<T_{9 \, \text{km}}<0.65$) can be almost entirely corrected, 
although a residual bias of about 10-15\% persists for Method~II. Only in this transmission bin does Method~I clearly outperform Method~II.
Depending on the scientific goal of the analysis, data in the lower transmission bins might still be usable under the assumption of increased systematic uncertainties.

The correction of data with even worse conditions, $T_{9 \, \text{km}}\!\lesssim\!0.5,$ has not been considered because of the lack of such data. Imaging Atmospheric Cherenkov Telescope normally do not collect data under such bad atmospheric conditions.


Finally, we discuss the systematic uncertainties and limitations of the presented LIDAR corrections. 
We find that systematics on $(1-\overline{T}_\mathrm{aer})$ are much higher for clouds ($\sim$30\%) than for ground-layer dust intrusions ($\sim$8\%). The former can be attributed to the variable nature of clouds moving in and out of the (slightly displaced) FoVs of the MAGIC telescopes and the LIDAR, and the complications arising from a Cherenkov light emission profile crossed in the middle by a cloud layer. Moreover, an average \revone{flux} reconstruction bias of \revone{$\sim\! 24\%\cdot(T_{9\,\text{km}}-1)$} has been observed  for Method~II, though the combined overall spectra show an even higher negative bias. Method~I is almost bias-free. While these values are roughly compatible with other performance evaluations found in the literature, this is the first time that IACT data corrections for the case of clouds have been assessed using real data. 

Major improvements of the method can be achieved in the future by following and characterizing the exact FoV observed by an IACT array (e.g., with a stellar photometer). Additionally, replacing data-based corrections with time-interval-wise simulations of IRFs with observation times grouped into those of comparable aerosol extinction profiles can provide further improvement. 

\section*{Acknowledgements}

This work would have been impossible without the support of our colleagues from the MAGIC collaboration. We are especially grateful to the many shifters who have helped to debug and improve the LIDAR system. We would also like to thank the Instituto de Astrof\'{\i}sica de Canarias for the excellent working conditions at the Observatorio del Roque de los Muchachos in La Palma. We thank the PI \'{A}frica Barreto for her effort in establishing and maintaining the AERONET station at the ORM site. \revone{We thank the anonymous journal referee for the detailed and valuable comments which have helped to improve the paper.}
The financial support of the Spanish grant PID2019-107847RB-C42, funded by MCIN/AEI/~10.13039/501100011033, the German BMBF and MPG, and by the Croatian Science Foundation (HrZZ) Project IP-2016-06-9782 and the University of Rijeka Project uniri-prirod-18-48 is gratefully acknowledged.

\bibpunct{(}{)}{;}{a}{}{,} 
\bibliographystyle{aa}
\bibliography{main}

\begin{appendix}

\section{Unfolding the energy spectra}
\label{sec:unfolding}
In order to unfold the obtained spectra, we modify the approach outlined in~\cite{unfolding}.

 We assume that the reconstructed and corrected energy, $E_\mathrm{corr}$, has a non-negligible spread around the true energy of the gamma-ray, $E_\mathrm{true}$.
 We further assume that events, which would have been reconstructed with the corrected energy $E_\mathrm{corr}$, have migrated to lower reconstructed energies $E_\mathrm{rec}$, due to the non-simulated excess aerosol extinction. The event-by-event fluctuations are largely dominated by the physical shower fluctuations, and fluctuations during the absorption process of light by aerosols play a minor role, at least if a typical lower cut on the number of photo-electrons in the shower image is made.

Following the notation of~\citet{unfolding}, we introduce the following vectors: 

\begin{description}
  \item[$Y_i$], the number of events in bin $i$ of $E_\mathrm{rec}$ ($i = 1\ldots  n_a$),
and    \item[$S_k$], the number of events in bin $k$ of $E_\mathrm{true}$ ($k = 1\ldots  n_b$),
\end{description}
\noindent
where the first vector, $\vec{Y}$, denotes the analysis outcome, without excess aerosol energy correction, and the second vector, $\vec{S}$, the outcome of the unfolding procedure (i.e., an estimator for the true energy spectrum) 

The energy migration matrix consists now of two elements, namely 
\begin{description}
  \item[$M_\textit{ij}$], which describes the migration of events from $E_\mathrm{true}$ to $E_\mathrm{corr}$ ($i = 1\ldots  n_a$, $j = 1\ldots n_b$) and
  \item[$H_\textit{jk}$], which describes the migration of events from $E_\mathrm{corr}$ to $E_\mathrm{rec}$ ($j = 1\ldots n_b$, $k = 1\ldots  n_b$).
\end{description}

 Here, the first matrix $\boldsymbol{M}$ is identical to the one use in~\citet{unfolding} and is obtained from simulations with the standard atmosphere and no excess aerosol. The matrix $\boldsymbol{H}$ is new and describes the probability that an event with reconstructed energy $E_\mathrm{corr}$ has further migrated to $E_\mathrm{rec}$ due to the excess aerosol and must be obtained from real events: 
\begin{eqnarray}
  H_\textit{jk} &=& \frac{N(E_\mathrm{rec}^k,E_\mathrm{corr}^j)}{\sum_{k=1}^{n_b} N(E_\mathrm{rec}^k,E_\mathrm{corr}^i)} \\
  \Delta H_\textit{jk} &=& \sqrt{ H_\textit{jk} \cdot (1 - H_\textit{jk} ) \cdot \frac{1}{\sum_{j=1}^{n_b} N(E_\mathrm{rec}^j,E_\mathrm{corr}^k)}  }
.\end{eqnarray}
\noindent
That matrix now contains the full information about the statistical occurrence of excess aerosol-induced energy migration. Both matrices, $M_\textit{ij}$ and $H_\textit{ki}$, are normalized:
\begin{equation}
  \sum_{i=1}^{n_a} M_\textit{ij} = \sum_{j=1}^{n_b} H_\textit{jk} = 1 \quad.
\end{equation}
\noindent
We further define 
\begin{equation}
  M_\textit{ik}'  = \sum_{j=1}^{n_b} M_\textit{ij} \cdot H_\textit{jk } \quad,
\end{equation}
\noindent
whose elements describe the fraction of events in bin $i$ of true energy that move first down to bin $j$ in true energy, due to additional aerosol extinction, and then further to bin $k$, due to the finite experimental energy resolution. The definition of the migration matrix in this way assumes that the energy resolution of a shower whose Cherenkov light has been previously absorbed, is similar to the one of a shower of lower energy, observed during clear nights and which yields the same image size:
\begin{equation}
  Y_i = \sum_{k=1}^{n_b} M_\textit{ik}' \cdot S_k 
.\end{equation}
\noindent
The different unfolding methods~\citep[e.g.,][or forward unfolding methods]{tikhonov,schmelling1994,Bertero} to retrieve $S_k$, based on the knowledge of $Y_i$ and $M_\textit{ik}'$ can then be carried out along the lines described in~\citet{unfolding}, by simply replacing $M$ by $M'$. 

Eq.~\ref{eq:aeff} is then adapted to the case of true energy, namely

\begin{equation}
  \avg{A_{i}} = \frac{N_i}{ \sum_{j=0}^{N_i}{\frac{1}{A_{i-\delta_j}}} } =  \frac{1}{ \sum_{j=0}^{n_b} H_\textit{ij} / A_j}  ~ .   
  \label{eq:aeffH}
\end{equation}

The bin-wise collection area defined in
Eq.~\ref{eq:aeffH} can also be used for re-weighting the effective collection area for different assumed energy spectra and/or further dependences on external parameters, like the zenith angle, \citep[see Eqs.~28 and 31 of][]{unfolding}.

\section{Example reference spectra}
\label{sec:unfolding}
\begin{figure}[h!]
\centering
\resizebox{0.88\hsize}{!}{\includegraphics{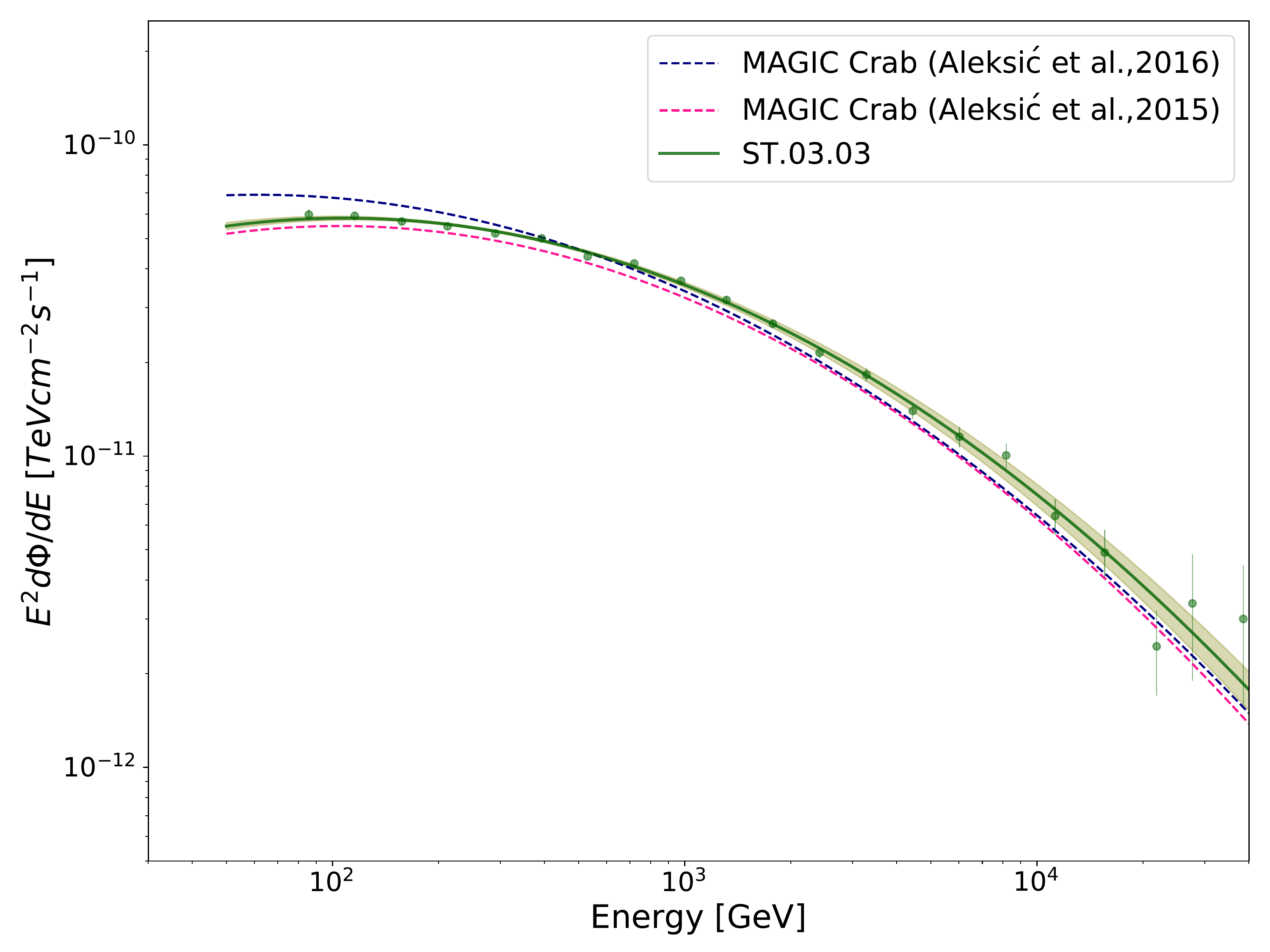}}
\resizebox{0.88\hsize}{!}{\includegraphics{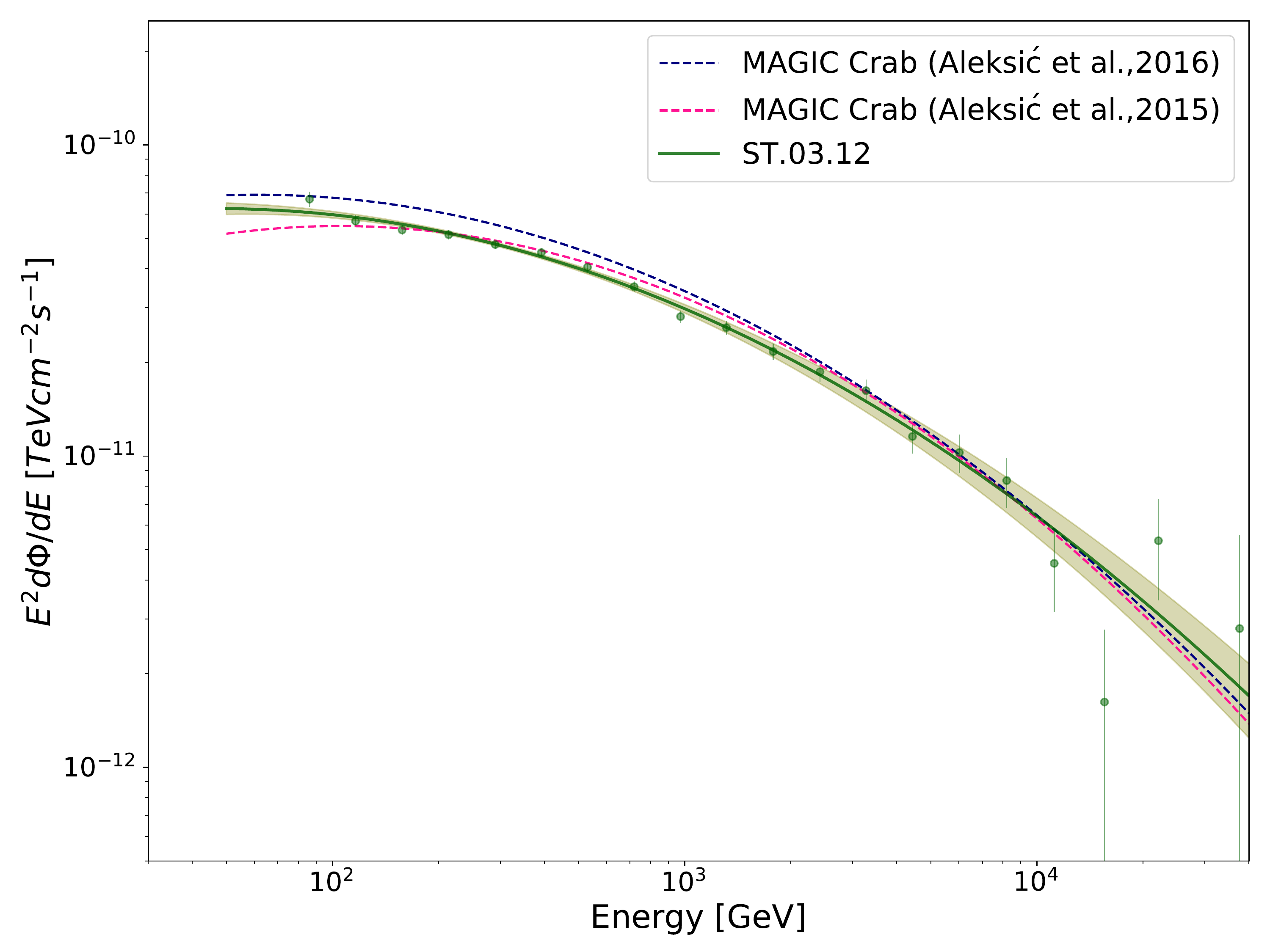}}
\caption{Example reference Crab Nebula spectra from the first (ST.03.03; top) and last (ST.03.12; bottom) analysis period used in this study, in comparison with two spectra published by the MAGIC collaboration~(\protect\citealt{aleksic:2016} and~\protect\citealt{aleksic:2015}). \revtwo{The shaded bands show the one sigma uncertainty band of the spectral fits.}}
\label{fig:example_ref}
\end{figure}


\end{appendix}

\end{document}